# A Comprehensive Survey of the Tactile Internet: State-of-the-art and Research Directions

Nattakorn Promwongsa, Amin Ebrahimzadeh, Diala Naboulsi, Somayeh Kianpisheh, Fatna Belqasmi, *Member* IEEE, Roch Glitho, *Senior Member* IEEE, Noel Crespi, *Senior Member* IEEE, and Omar Alfandi

*Abstract* — The Internet has made several giant leaps over the years, from a fixed to a mobile Internet, then to the Internet of Things, and now to a Tactile Internet. The Tactile Internet goes far beyond data, audio and video delivery over fixed and mobile networks, and even beyond allowing communication and collaboration among things. It is expected to enable haptic communication and allow skill set delivery over networks. Some examples of potential applications are tele-surgery, vehicle fleets, augmented reality and industrial process automation. Several papers already cover many of the Tactile Internet-related concepts and technologies, such as haptic codecs, applications, and supporting technologies. However, none of them offers a comprehensive survey of the Tactile Internet, including its architectures and algorithms. Furthermore, none of them provides a systematic and critical review of the existing solutions. To address these lacunae, we provide a comprehensive survey of the architectures and algorithms proposed to date for the Tactile Internet. In addition, we critically review them using a well-defined set of requirements and discuss some of the lessons learned as well as the most promising research directions.

**NOMENCLATURE**

| | |
|---|---|
| AI | Artificial intelligence |
| ANN | Artificial neural network |
| AR | Augmented reality |
| AWGN | Additive white Gaussian noise |
| BANN | Body area nano network |
| BER | Bit error rate |
| BS | Base station |
| CO | Central office |
| CRAN | Cloud-based radio access network |
| CSI | Channel state information |
| D2D | Device-to-device |
| DCF | Distributed coordination function |
| DCT | Discrete cosine transform |
| DNA | Dynamic network architecture |
| DoF | Degree of freedom |
| DPCM | Different pulse-code modulation |
| EDCA | Enhanced distributed channel access |
| eMBB | Enhanced mobile broadband |
| EPON | Ethernet passive optical network |
| FDD | Frequency division duplex |
| Fi-Wi | Fiber-wireless |
| FOLP | First-order linear predictor |
| H2H | Human-to-human |
| H2M/R | Human-to-machine/robot |
| HCCA | Hybrid coordination function-controlled channel access |
| HetNet | Heterogeneous network |
| HITL | Human-in-the-loop |
| HMM | Hidden Markov model |
| ILP | Integer linear programing |
| IoT | Internet of thing |
| IR-HARQ | Incremental redundancy-hybrid automatic repeat request |
| JND | Just-noticeable difference |
| LTE | Long-term evolution |
| LTE-A | Long-term evolution-advanced |
| LTE-U | Long-term evolution in an unlicensed band |
| M2M | Machine-to-Machine |
| MAP | Mesh access point |
| MEC | Mobile edge computing |
| MIMO | Multiple-input multiple-output |
| MISO | Multiple-input single-output |
| ML | Machine learning |
| MLP | Multi-layer perceptron |
| MMT | Model-meditated teleoperation |
| mMTC | Massive machine-type communication |
| MPP | Mesh portal point |
| MPR | Multiple-plane routing |
| NFV | Network function virtualization |
| NG-EPON | Next-generation Ethernet passive optical network |
| NOMA | Non-orthogonal multiple access |
| OFDMA | Orthogonal frequency division multiple access |
| OLT | Optical line terminal |
| OMA | Orthogonal multiple access |
| ONU | Optical network unit |
| OSPF | Open shortest path first |
| PCF | Point coordination function |
| PD | Perceptual deadband |
| PON | Passive optical network |
| QoE | Quality of experience |
| QoS | Quality of service |
| QSI | Queue state information |
| RAN | Radio access network |
| RB | Resource block |
| RLNC | random linear network coding |

N. Promwongsa A. Ebrahimzadeh and S. Kianpisheh are with CIISE, Concordia University, Montreal, QC H3G 1M8, Canada (e-mail: n_promwo, a_ebrahimzaleh, s_kianpi @encs.concordia.ca).

D. Naboulsi is with the École de Technologie Supérieure, Montreal, QC H3C 1K3, Canada (e-mail: diala.naboulsi@lacime.etsmtl.ca).

R. Glitho is with CIISE, Concordia University, Montreal QC H3G 1M8, Canada, and also with the Computer Sciene Programme, University of Western Cape, Bellville 7535, South Africa (e-mail: glitho@ece.concordia.ca).

N. Crespi is with Institut Mines-Telecom, Telecom SudParis, Evry, France (e-mail: noel.crespi@it-sudparis.eu).

F. Belqasmi and O. Alfandi are with Zayed University, Abu Dhabi 144534, United Arab Emirates (e-mail: {fatna.belqasmi, omar.alfandi}@zu.ac.ae).

| | |
|---|---|
| SCMA | Sparse code multiple access |
| SCTP | Stream Control Transmission Protocol |
| SDN | Software defined network |
| SER | Symbol error rate |
| SPS | Semi-persistent scheduling |
| TCP | Transmission Control Protocol |
| TDD | Time division duplex |
| TDPA | Time-domain passivity approach |
| TTI | Transmission time interval |
| UDP | User Datagram Protocol |
| UE | User equipment |
| URLLC | Ultra-reliable and low-latency communication |
| V2I | Vehicle-to-infrastructure |
| V2V | Vehicle-to-vehicle |
| V2X | Vehicle-to-everything |
| VNF | Virtual network function |
| VR | Virtual Reality |
| WBAN | Wireless body area network |
| WLAN | Wireless local area network |
| ZOH | Zero-order hold |

I. INTRODUCTION

TODAY'S Internet is ushering in a new era of communications, where interactive cyber-physical systems can exchange not only the conventional triple-play data (i.e., audio, video, and text), but also haptic control messages in real-time through the so-called Tactile Internet. The Tactile Internet is a natural evolution of the Internet, which went from a fixed and text-based Internet to a multimedia mobile Internet, and then to an Internet of Things (IoT). The Tactile Internet is geared towards haptic communication over the network. Using the Tactile Internet, a physician may be able to perform a remote surgical operation on a distant patient by perceiving not only real-time visual, auditory, but also touching senses of the distant environment. These new possibilities will revolutionize the set of applications and services provided to date by the Internet and take the next-generation systems to an unprecedented level of human-like communication. The Tactile Internet will revolutionize and combine machine-to-machine and human-to-machine interactions [1].

The term Tactile Internet was first coined in 2014 by Fettweis in his seminal paper [2]. Fettweis defined the Tactile Internet as the enabling technology for control and steering of real and/or virtual objects through the Internet by requiring a very low round-trip latency [2]. Not long afterwards, in March 2016, the IEEE P1918.1 standards and working group was formed, which aims to define the reference architecture of the Tactile Internet framework [3]. The IEEE P1918.1 standard working group defines the Tactile Internet as "*A network, or a network of networks, for remotely accessing, perceiving, manipulating, or controlling real and virtual objects or processes in perceived real-time*". Some of the key use cases mentioned in IEEE P1918.1 are remote robotic surgery, autonomous driving, and haptic-enabled virtual reality.

Tactile Internet applications, such as tele-surgery mostly require ultra-low latency and a high level of reliability and security to function correctly and safely. Latency requirements of Tactile Internet applications may vary, depending on the type of application and dynamicity of the environment. To be more specific, while the latency requirements of Tactile Internet applications may vary from <10 ms up to tens of milliseconds, the Tactile Internet targets an ultra-low end-to-end round-trip latency of 1 ms. This requires keeping tactile applications relatively close to the applications' end points (e.g., keeping the robotic surgery application relatively close to the surgeon's console and to that of the robot performing the surgery) [4]. Considering the best case scenario, where the propagation delay imposed by the limited speed of light (300 km/ms) is the only limiting factor, the distance between the end-points of any given tactile application is upper-bounded by 150 km to ensure that the 1 ms round-trip latency requirement of the Tactile Internet is met. In real-world scenarios, however, this upper-bound becomes even smaller, as the queueing delay, channel access delay, and transmission delay, among others, may also contribute to the overall latency experienced by the haptic packets. Similar to the latency requirement, reliability requirement of the Tactile Internet can also vary, depending on the given application. It can start from a failure rate of $10^{-3}$ [3][5]. The most critical Tactile Internet applications (e.g., tele-surgery) require a reliability of up to $10^{-7}$ [5].

The following sub-sections introduce the application domains and challenges of the Tactile Internet and discuss the differences between this survey and other surveys and tutorials on this topic. The paper's scope and organization, as well as a reading map are presented in the last subsection.

*A. Application Domains and Challenges*

A Tactile Internet is expected to allow the provisioning of a broad range of new applications that will enrich the set already offered by today's Internet. Interactive real-time haptic applications will comprise the bulk of these, and they will play a pivotal role in enhancing everyday life. Remotely-controlled systems may become key to the functioning of several very important and sensitive domains, such as healthcare (e.g., tele-diagnosis, tele-surgery), industry (e.g., dangerous and difficult-to-reach environments), virtual and augmented reality (e.g., a firefighter training system), road traffic (e.g., automated and cooperative driving), education and serious gaming (e.g., games for personalized cardio-training), and many others [1][6].

Haptic applications bring a set of stringent requirements in terms of latency, reliability, and security. An ultra-low latency of 1 ms is required to ensure the timely delivery of control messages (e.g., a surgeon's hand movement) and avoid cybersickness (e.g., in the case of virtual reality). Cybersickness occurs when multiple senses (audio, video and touch) participate in an interaction but the feedback of the different senses are notably unsynchronized (e.g., the time-lag between tactile and visual movement exceeds 1 ms) [7]. When a Tactile Internet signal includes the synchronization between several senses, this implies a multi-modal signal.

On another note, ultra-high reliability and security are required for several of the applications envisioned for the Tactile Internet. For instance, it is inconceivable that a data communication error (e.g., malicious or network-based) might occur during remote open-heart surgery or when controlling an airplane in flight.

Given the targeted applications and their requirements, the realization of the Tactile Internet clearly brings many challenges. As captured in [8], these challenges can be categorized under four streams: haptics, intelligence, computing and communication. Each of these are briefly discussed below.

**Haptic challenges:** These cover haptic information acquisition, storage, transmission and delivery. Haptic devices capable of sensing and delivering both cutaneous and kinesthetic feedback are needed at both ends [9]. Multimodal encoding schemes must be defined to support the different modalities (i.e., vocal, visual, and haptic) of Tactile Internet information, without increasing the end-to-end latency [10]. Steinbach *et al.* [10] report that visual and auditory information encoding have been studied extensively, but not haptic encoding. Meeting the challenge of haptics will also require customized interfaces to support and distinguish between human-to-machine, machine-to-human and machine-to-machine interactions.

**Intelligence challenges:** These are related to the intelligence required for haptic information prediction (e.g., to predict a surgeon's next move). This covers the actions, movements, and feedback predictions to enhance the latency and reduce communication delays and interruptions. Predictions may fill a critical gap; for instance, if the command signal for the next action is not received on time, due to either the distance (more than 150 km), or a connection problem.

**Computation challenges:** These deal with all the computational aspects of haptic communication, including processing haptic feedback for a real-time interaction 'feel' and running the prediction intelligence program. While computations may be resource- and time-consuming, they should not affect the end-to-end latency requirement.

**Communication challenges:** This category encompasses the requirements for ultra-low latency, ultra-high reliability, a high exchange rate, and the support of the extra communication with the cloud and edge computing components that can be used for computing. A communication infrastructure that meets all these requirements remains to be developed. Security mechanisms and multiplexing schemes combining multiple modalities and respecting the 1ms latency are also needed [7].

### B. Existing Surveys and Tutorials on the Tactile Internet

The Tactile Internet has spurred a great deal of interest in both academia and industry and is gaining more and more momentum. Back in 2014 when the term Tactile Internet was coined, only one paper was published on the topic, and by early 2020, the number has increased to 135. As mentioned earlier, dedicated standards are being developed by the IEEE 1918.1 working group. Reference [3] gives an overview of this standardization effort. The first goal is to produce a baseline standard that includes the terminology, application scenarios, requirements, and even the basic architecture(s). More specific standards, such as haptic codecs, are also under being developed [10].

There are several surveys and tutorials among the numerous papers published so far on the Tactile Internet. However, to the best of our knowledge, none of them comprehensively covers both architectures and algorithms. Furthermore, none of them presents a systematic and critical review of the existing solutions, using clearly defined requirements. Moreover, no survey has dealt with architectures in a comprehensive manner. The same applies to algorithms. There are general tutorials aimed at the topic at large, and there are surveys/tutorials that focus on specific aspects of the topic.

*1) General Tutorials*

Many general tutorials give a general introduction to the Tactile Internet, explaining its key vision, enablers and challenges (e.g. [2]-[4][11]-[15]). For example, reference [2] defines the Tactile Internet, discusses application scenarios and presents the key challenges ahead. Reference [4] presents the Tactile Internet vision and its predicted impact, discusses its similarities and differences with the mobile Internet and the IoT, introduces the related research challenges, and surveys the advances in the enabling technologies (e.g., software-defined networks, cloudlets and edge computing). Reference [11] identifies the research challenges for the Tactile Internet and gives hints about how they might be addressed. As the reader can easily surmise, none of these general tutorials offers the detailed treatment this comprehensive survey sets out to provide.

Some other general tutorials focus on providing a Tactile Internet in a 5G environment and discuss the inherent support provided by such environments (e.g., [6][16][17]). Some of these papers also analyze potential improvements to enhance such support. As an example, Simsek *et al.* in [6] investigate the commonalities of 5G and the Tactile Internet, discuss Tactile Internet applications (e.g., robotics, healthcare) and architectural components (master, slave, and network domains), and overview projected 5G technologies' developments and how they can support the Tactile Internet architecture. A few other papers simply assume the use of 5G; one such example is [18]. It introduces a Tactile Internet architecture, provides a discussion of potential applications, and discusses the open challenges. These tutorials follow the same path as those discussed in the previous paragraph, neglecting to provide detailed treatment of both architectures and algorithms. This comprehensive survey is a clear contrast to such approaches.

A tutorial [19] that focuses on providing a Tactile Internet in beyond 5G era has been published very recently. It introduces a general framework for wireless Tactile Internet, reviews three Tactile Internet domains (i.e., haptic communications, wireless augmented/virtual reality), discusses security/privacy and identifies open research challenges. Although this paper provides more details than all the other general tutorials discussed in the previous paragraphs, it still does not provide a detailed treatment of both architectures and algorithms. For instance, none of the many existing works on architectures is discussed. Moreover, none of the algorithmic works introduced in the paper is critically reviewed. Several existing algorithmic works are also not discussed. Some examples are the algorithms for edge resource allocation and the algorithms for bandwidth allocation in passive optical networks.

*2) Tutorials/surveys that focus on specific aspects*

Many tutorials and surveys are dedicated to haptic communication. A few others deal with other topics. None of these tutorials and surveys targets the Tactile Internet at large as this manuscript does. They are briefly introduced below and the differences between them and this comprehensive survey are elaborated.

**Tutorials and surveys on haptic aspects:** Reference [20] discusses haptic communications, but with no focus on the Tactile Internet. It reviews the state of the art from psychological and technical perspectives. On the other hand, Van Den Berg *et al*. [7] focuses on the challenges in haptic communications over a Tactile Internet. The authors propose some requirements and use them to critically review the existing haptic communications technologies. However, the paper reviews no other aspects of the Tactile Internet. Reference [10] is even more limited in scope, as it focuses solely on haptic codecs for a Tactile Internet.

Reference [21] provides a survey on haptic communications over the 5G Tactile Internet. It introduces the basics of teleoperations and discusses haptic communications over the Internet, data reduction/compression, haptic control system approaches, and haptic communications over 5G. The survey is indeed comprehensive in terms of haptic communications. However, like all tutorials/surveys that do not take a holistic view to the Tactile Internet, it does miss a few key aspects. One example is resource allocation in the Tactile Internet system. A plethora of algorithms have been proposed in the literature. Our survey reviews these algorithms while reference [21] does not. The same applies to the many architectures that have been proposed for Tactile Internet systems.

**Tutorials and surveys on other aspects:** There are tutorials/surveys on many other aspects of the Tactile Internet: communications (with no emphasis on haptics), applications, security and integration with other technologies. Reference [22] reviews the use of a specific radio communication technology, non-orthogonal multiple access (NOMA) for Tactile Internet systems. The review does not go beyond that technology. References [5] and [23] are devoted to low latency and ultra-reliable communications, a critical aspect of Tactile Internet systems. The former provides an in-depth overview of 5G functionality for low latency and ultra-reliability, while the latter gives an overview of the novel physical layer technologies. The treatment in both of these papers fails to go beyond the physical and MAC layers. It should be noted that a few references (e.g., [24]) go beyond the physical and MAC layers in their discussions of low latency. However, these papers still do not go beyond communications issues.

The Tactile Internet applications reviewed so far in the literature include cooperative driving [25], industries at large [26], tactile robots [27] and robotic surgery [28][29]. All of these papers are application-focused and do not review the Tactile Internet at large. A few papers are also solely devoted to security issues (e.g., [30][31]). The integration or use of the following technologies with the Tactile Internet has been discussed in several tutorials: fog computing [32], edge [33], software-defined networks [34], IoT and edge [8], and 5G and edge computing [9]. However, none of these articles provides a general review of the Tactile Internet. Table I provides a summary of these works. The first column lists the categories of the relevant tutorials/surveys, the second column summarizes the essential features of each category, and the third column indicates the weaknesses of each category.

*C. Paper's Scope, Organization, and Reading Map*

This paper provides a comprehensive survey of the existing architecture and algorithm-related solutions for the Tactile Internet. The solutions are critically reviewed using a set of well-defined requirements. In this subsection, the sources of the works reviewed in this survey are discussed first. It should be noted that although the survey aims at the Tactile Internet at large, some works are excluded from its scope. The excluded works are made explicit after the discussion of the sources. This is followed by a classification of the reviewed works. The subsection then ends with the paper organization and reading map.

*1) Sources of the Works Reviewed in this Survey*

In this survey, we review the architectures and algorithms presented in peer-reviewed venues, as well as the architectural solution proposed by the IEEE P1918.1. Tactile Internet standard working group. In the group, no standard algorithm has been put forward so far to the best of our knowledge. It should be also noted that the proposed architectural solution is still not yet available in the public domain. However, its essence is discussed in a peer reviewed paper entitled "The IEEE 1918.1 'Tactile Internet' Standards Working Group and its Standards" published in February 2019 [3]. Our review of this standard solution is therefore based on this third-party source.

The IEEE P.1918.1 standard working group [35] is actually nowadays the only group which has produced concrete specifications on Tactile Internet. The ITU-T has also published a Technology Watch report [1] which introduces the Tactile Internet at a high level, but it has developed no standard so far. Yet another example of body that has produced no concrete standard specification so far is the TACNET 4.0 consortium [36], which aims at providing specification for a standard system. It focuses on Tactile Internet for production, robotics, and digitalization of the industry. In addition, the European Standard Institute (ETSI) has started a work item on IPv6-based Tactile Internet, but the work item has now been stopped [37].

The reader should note that no solution from industry is considered in this survey given that there is no work so far available in the public domain (or adequately described in third-party sources). The Tactile Internet will certainly play a key role in industry as shown by reference [26]. However, the papers originated from the industry itself and available in the public domain are limited to blog papers such as reference [38]. These blog papers generally introduce the Tactile Internet at a very high level, and it is not worth reviewing them in a peer reviewed survey such as this one.

*2) Works Excluded from the Scope of this Survey*

The following aspects of the Tactile Internet (including standards, if any) are excluded from the scope of this survey: haptic challenges, security issues, and performance analysis. Haptic challenges are excluded, as they have already been exhaustively discussed in surveys such as reference [21]. Our focus is therefore on the intelligence, computation, and communication challenges.

While papers (e.g., [39][40]) do focus on the security aspects, security issues are now generally reviewed in dedicated surveys (e.g., [30][31]). As for performance analysis publications, we decided not to include them because they do not put forward any new solutions. Indeed, the main objective of those works is to demonstrate that existing solutions are capable of supporting Tactile Internet applications. For instance, Li *et al.* [17] study the suitability of 5G new radio and long-term evolution (LTE) ultra-reliable and low-latency communications (URLLC) for the Tactile Internet. Adhami *et al.* [41] conduct extensive simulations to investigate the performance of existing WiMAX, 3G and Wi-Fi technologies for the Tactile Internet. Another example is [42], which evaluates the performance of existing real-time video communication solutions for the Tactile Internet.

TABLE I
Comparison of existing surveys and tutorials on the Tactile Internet with our survey

| Category | Essential Features | Weaknesses (Compared to our survey) |
|---|---|---|
| **General tutorials** [2]-[4], [6],[11]-[19] | • Provide a general introduction to the topic (e.g. vision, enabler, key concepts). | • Do not systematically and critically review existing solutions. <br> • Do not cover both architectural and algorithmic aspects. |
| **Tutorials / surveys on specific aspects (haptics)** [7],[10],[20],[21] | • Focus on haptics and might provide very detailed treatment of haptics. | • Do not tackle the Tactile Internet at large. <br> • Focus solely on the haptic communications aspects. |
| **Tutorials / surveys on other aspects than haptics** [5], [8], [9], [22]-[34] | • Focus on aspects other than haptics (e.g. radio communications, applications, integration with other technologies). <br> • Might provide very detailed treatment of that aspect. | • Do not tackle the Tactile Internet at large. <br> • Focus on specific (non-haptics related) aspects |
| **Our survey** | • Provides a comprehensive survey of the architectural and algorithmic solutions proposed so far for the Tactile Internet: <br> - Unlike the general tutorials, it provides a systematic and critical review of existing solutions and covers both architectural and algorithmic solutions. <br> - Unlike the tutorials and surveys which focus on specific aspects, it covers the Tactile Internet at large. | |

*3) Literature Classification*

We have identified a grand total of seventy-five papers to be reviewed in this survey, without taking into consideration the papers that were explicitly excluded in the previous subsection. We have selected the papers that meet any of the two following criteria: (i) the term Tactile Internet is specifically mentioned in the title, (ii) the Tactile Internet is considered explicitly in the paper as the main motivation of the work. These papers are classified into four main categories: (i) architectures, protocols, and intelligent prediction, (ii) radio resource allocation algorithms, (iii) non-radio resource allocation algorithms on lower layers (i.e., PHY and MAC layers), and (iv) non-radio resource allocation algorithms beyond lower layers for the Tactile Internet.

The intelligent prediction covers the works that aim at tackling the intelligence challenges. We include them in the same category as architectures and protocols given that they use machine learning techniques to intelligently predict tactile/haptic information (e.g., control commands and tactile/haptic feedbacks), which is expected from a remote functional entity and not received on time. The reader should note that some of the algorithms presented in this survey do use machine learning techniques for other purposes than predicting the expected tactile/haptic information that is not received on time (not tackle the intelligence challenges). They are reviewed as part of the algorithmic categories to which they belong.

A whole category is devoted to radio resource allocation, mainly because it is the subject of most of the algorithms proposed to date. This survey does not review the many papers (e.g., [43]-[47]) that investigate virtual network function placement for low latency and high reliability, as those papers do not generally have the very stringent low latency and high reliability required by the Tactile Internet as their objectives. Most of them do not even mention the term Tactile Internet.

The proposed taxonomy of the reviewed papers is illustrated in Figure 1. The first category comprises of the following four subcategories: (i) architectures for non-specific applications, (ii) architectures for specific applications, (iii) protocols, and (iv) intelligent predication schemes. In the second category (i.e., radio resource allocation algorithms), four subcategories are identified: (i) uplink transmission, (ii) downlink transmission, (iii) joint uplink and downlink transmission, and (iv) network slicing. The third category (i.e., non-radio resource allocation on lower layers), is further classified into three subcategories: (i) resource allocation in wireless local area networks (WLANs) and wireless body area networks (WBANs), (ii) wavelength and bandwidth allocation in passive optical networks (PONs), and (iii) PHY layer. Finally, the fourth and last category (i.e., non-radio resource allocation beyond lower layers) consists of four subcategories: (i) edge resource allocation, (ii) co-design, (iii) routing, and (iv) switching.

*4) Paper Organization and Reading Map*

Figure 2 shows the structure of the survey. In Section II, after a discussion of the enabling and related technologies for the Tactile Internet, we present three use cases (i.e., remote robotic surgery, autonomous driving, and remote phobia treatments) in detail and then propose our evaluation criteria for a given Tactile Internet system. This section is followed by four sections that conduct a critical review of the solutions proposed to date for the Tactile Internet. Section III focuses on the architectures. It deals with application-agnostic architectures, application specific architectures, as well as the protocols and intelligent prediction schemes. Section IV reviews the radio resource allocation algorithms. Section V surveys the non-radio resource allocation algorithms that focus mainly on the lower layers (i.e., PHY and MAC layers), while Section VI discusses the non-radio resource allocation algorithms that go beyond the lower layers in the protocol stack. The reader should note that summary, insights, and lessons learned are presented at each one of each section.

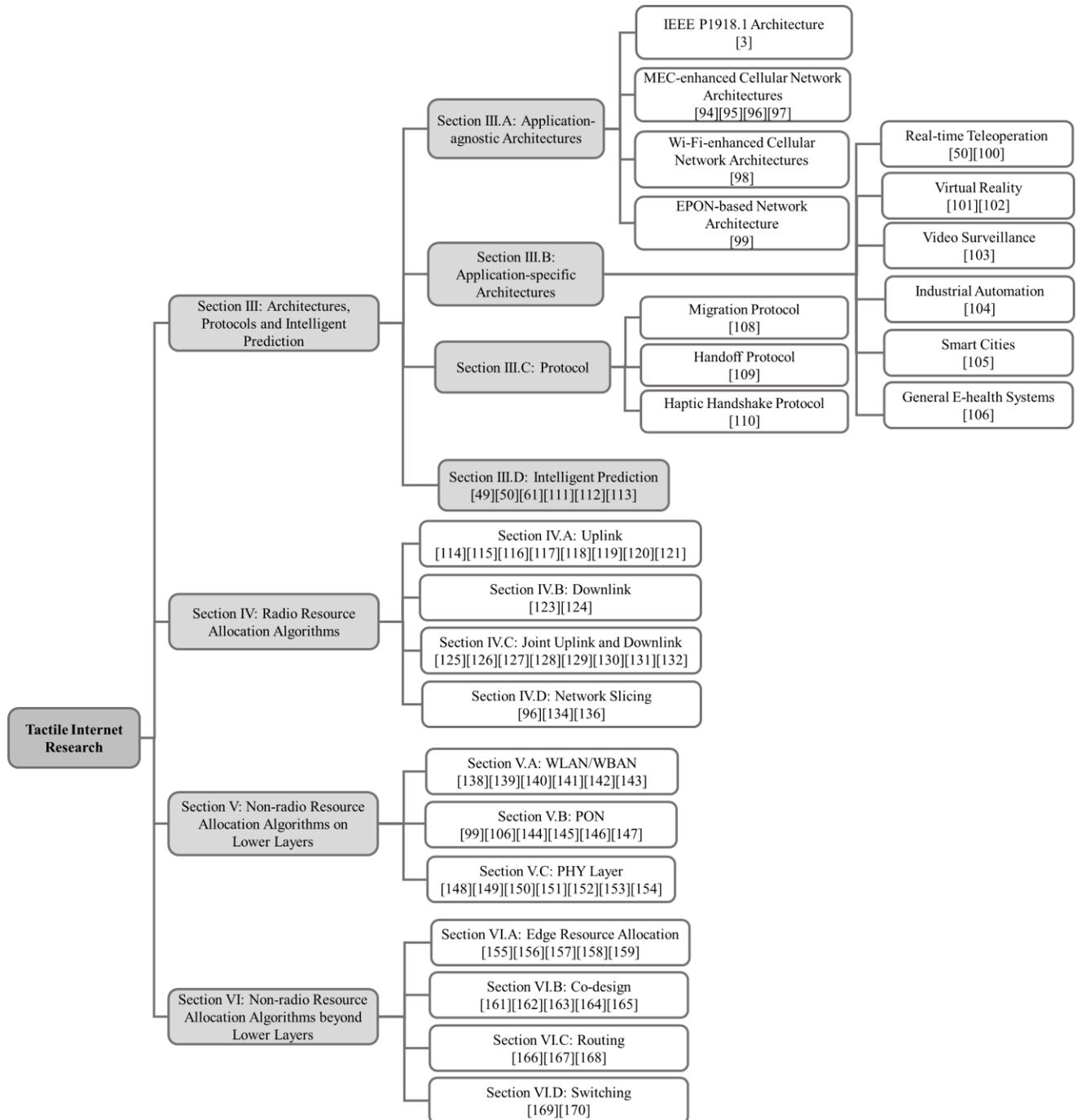

Fig. 1. Overview of the surveyed research works according to the proposed taxonomy.

Section VII then discusses the potential research directions. Finally, we conclude the paper in Section VIII.

This paper targets several types of readers and can be read in an "a la carte" manner. Figure 3 provides a reading map. The readers interested in a general overview of the Tactile Internet could focus solely on Sections I, II, III.E, IV.E, V.D, VI.E and VIII. When it comes to the readers interested in in-depth discussions on architectural and protocol solutions, we suggest their reading on Sections I, II, III, VII.A, VII.B, VII.C and VIII. Finally, as for the readers with a high interest in discussions on the algorithmic aspect, Sections I, II, IV, V, VI, VII.D and VIII will be the most appropriate.

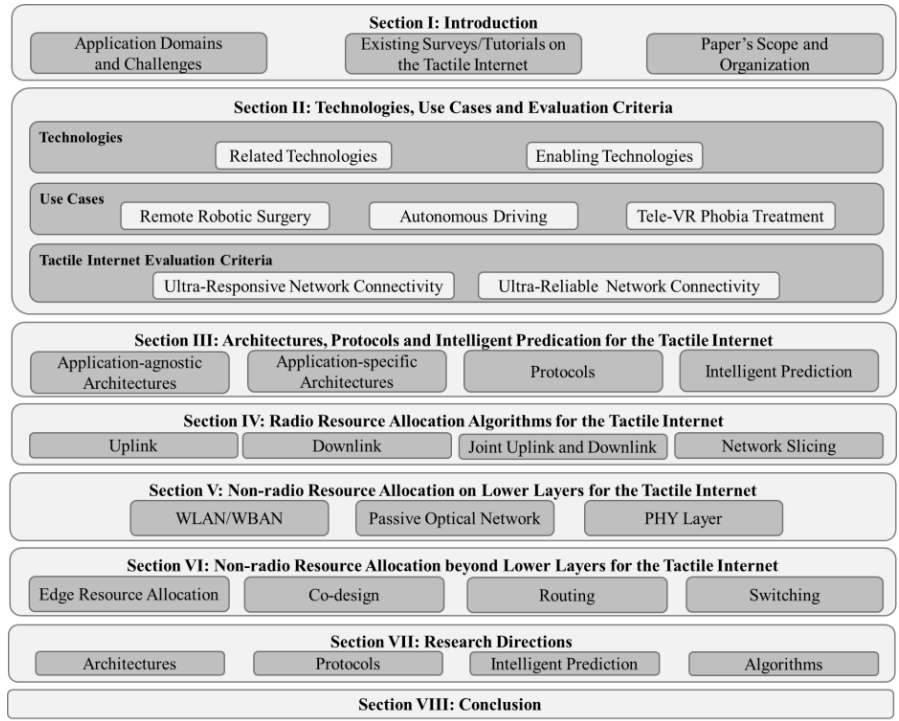

Fig. 2. The structure of the survey

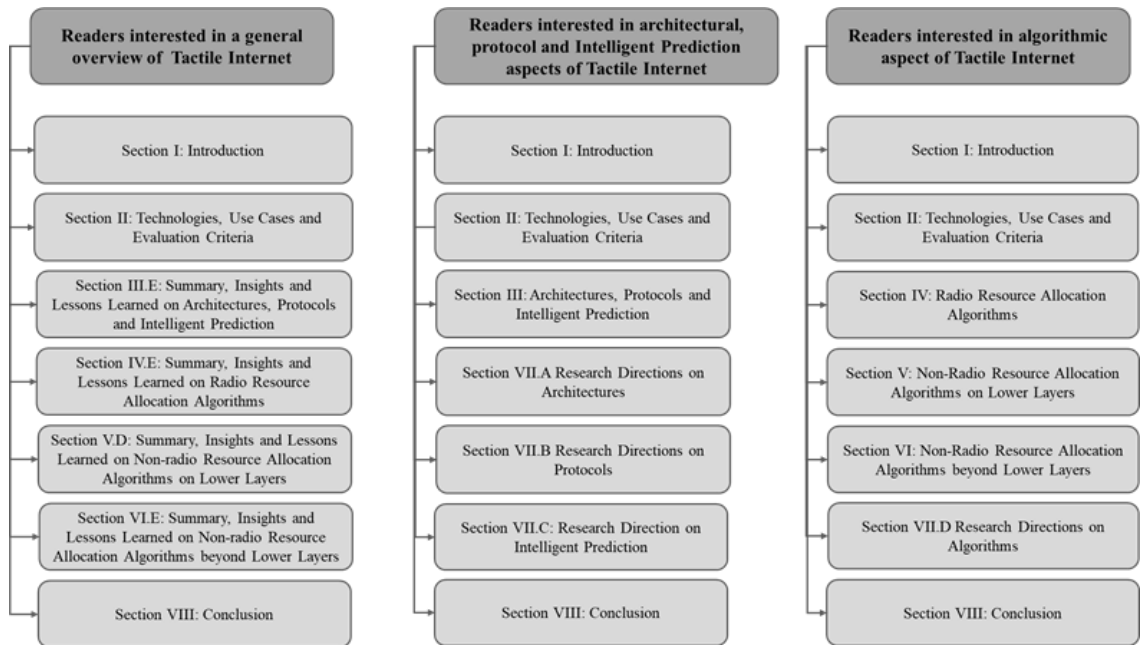

Fig. 3. Reading map

## II. TECHNOLOGIES, ILLUSTRATIVE USE CASES AND EVALUATION CRITERIA

This section starts with a discussion of the related and key enabling technologies for the Tactile Internet. After that, we introduce illustrative Tactile Internet use cases and derive evaluation criteria for evaluating the solutions proposed for Tactile Internet systems.

### A. Technologies

In this subsection, we discuss the commonalities and differences between the Tactile Internet and the related technologies as well as the key enabling technologies of the Tactile Internet.

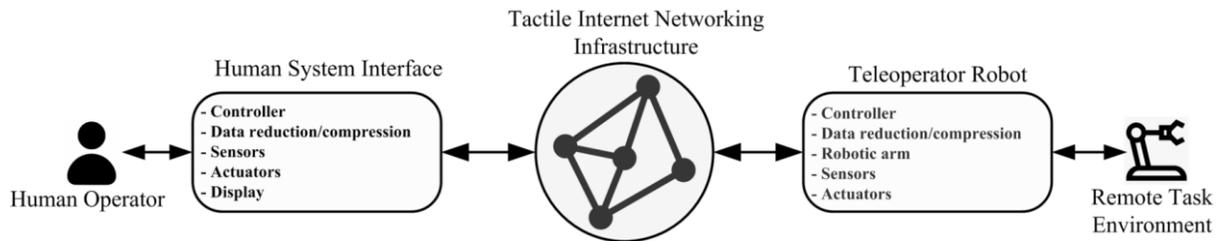

Fig. 4: Teleoperation system based on bidirectional haptic communications between a human operator and a teleoperator robot.

*1) Related technologies*

To better understand the vision of the Tactile Internet, it may be helpful to elaborate on the commonalities and subtle differences between the Tactile Internet and the emerging IoT, and mobile networks (e.g., 2G/3G/4G/5G). To begin with, we note that conventional pre-5G mobile networks focused on enhancing human-to-human (H2H) communications, with the main focus on the transmission of triple-play traffic (i.e., voice, video, data), whereas the IoT relies on its machine-to-machine (M2M) communications, which may be useful to enable the automation of industrial and machine-centric processes [48] . In contrast, the Tactile Internet involves the inherent human-in-the-loop (HITL) nature of human-to-machine/robot (H2M/R) interaction, thus mandating the need for a human-centric design approach. This  opens up a plethora of exciting research directions and adding a new dimension to the H2M/R interaction via the Internet [4][49]. To better grasp the full potential of the emerging robotics, it is important to note that the ultimate aim of the Tactile Internet is not to substitute humans by machines, but rather to augment humans by extending their capabilities [50]. Unlike mobile networks and the IoT, in Tactile Internet there will be a significant emphasis on the role of HITL systems, which inherently relies on human expertise from time to time. An interesting example of low-latency, HITL Tactile Internet applications is robotic telesurgery, where the underlying robotic system extends the capabilities of the surgeon, as opposed to a fully-autonomous robotic surgery system, where the operation is done by a robot, and the human is left out of the loop [9].

5G networks was initially known as a promising candidate to enable the Tactile Internet, though the current deployments of these networks prove otherwise. Although 5G is capable of providing a peak data rate of up to 20 Gbps, it falls short of satisfying the stringent requirements of the Tactile Internet due to the following reasons. Firstly, it is expected that the emerging immersive Tactile Internet applications will require data rates on the order of Tbps. As an example, we note that a typical VR headset requires a communication link of multiple Gbps. Within 5G networks, this can be realized only via a cable connection to a PC, thus posing serious limitations on the mobility of users. Clearly, these requirements may become even higher for Tactile Internet applications, where there is a need to transmit multi-sensory contents in real-time [51]. Secondly, it is evident that like the previous generation mobile networks, 5G is mainly driven by content- and machine-centric design approaches to realize H2H and M2M traffics, respectively. By contrast, the Tactile Internet aims to realize human-machine interaction via H2M/R communications, which is different from the conventional H2H and M2M traffics in many ways.

The first key difference between the conventional H2H and M2M traffics and H2M/R traffic lies in the underlying type of communication, which stems from the HITL nature of the latter one. Central to the H2M/R traffic is haptic communication, which enables a human operator to couple with a remote teleoperator robot via a truly bidirectional communication, as opposed to the conventional triple-play traffic that relies on either unidirectional or 2x-unidirectional communication. For better understanding, Figure 4 depicts an example of a typical teleoperation system based on bidirectional haptic communications for real-time exchange of control commands from a human operator and haptic feedbacks from a teleoperator robot. We first discuss the components of the system and then contrast its haptic aspects. The key components of the teleoperation system shown in Figure 4 are human operator, human system interface, networking infrastructure, and teleoperator robot. A local human system interface device is used to display haptic interaction with the remote teleoperator robot to the human operator. Note that the number of independent coordinates required to completely specify and control/steer the position, orientation, and velocity of the teleoperator robot is defined by its degrees of freedom (DoF). Currently, available teleoperation systems involve robotic arms with 1 to 20 DoF, or even higher. For instance, a 6-DoF robotic arm allows for both translational motion (in 3D space) via force and rotational motion (pitch, yaw, and roll) via torque. The local controllers on both ends of the teleoperation system ensure the tracking performance and stability of the human system interface and teleoperator robot. The properties of human haptic perception can be exploited by advanced perceptual coding techniques in order to substantially reduce the haptic packet rate. Table II briefly summarizes the comparison between the haptic aspects of the Tactile Internet teleoperation systems (for more detailed discussions, the interested reader is referred to [10], [20], [21]).

Another subtle difference that sets the H2M/R traffic aside from the conventional H2H and M2M traffics can be expressed in terms of the packet arrival process. While the Poisson and Pareto packet arrival models have long been used to describe the H2M and M2M traffics, these distributions cannot capture the unique characteristics of H2M/R traffic, which is dominated by control (i.e., command) and feedback signals. Recently, the authors of [49] have shown in a trace-driven study that the command and feedback paths of teleoperation systems can be jointly modeled by generalized Pareto, gamma, or deterministic packet interarrival time distributions, depending on the given value of the respective deadband parameters. In another study, the authors of [61] and [62] have shown that the packet interarrival times of a haptic-enabled VR system can be modeled by generalize Pareto distribution, which resonates with the findings of [49].

TABLE II
Comparison of the haptic aspects of the Tactile Internet.

| Component | Approach | | Contribution | Reference(s) |
|---|---|---|---|---|
| Data reduction /compression techniques | Statistical approaches | Predictive models (e.g., adaptive DPCM) | Using previously quantized samples to generate a predicted value for the current sample | [52],[53] |
| | | Lossy compression (e.g., DCT) | After transforming the time series haptic data into frequency domain using DCT, the DCT coefficients are quantized | [54] |
| | Perceptual approaches | Non-predictive coding (e.g., based on Weber's law) | To transmit only the samples that change more than a given threshold with respect to the previously transmitted sample. | [55] |
| | | Predictive coding (e.g., ZOH, linear predictor, FOLP, third order auto-regression, geometry-based predictor, and hybrid methods) | Prediction models of haptic signals can be used to estimate future haptic samples from previous data | [56] |
| Control systems | Passivity-based control | | Built on the idea that bilateral control systems must be passive and therefore stable | [57] |
| | Wave-variable control | | Combination of scattering transformation, network theory and passivity control | [58] |
| | Time-domain passivity control | | System's passivity is retained using adjustable damping elements | [59] |
| | Model-mediated teleoperation | | Instead directly sending back the haptic signals, the parameters of the object model which approximate the remote environment are estimated and transmitted back to the human operator in real-time | [60] |

Last but not least, the Tactile Internet mandates the need for an end-to-end design of communication, control, and computation functionalities, which are mostly neglected in 5G. The discussion above clearly shows that even though 5G might be an enabler of URLLC applications in general, its design goals were not tailored to the unique features and requirements of the Tactile Internet with its inherent HITL nature. Such shortcomings of 5G have recently attracted a great deal of interest from both academia and industry to define next-generation 6G systems, which is envisioned as an enabler of a variety of disruptive applications ranging from haptics to extended reality [48][51].

Beside the differences between the Tactile Internet and 5G discussed above, it is worth mentioning that the Tactile Internet should not be restricted to the context of 5G. In fact, as will be discussed later on, there are several research efforts that aim to realize the Tactile Internet over other technologies (e.g., WLAN and WBAN technologies, sub-GHz technology, and combinations of these technologies). The HITL-centric nature of the Tactile Internet also opens up a plethora of possibilities to leverage on artificial intelligence to decouple the human perception from the impact of excessive propagation delays, as typically encountered in wide area networks [9][49], which is not a focus of 5G [3]. Furthermore, according to the Technology Watch Report on the Tactile Internet, published by the ITU-T in 2014, both wired and wireless access networks are essential to meet the stringent latency as well as carrier-grade reliability requirements of Tactile Internet applications [1]. Wired access networks may be particularly useful for those Tactile Internet use cases that do not necessarily require mobility at all times and therefore can be also realized in fixed networking infrastructures.

Despite their differences, the Tactile Internet, the IoT, and mobile networks share the following set of common design objectives: (*i*) guaranteed availability, (*ii*) coexistence of different channel access technologies (e.g., cellular and Wi-Fi), and (*iii*) security [49].

2) *Enabling technologies*

5G is one of the key enablers for the Tactile Internet, given its inherent objectives of extremely low latency, high data rates, high availability and high-level security although it is not naturally designed to capture the HITL nature of the Tactile Internet. It will provide a promising infrastructure for Tactile Internet realization and for addressing the communication challenges [7]. A comprehensive survey on 5G is given in [63]. Edge computing [64], software defined networks (SDNs) [65] and network function virtualization (NFV) [66] are three other important concepts that can be leveraged to address the anticipated computation and communication challenges. Edge computing pushes computing and storage to the edge network, thereby offering a substantial reduction in communication latency [67]. The concept has been addressed in the literature from several aspects; examples of edge computing paradigms include mobile-edge computing, fog computing, cloudlets, and mobile cloud computing [68].

SDN refers to a networking architecture where the network control plane is separated from the data plane [65]. The control plane is responsible for making decisions on how to handle the network traffic, while the data plane performs the actual traffic forwarding based on those decisions. The networking devices (e.g., switches, routers) are managed by a centralized controller, thereby simplifying the whole management process (e.g., network configuration and policy enforcement).

NFV takes the process of network building and operating to a new level of flexibility by decoupling the network functions from the hardware infrastructure running those functions. NFV relies on standard virtualization technologies to virtualize network functions and elements, creating virtual network functions (VNFs) [66]. These VNFs are instantiated, chained, and run as needed to provide targeted services, thereby shortening the life cycle of provisioning new services.

The different networking enablers can be combined to create even more benefits. For example, [68] and [69] investigate how edge computing and 5G, respectively, can benefit from SDN. Aijaz *et al.* [11] discuss a new architecture leveraging 5G, edge computing, SDN and NFV to provide a general-purpose architecture that can support the Tactile Internet. Their architecture uses the virtualization concept of NFV to allow dynamic instantiation of an end-to-end network using the targeted applications' requirements. Edge computing and SDN can foster latency reduction by bringing the computing capabilities close to the user equipment and by using the centralized controller of an SDN (i.e., limiting the number of intermediate nodes) [70].

Artificial intelligence (AI) is another key enabler for overcoming the 1 ms latency challenge. Indeed, while the previously discussed enablers can contribute to reducing the end-to-end latency, the scope of their efficiency is limited to a maximum of 150 km, as they are bound by the speed of light. The only approach that can address the 1ms limitation while eliminating the 150 km constraint is to use predictions [7].

*B. Illustrative Use Cases*

We discuss three use cases in this survey: remote robotic surgery, autonomous driving and remote phobia treatment. Remote robotic surgery and autonomous driving are often mentioned in the literature (e.g., [1][6][13]), but generally not discussed in as much depth as we do here.

*1) Remote Robotic Surgery*

Remote robotic surgery will make surgery available almost anywhere, no matter where surgeons and patients are located. This will offer multiple potential benefits to human society, including reducing the risks and delays associated with long-distance patient travel and making surgery available to patients who live in underserved regions. In short, it will enable the proliferation of advanced surgery skills [71][72].

Figure 5a shows a 5G-based Tactile Internet system for remote robotic surgery. It relies on the edge-based architecture hinted at by several references (e.g., [6][11]). The system can be divided into a surgeon-side edge, a patient-side edge and its core network. The surgeon-side edge provides wireless connectivity to a master surgical console through which surgeons can remotely manipulate a surgical robot and receive real-time auditory, visual and haptic feedback. On the patient-side edge, the system provides wireless connectivity to the controlled surgical robot that operates on the patient according to the commands received from the surgeons and which generates the corresponding feedback. Both edges are equipped with edge clouds (e.g., fog computing [73]) on which a tactile artificial intelligent engine resides to facilitate haptic information prediction. The core network acts as a medium to provide bilateral communications between both edges.

Let us consider that surgeons in one hospital need to operate on a patient who is located in a different hospital. In general, a robotic surgery procedure consists of four phases: (i) access to the body cavity, (ii) tissue dissection, (iii) tissue destruction, and (iv) tissue reconstruction (e.g., suturing, knot-tying, etc.) [74]. Specific tasks need to be performed in each phase. To illustrate this use case, a knot-tying task, which is performed during the last phase, is considered. During this task, surgeons need to perform multiple actions/gestures such as picking up needles with both hands (A1), making a C loop around the right-side needle (A2), picking up a suture (A3), and pulling the suture with both hands to tie a knot (A4) [75], all via the master surgical console. These actions will be translated into a control message and sent to the surgical robot. Once they are received, the controlled surgical robot will perform these actions on the patient and generate the corresponding feedback. The feedback will then be sent back to the master surgical console, and finally the master surgical console presents that feedback to the surgeons. It is obvious that these communications must be very rapid and reliable enough to ensure the stability of real-time interaction. Otherwise, surgeons will notice a time lag between their actions and the corresponding feedback, preventing them from being sure about what to do next because the feedback of a previous action influences their next action.

The tactile artificial intelligence engine will play an important role in addressing this issue. For example, assume that surgeons perform an action A1, and then based on the corresponding feedback (F1), they will decide to do either A1 again or A2 for the next action. Next, we can consider a scenario where A1 is successfully transmitted to the controlled surgical robot. However, F1 is delayed or gets lost on its way to the master surgical console. In this scenario, the tactile artificial intelligence engine on the surgeon's side will predict feedback that is very similar to F1 and send it to the master surgical console. This will allow the surgeons to experience immersive real-time interaction and know what to expect next, even though the actual feedback is still travelling through the network.

*2) Autonomous Driving*

Autonomous driving is another potential application that can be realized by the Tactile Internet. It is expected to be a key contributor to reducing traffic congestion, accidents and greenhouse gas emissions in the future [76]. Generally, an autonomous driving system can be divided into three main subsystems: perception, planning and control [77]. The perception subsystem is responsible for collecting information about location, speeds, road conditions, nearby vehicles and surroundings from sensors, and to continually use that constantly-upgraded information to create a dynamic environment model. The planning subsystem makes driving decisions based on that environment model and on feedback from the control subsystem. The control subsystem then manipulates vehicles according to the driving decisions -- adjusting their speed, steering angle and acceleration, etc. Nowadays, most of the current autonomous vehicles rely exclusively on onboard sensors; however, these may not be sufficient for making safe driving decisions due to their limitations, including their restricted perception capacities and limited perception range [76]. Autonomous vehicles therefore need to communicate with each other as well as with various infrastructure components in order to increase the environmental perception and thereby improve autonomous driving decision making [78]-[80]. This aspect refers to vehicle-to-everything (V2X) communication.

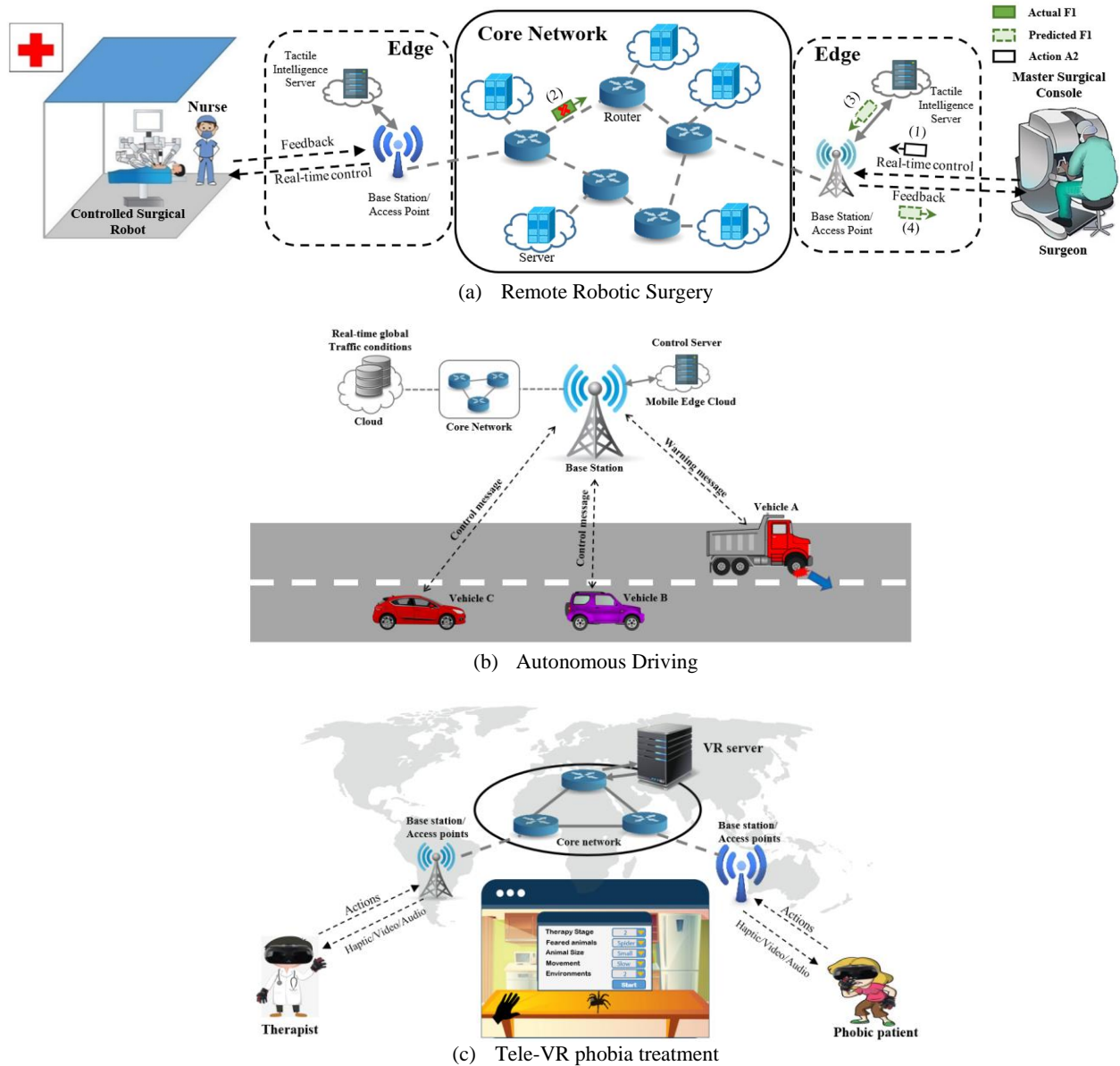

Fig. 5. Illustrative use cases of the Tactile Internet. (a) Remote robotic surgery for performing a knot-tying task, (b) Autonomous driving for collision avoidance and (c) Tele-VR phobia treatment for spider phobias.

With the support of V2X communication, autonomous driving use cases may be categorized into two groups: predictable behaviors (e.g., short-distance platooning, cooperative lane changes) and short reaction times (e.g., cooperative emergency maneuver for collision avoidance) [78]. In our use case, we focus on cooperative emergency maneuvers for collision avoidance. Figure 5b illustrates a Tactile Internet system for autonomous driving use cases. This system provides connectivity to support V2X communications for autonomous driving. It also offers computational and storage resources for both core and edge networks; these can be used to run a control server and store real-time global traffic conditions.

Next, we consider a scenario where autonomous vehicles are moving on a highway at medium speed as shown in Figure 5b. Vehicle A gets a flat tire and suddenly changes to another lane. We assume that the driving decisions of all autonomous vehicles are made by a control server running on a mobile edge cloud in close proximity to the vehicles [80]. The control server makes driving decisions based on the information received from all the vehicles concerned and sends control messages to all of them. In addition, all of these vehicles periodically exchange information with the control server. To avoid an accident in this scenario, Vehicle A first sends a warning message consisting of its status, position, speed and direction to the control server. The control server then processes the received information to make driving decisions for Vehicle B and C and sends control messages to both vehicles. Finally, based on the control messages, Vehicle B slows its speed while Vehicle C changes to another lane to avoid a collision.

*3) Tele-VR Phobia Treatment*

A specific phobia is one of the most common mental health disorders [81][82]. As reported in [82], approximately 8.7% of Americans suffer from a specific phobia. One technique for phobia treatment is vivo exposure, in which a patient is presented with the feared stimulus in order to overcome their phobia with a therapist's guidance [83]. A patient who has cockroach phobia may, for instance, first look at a picture of a cockroach. Eventually, the patient may need to see, and even perhaps touch a real cockroach. However, this technique might not work for everyone. As shown in [84], this technique has drop-out rates of up to 45%. Recently, virtual reality (VR) has proven to be an effective technique for phobia treatment [81][82][85]. With this technique, therapists can treat phobic patients by gradually and systematically exposing patients to their feared objects (e.g., snakes or spiders) via a VR environment without any danger. Unlike other phobia treatment techniques, therapists can fully control the feared stimulus based on their patient's fear level by adjusting the size and movement of the feared objects in virtual environments [82][85].

The Tactile Internet will bring more benefits to traditional VR phobia treatment. It will allow therapists to treat their phobic patients anywhere and anytime regardless of their location. Therapists and patients are able to collaboratively touch and see virtual feared objects, feeling the real-time sensations of touch through a shared haptic VR environment without the need to be in the same room. Not having to travel to see their therapists for this treatment will save patients a considerable amount of time (and, most likely cost). Figure 5c shows a Tactile Internet system for a tele-VR phobia treatment use case. Both the therapist and the patient need to communicate with a VR server in order to interact with a shared haptic VR environment. Both are equipped with VR headsets, haptic devices (e.g., haptic gloves) and tracking devices to receive real-time visual, auditory and haptic feedback, and to move freely in the VR environment. The Tactile Internet system will provide the required low-latency communication between therapists, patients and VR servers.

Tele-VR phobia treatment could easily be divided into three stages: (i) seeing and hearing feared objects without any touch, (ii) touching feared objects with the help of a therapist, and (iii) touching without any help from a therapist. For an illustration of this use case, the second stage is our focus. We consider a scenario where a therapist at a hospital needs to treat an arachnophobic patient who is at home. We assume that the patient has already completed the first stage. In the second stage, the therapist will expose the patient to a virtual spider by allowing the patient to feel the sensation of touching a virtual spider. However, the patient may be too scared to touch the virtual spider by himself. Therefore, in this stage, the therapist is the person who touches the virtual spider, while both therapist and patient can experience the sensation of touch.

In a therapy session, the therapist first attempts to touch a virtual spider (e.g., the spider's head). Tracking devices will then detect the therapist's action and send it to the VR server. After that, based on the received action, the VR server will then perform VR frame processing/rendering and generate haptic feedback, sending the corresponding video frame and haptic feedback to both therapist and patient. Finally, the patient will experience immersive VR and feel the texture of the spider's skin through a VR headset and haptic gloves, respectively, without touching. This stage may be repeated several times until the patient gets more familiar with the virtual spider. After that, the patient can move to the last stage where he touches the virtual spider by himself.

### C. Tactile Internet Evaluation Criteria

In this subsection, we propose a set of criteria to evaluate the architectures, protocols, intelligent prediction schemes and algorithms proposed to date for the Tactile Internet. Derived from the three use cases discussed in the previous subsection, these criteria are designed to cover all possible Tactile Internet applications. We mainly focus on the key requirements related to latency and reliability, given that security is outside the scope of this paper. The reader should note that while many papers (e.g., [4][6][7]) do identify these very same requirements, they do not usually give their full particulars and do not support them with detailed use cases.

The first criterion ($C_1$) is the need to provide ultra-responsive network connectivity (ultra-low latency). The Tactile Internet is expected to enable real-time tactile/haptic interaction. This interaction requires that the time-lag between actions and feedback should be less than all human reaction times to allow all human senses to interact with machines. For instance, in the remote robotic surgery use case, whenever a surgeon manipulates a robot to cut a patient's tissue, the surgeon should receive force feedback within normal human reaction time for manual interaction (i.e., 1 ms). This will allow the surgeon to feel the real-time sensation of touch. Similarly, for the tele-VR phobia treatment use case, high-fidelity interaction between VR users and haptic VR environments can be achieved only if the end-to-end latency is on the order of a few milliseconds [1]. Otherwise, VR users will experience cybersickness. Moreover, to ensure the consistency of the VR view for all VR users in a shared haptic VR environment, the communication latency between the VR users and the VR server should be very low.

Making real-time tactile/haptic interaction mimic reality requires end-to-end latency on the order of 1 ms [6]. As defined in [6], this end-to-end latency (round-trip latency) in technical systems includes the uplink delay for transmitting information from a sensor (human) to a control/steering server, the delay for processing the received information and generating a response, and the downlink delay for transmitting the response back to the actuator (human). This definition can also refer to a round-trip latency/a response time delay [6], depending on specific use cases. Either of these latency metrics will be used to evaluate solutions for this criterion, depending upon the type of the latency they are targeting. For instance, the round-trip latency will be applied to an algorithmic solution that focuses on edge resource allocation for computation offloading, as it aims to minimize the total time spent on offloading a task from an end-user to an edge computation unit, processing the task, and then sending the response back to the end-user. Meanwhile, the end-to-end delay will be used to evaluate an algorithmic solution in lower layers (i.e., resource allocation in RAN, WBAN and WLAN) as its main objective is to minimize the resultant end-to-end packet delay. Therefore, in order to meet the ultra-low latency criterion, solutions in a Tactile Internet system should be able to achieve either an end-to-end latency or round-trip latency (response time) of 1 ms.

The second criterion ($C_2$) is the need to provide ultra-reliable network connectivity (ultra-high reliability), as many Tactile Internet applications will have a huge impact on human lives. In remote robotic surgery, the most critical Tactile Internet example, surgeons generally perform each subsequent surgical action based on the feedback received from the previous action. If the feedback gets lost during an operation, the surgeon could make a wrong decision on the next action, which could have life-threatening consequences. Likewise, in the autonomous driving use case, autonomous vehicles need to keep exchanging warning and control messages with the control sever. If control messages get lost during their transmission to the autonomous vehicles, this could lead to serious chain collisions. In general, remote robot surgery and autonomous driving require a failure rate of up to $10^{-7}$ to guarantee reliable data transmission, which translates into a few seconds of outage of the system per year [5]. This rate can also be quantified in terms of a packet loss rate of $10^{-7}$. Furthermore, the ultra-high reliability can be interpreted in terms of prediction error of $10^{-7}$ when it comes to intelligent prediction solutions that aim to fulfil a gap, where the data does not arrive on time due to either distance (greater than 150 km) or a connection problem, as discussed in the remote robotic survey use cases. To ensure reliable tactile/haptic interaction, the differences between predicted and actual data should be kept extremely small. Similar to the first criterion, either of the aforementioned reliability metrics will be used to evaluate different solutions. While the system outage (or equivalently interpreted as system availability) will be mainly applied to architectural solutions, the packet loss will be used to evaluate algorithmic solutions. In addition, the prediction error will be specifically applied to solutions that tackle the intelligence challenges.

The Tactile Internet is considered as a system that provides ultra-low latency and ultra-high reliability. To evaluate the end-to-end latency and reliability performance of the Tactile Internet system, all architecture, protocol and algorithmic design choices in any part of this system should be considered. However, most studies aim to propose a single solution, which is either an architecture, an algorithm or a protocol, to address some specific part of it. Given that reality, this survey instead seeks to evaluate each solution to determine if it can meet either the ultra-low latency criterion ($C_1$), ultra-high reliability criterion ($C_2$) or both. This evaluation could thus provide some useful guidelines with which to determine if a solution would be suitable for the Tactile Internet system. The evaluation of a solution is justified based on its best achieved performance reported in the corresponding paper for the given deployment scenario and network parameter setting specified therein. As for the algorithmic approaches, a given algorithm is considered to meet a given criterion if that particular criterion is accounted for within the constraints that the algorithm should satisfy. Moreover, a given solution is considered to partially meet a given criterion if it focuses only on either uplink or downlink directions, without explicitly taking the end-to-end connectivity into account. Furthermore, as for the solutions that tackle the intelligence challenges, we assume that they always meet the ultra-low latency criterion ($C_1$) due to the fact that they can flexibly adjust a bounded latency (which is usually set to 1 ms) for performing predictions.

The readers should also note that some papers in the literature focus on the details of how to design the Tactile Internet application itself and do somehow assume that the infrastructure which meets the requirements is available. We do not review them in this survey since the focus is rather on the works done to make this infrastructure a reality. However, we discuss very briefly in the next paragraphs their essence.

The vast majority of these papers (e.g., [86]-[91]) deal with multi-robot applications in a setting where there are still human operators. References [86]-[88] for instance deal with allocation of physical and/or digital tasks to robots, while references [89][90] study the task migration from human to robots and robots to agents. When it comes to reference [91], it focuses on scheduling and assignment of delay constrained teleoperation tasks to the human operators with the assumption of a low-latency reliable Tactile Internet networking infrastructure. Examples of papers which do not belong to this vast majority are references [92] and [93]. They aim at utilizing the Tactile Internet infrastructure (which is assumed to be available) to enable autonomous vehicle and unmanned aerial vehicle flock control applications, respectively.

## III. ARCHITECTURES, PROTOCOLS AND INTELLIGENT PREDICTION FOR THE TACTILE INTERNET

Several works have introduced architectures for the Tactile Internet. All of these architectures are either application-agnostic or application-specific. Some works have also focused exclusively on the protocol aspects of architectures without proposing full-blown architectures. In addition, some of them have proposed intelligent prediction schemes to tackle the critical issue of haptic/tactile information (e.g., control command, haptic/tactile feedbacks) that is expected by remote functional entities and does not arrive on time.

In this section, we first review the various application-agnostic architectures and then the application-specific architectures, followed by the protocols and the intelligent prediction schemes. It should be noted that the architecture proposed by the IEEE P1918.1 standard [3] is discussed in the subsection on the applications-agnostic architecture especially given that it can be mapped to all Tactile Internet applications. The section then ends with summary, potential insights, and lessons learned.

### A. Application-agnostic Architectures

The proposed IEEE P1918.1 architecture [3] focuses on (as most standards) defining functional entities and interfaces to ensure interoperability. Meanwhile, most of the application-agnostic architectures presented in peer-reviewed venues aim at enhancing cellular networks so that they can meet the stringent Tactile Internet requirements. They leverage either mobile edge computing (MEC) [94]-[97] or Wi-Fi technology [98] to offer offloading capability and increase network capacities. In addition, one application-agnostic architecture relies on EPON technology due to its high capacity and flexibility [99] to support Tactile Internet applications. The reader should note that references [96] and [99] address both architectural and algorithmic aspects. While their architectural contributions are discussed in this section, their algorithmic contributions will be presented later in the algorithm section (see Section IV.D for [96] and Section V.B for [99]). This section first discusses the IEEE P1918.1 standard architecture followed by the MEC-enhanced cellular network architectures, the Wi-Fi enhanced cellular network architectures and the EPON based-network architecture.

*1) IEEE P1918.1 Architecture*

The proposed IEEE P1918.1 architecture [3] is generic and can be mapped to all types of connectivity (i.e. local area, wide area, wired, wireless and hybrid wired/wireless) in addition to being application agnostic. It relies on a few well thought design principles (e.g., modularity, flexibility, separation of control and data planes, leveraging on cloud and edge resources). It is made up of network domains and edges domains.

An example of functional entity is the tactile device located in the edge domain. It consists of sensors, actuators, and has networking capabilities. It also has processing capabilities, although limited. A second example is the gateway node which may be located in the edge domain or in the network domain. It enables interactions between the network and edge domains. Yet another example is the support engine. It may be located at the edges or in the networks domains or spread over the edges and network domains. It is a critical component because it may play several roles. For instance, it can offer additional processing capabilities to tactile devices and caching capabilities at the edge. It may also run algorithms to predict packets that might get lost in the network or reach the edge too late.

Moreover, several interfaces are defined to enable interactions between domains, and between functional entities. Access (A) interface for instance allows interactions between the edge and network domains. When it comes to Tactile (T), it enables interactions between the entities within the edge domains. The messages exchanged at these interfaces are defined and key performance indicators are also specified. These indicators include reliability, latency, and scalability.

The proposed architecture is very promising when it comes to interoperability. However, the architecture itself does not meet any of the two criteria we set in this paper. The key reason for this is that as all standards, it leaves to implementation how the specific functional entities will be realized. Meeting the ultra-low latency criterion ($C_1$), for instance, will depend on the specific prediction algorithm that will run on the support engine, and this is clearly outside the scope of the standardization.

*2) MEC-enhanced Cellular Network Architectures*

The first application-agnostic architecture which utilizes MEC to enhance a cellular network is introduced by Ateya *et al.* [94]. The authors propose a multi-level cloud system to provide offloading capability in the cellular network to support Tactile Internet applications. This system aims to improve the latency and reduce the network congestion in the core network. It consists of three hierarchical cloud levels, with micro-cloud, mini-cloud and core-cloud units at the lowest, middle and highest levels, respectively. The micro-cloud units, at the lowest level, are connected to every base station (BS) via fiber links. They handle tasks that are offloaded directly by mobile users. The mini-cloud units, at the middle level, are located in the BSs. Each of these units is connected to a group of micro-cloud units and have higher capacities than individual micro-cloud units. They manage their micro-cloud units and handle the tasks that cannot be done by the micro-cloud units under their control. The core cloud unit is at the top of the cloud hierarchy. It has the most powerful capacities and resides in the core network. It controls all the mini-cloud units in the network and serves as a gateway to/from remote public clouds. In addition, the tasks that cannot be handled by the mini-cloud units will be offloaded to this cloud level. To evaluate the performance of the proposed system, the authors carry out simulations with several scenarios. The results show that the proposed system can achieve round-trip latency on the order of 1 ms. However, only a low number of users is considered in the simulation. Thus, this paper can only meet the ultra-low latency criterion ($C_1$) in small-scale scenarios. Since no evaluation metrics related to the reliability are considered, it does not cover the ultra-high reliability criterion ($C_2$).

In addition to deploying the multi-level cloud system proposed in [94], Ateya *et al.* [95] leverage a SDN technology in core networks to provide flexibility in traffic routing to improve communication latency. In this architecture, the core network is formed by access switches (e.g., Open vSwitch), OpenFlow switches, Middleboxes and a centralized SDN controller. The access switches in the core network are also connected to BSs in a radio access network (RAN). In addition, all nodes in the network are controlled and managed by the centralized SDN controller through the OpenFlow protocol. The centralized SDN controller is aware of the global network status of all nodes. It can establish effective routing paths to ensure the quality of service (QoS) requirements of end-users (i.e., limiting the number of intermediate nodes). The authors conduct simulations with two scenarios to evaluate the end-to-end delay performance of the proposed architecture. In the first scenario, BSs serving both end-users are connected to the same mini-cloud unit, whereas in the second scenario, the end-users are connected to separate mini-cloud units. The results indicate that end-to-end delay on the order of 1 ms can be achieved in both scenarios. In addition, the end-to-end delay can be improved further by increasing the network bandwidth. Therefore, this work meets the ultra-low latency criterion ($C_1$). Nevertheless, the performance of the proposed architecture with higher numbers of users needs to be further investigated. This work does not cover the ultra-high reliability criterion ($C_2$), as there is no discussion regarding the reliability aspect.

Another application-agnostic architecture that relies on the MEC concept is proposed by Shafigh *et al.* [96]. The authors propose a flexible cloud-based radio access network (FRAN) for the Tactile Internet. Their architecture combines a dynamic network architecture (DNA) with a cloud-based radio access network (CRAN) to offer offloading capability to network operators. The DNA is a network in which users' devices share their connectivity and act as wireless access points to other users. The proposed architecture allows network operators to offload non-tactile users temporarily onto the DNA network if their CRAN network does not have enough resources to guarantee the QoS requirements of tactile users. In this architecture, the CRAN network consists of remote radio heads and baseband units. The remote radio heads are radio-frequency processing units that provide wireless connectivity to end-users. These remote radio heads are also connected to a pool of baseband units running on a cloud platform via fronthaul links for baseband processing. In addition, the DNA network comprises several dynamic access points and a DNA-OpenFlow controller. The dynamic access points are users' devices that are willing to share their connectivity and unused resources. All the dynamic access points are controlled by the DNA-OpenFlow controller through the OpenFlow protocol. The DNA-OpenFlow controller also handles all of the signaling in a control plan between the CRAN and DNA networks for pricing, reimbursement and offloading purposes. Interestingly, the authors claim that the proposed FRAN can be applied without any hardware/software modification in the network. In addition to the proposed architecture, the authors also introduce an efficient resource allocation framework in the FRAN to optimize the network operators' profits, as the network operators will be charged whenever they offload users onto the DNA network. This resource allocation framework will be discussed in Section IV.D. Shafigh *et al.* [96] do not evaluate the performance of their proposed FRAN separately from the resource allocation framework, and so it does not cover any of the criteria.

To provision Tactile Internet services in NFV-based MEC systems, Xiang *et al.* [97] propose a new design approach called chain-based low latency VNF implementation (CALVIN) to implement low-latency service function chains. In contrast to the existing centralized approach where all VNFs in a service function chain are installed in a single virtual machine, and both kernel and user spaces are utilized, the ALVIN implements each VNF in a separate VM. It also implements each VNF either purely in the kernel space or in the user space of a virtual machine, depending on the VNF type. These design choices aim at eliminating the context switching overhead thanks to a shared virtual CPU between the kernel and the user spaces, which generally occurs in the centralized approach, thereby reducing the latency. In the CALVIN, VNFs are classified into three types: (i) elementary, (ii) basic and (iii) advanced. The first type covers VNFs that retrieve packets from an ingress virtual interface and directly forward the packets to an egress virtual interface, without performing any operation. (e.g., an elementary forwarding function). This elementary type can be implemented in both spaces. The second (basic) type covers VNFs that require low computational intensity (only the header is processed, such as the network address function). This type can only be implemented in kernel space. The advanced type covers VNFs that require high computational intensity (both header and payload are processed, as with a network coding function). This last type can only be implemented in the user space. The architectural design, software implementation choices and CALVIN configurations are suggested for practical use. A testbed with two computer nodes is implemented to evaluate the proposed CALVIN, where the first one runs UDP client and server while the second one runs VNFs to create a service function chain. The results indicate that the CALVIN can significantly improve the latency in comparison with the centralized approach, although it has a worse performance in terms of the throughput. The CALVIN can achieve an average round-trip latency of 0.32 ms with 1400-byte packet size in the case of two elementary VNFs in the chain. It can also provide an average round-trip latency of 0.35 ms when three computation-intensive VNFs (e.g., network coding and encryption functions) are considered in the chain. Therefore, this work meets the ultra-low latency criterion ($C_1$). However, it does not cover the ultra-high reliability criterion ($C_2$) since there is no discussion about the reliability aspect.

*3) Wi-Fi-enhanced Cellular Network Architectures*

Beyranvand *et al.* [98] propose a fiber-wireless (Fi-Wi)-enhanced LTE-Advanced (LTE-A) heterogeneous network (HetNet) architecture for the Tactile Internet. This architecture aims to enhance LTE-A HetNets with Fi-Wi access networks to provide high-capacity fiber backhaul and Wi-Fi offloading capabilities. It utilizes an Ethernet passive optical network (EPON) technology to form backhaul links between core and edge networks, and a Gigabit WLAN technology to build Wi-Fi access networks. This architecture considers both time division multiplexing and wavelength division multiplexing EPONs. The EPON fiber backhaul consists of an optical line terminal (OLT) connected to multiple optical network units (ONUs) via fibers. The OLT serves as a gateway to/from the core network, while ONUs reside on the network edge. Three different types of ONUs are covered. The first type makes it possible to serve fixed users. The second type, connected to cellular BSs, allows mobile users to be served. The third type is equipped with mesh portal points (MPP) that communicate wirelessly to mesh access points (MAPs) to form Wi-Fi access networks and serve the offloading mobile users. This architecture also offers backhaul redundancy strategies. Both direct-wired and wireless communication among ONUs are considered. In addition, subsets of ONUs have additional backup links to the OLT.

The authors of [98] derive analytical expressions of the Fi-Wi connectivity probability, the end-to-end delay, and the maximum aggregate throughput to evaluate this architecture's performance. Simulations were also conducted to validate the analytical expressions. First, they show that the analytical results closely match the simulation results. Second, the redundancy mechanisms clearly offer a flat connectivity probability to users with respect to the EPON fiber link failure probabilities that are lower than 0.1. Third, they show that higher offloading ratios significantly improve the system's throughput and delay performance. Finally, an average end-to-end delay for data transmission between end-users in the network on the order of 1 ms can be achieved despite the fiber faults. This work therefore meets the ultra-low latency criterion ($C_1$). For the reliability aspect, the proposed architecture can almost guarantee Fi-Wi connectivity for fixed users, which reflects the reliability of the system. However, this is not the case for mobile users. Consequently, it is judged as meeting the ultra-high reliability criterion ($C_2$) for fixed users.

*4) EPON-based Network Architecture*

Neaime *et al.* [99] introduce an optical cloud distribution network (OCLDN) architecture to provide high-speed, programable and scalable optical access networks to support low latency services. The OCLDN relies on a next-generation EPON (NG-EPON) that allows ONUs to transmit traffic over multiple wavelength channels simultaneously, thereby increasing the network capacity. In the OCLDN, ONUs are distributed to provide access points (APs) for end-devices (e.g., IoT devices). The ONUs are also equipped with either fixed or tunable transceivers to enable simultaneous transmission over multiple uplink channels. In addition, each of them is connected to a single OLT residing at a central office (CO). To offer low latency services, a mini-data center is also deployed at the CO. The mini-data center includes control/steering tactile servers, cache servers, media servers and cloud servers. Furthermore, it relies on a central office re-architected as a data center architecture that utilizes the NFV and SDN technologies to enable agile, flexible and scalable service provisioning. In addition to the proposed OCLDN architecture, the authors also propose a resource allocation algorithm for uplink transmission. The proposed algorithm aims to ensure the latency requirement of tactile services when several types of services are being served simultaneously. The algorithm will be discussed later, in the algorithm section. We note that the performance of the OCLDN is not evaluated separately from the proposed resource algorithm. Therefore, this work does not cover either of the criteria.

*B. Application-specific Architectures*

Some architectural works target specific Tactile Internet applications such as real-time teleoperation [50][100], virtual reality [101][102], real-time video surveillance [103], industrial automation [104], smart cities [105] and general health applications [106].

*1) Real-time Teleoperation*

Real-time teleoperation (e.g., remote robotic surgery) is a prime application that will be realized by the Tactile Internet. This application has been considered in Maier *et al.* [50] and Holland *et al.* [100].

Maier *et al.* [50] introduce Fi-Wi-enhanced LTE-A HetNets proposed earlier in [98], along with AI-enhanced MEC servers located on ONU-BSs/MPPs to support HITL-centric teleoperation. The authors aim to demonstrate that this architecture can provide low latency communications by carrying out simulations based on real haptic traffic traces. To obtain these traces, they conducted teleoperation experiments in which two Phantom Omni devices were connected to create a bilateral teleoperation scenario. These experiments also considered a deadband coding technique. This technique exploits the limits of human haptic perception to reduce the data transmission by which a haptic sample is transmitted only if humans can notice any changes in haptic perception compared to the previously transmitted haptic sample. Based on the obtained traces, the authors ran simulations to create a scenario where human operators and teleoperator robots communicate with one another through the enhanced network architecture. They considered both local and non-local teleoperation cases. In the former case, a human operator and a teleoperator robot are in the same ONU, while in the latter, they are in different ONUs. Background traffic generated by fixed and mobile users was also considered. The results show that the proposed architecture can provide the 1 ms average end-to-end delay for the local case with deadband coding, whereas it cannot provide such delay for the non-local case. The results also reveal that the end-to-end delay for both cases can be bounded within 5 ms. Therefore, this work only meets the ultra-low latency criterion ($C_1$) for the local case. However, it does not cover the ultra-high reliability criterion ($C_2$), as it does not evaluate the reliability aspect.

In another real-time teleoperation work, Holland *et al.* [100] utilize sub-GHz technology to transmit tactile feedback and robotic control in a direct path without any retransmission mechanism, thereby minimizing the communication delay. This is possible because the sub-GHz transmission has very good wireless coverage. To compensate for the eventual loss in reliability due to the absence of a retransmission mechanism, the architecture allows the source and the destination to exchange channel control information over reliable mobile networks to dynamically adjust interleaving/coding schemes and frame duration to minimize transmission error. More precisely, the destination periodically monitors channel conditions, sending a request to the source over the reliable mobile network to adjust the coding schemes if needed to minimize transmission error, thereby improving the reliability. The authors implemented a prototype to prove the feasibility of the proposed concept. They used TV White Spaces as a sub-GHz medium and Carlson devices as transceivers. In their experiments, two devices, located in two different buildings in an area of England communicated with one another via line-of-sight paths. However, this prototype does not include coding schemes. The experimental results indicate that bit error rate (BER) distribution is relatively steady for a data rate of about 1 Mbps with a via line-of-sight distance of 7 km. Also, a signal-to-interference-plus-noise ratio of more than 15 dB is shown to be feasible within a via line-of-sight distance of 5 km and a transmission power of 36 dBm. However, the delay performance of the proposed concept was not evaluated, and thus it cannot meet the ultra-low latency criterion ($C_1$). Moreover, the best achievable BER of the proposed concept is still relatively high, at about 1.7 %. Thus, it does not meet the ultra-high reliability criterion ($C_2$).

*2) Virtual Reality*

The emerging Tactile Internet will enable new possibilities of VR applications. To date, two architectures have been proposed for these applications. Elbamby *et al.* [101] focus on multiplayer VR interactive gaming while Jebbar *et al.* [102] target remote VR phobia treatment.

Elbamby *et al.* [101] propose a system architecture for multiplayer VR interactive gaming in an indoor VR gaming arcade. The architecture utilizes a MEC network to reduce the computation delay for VR HD frame processing/rendering. It also leverages mmWave communications to increase the network capacity due to the generous available spectrum in this frequency range. In this architecture, VR players in an arcade are equipped with mmWave head-mounted VR displays (mmHMDs). They transmit their poses and actions to the MEC network for VR HD frame processing/rendering and receive the corresponding video frames to be displayed on their mmHMDs through mmWave band access points (mmAPs). The mmAPs operating at 60 GHz indoor band communicate wirelessly with the mmHMDs in line-of-sight paths. A deferred acceptance match algorithm is adopted to allocate mmWave transmission resources to the VR players. Dual connectivity is considered to solve the blockage and temporal disruption of the directional channel. In addition, the players' poses and actions can be predicted in the MEC network. Thus, some HD video frames corresponding to the anticipated and the most popular actions are proactively computed and cached in the MEC network, thereby improving the average computation delay. The authors conducted simulations to evaluate their proposed system. The results show that their system can provide an average total delay of less than 6 ms with up to 16 VR players. They also revealed that their system achieved a 99th percentile communication delay of about 8 ms. Lastly, their simulations showed that the average computation delay can be further improved by increasing the storage capacity in the MEC network. The authors clearly show that their proposed system can satisfy the delay requirement of single/multiple-user VR applications (e.g., 20 ms [101]). However, it does not satisfy the delay requirement of all Tactile Internet applications, and so it does not meet the ultra-low latency criterion ($C_1$). Moreover, the authors do not discuss the reliability aspect; therefore, it does not cover the ultra-high reliability criterion ($C_2$).

Jebbar *et al.* [102] propose a fog-based architecture for a remote VR phobia treatment application that allows a patient to have a remote therapy session with a therapist through a shared haptic VR environment. The proposed architecture consists of five main software components that are deployed in cloud/fog systems to enable the application. The first component is a haptic device manager that communicates with end-user devices such as tactile gloves, tracking devices and VR headsets. The second one is a zone detector that detects the user positions by collecting position data from the tracking devices via the haptic device manager. The third one is a VR sever responsible for updating and synchronizing VR views among users, based on the users' positions received from the zone detector. The fourth component is a feedback generator that generates corresponding tactile feedbacks, based on the user movement received from the zone detector. The last component is a conferencing server that establishes a voice and video call among users. All of the interfaces between the components are defined using representational state transfer technology. In the proposed architecture, all components except the conferencing server are deployed in the fog system to ensure low latency. The authors implement a prototype to evaluate the performance of the proposed architecture. They consider a use case where the patient is too scared to touch the animal by himself, and so that the patient can receive the corresponding haptic feedback through the therapist's touches. Based on this use case, the end-to-end delay, defined as the time difference between the virtual animal touched by the therapist and the haptic feedback received by the patient, is measured. In addition, the authors conduct experiments with several scenarios by varying the locations of the architecture components as well as those of the therapist and patient. The experimental results indicate that the end-to-end latency increases as the distance between the therapist and patient increases. Moreover, the lowest end-to-end latency that can be achieved when all components are running locally is 6 ms. Therefore, the work does not meet the ultra-low latency criterion ($C_1$). It does not cover the ultra-high reliability criterion ($C_2$) because there is no discussion about the reliability aspect.

*3) Video Surveillance*

A video surveillance system that requires an immediate response of 1 ms and a failure rate of $10^{-7}$ is considered in Grasso *et al.* [103]. The authors propose its use on unmanned aerial vehicles with the computation capacity to perform local data processing close to sensors and actuators and thus reduce the delay. In this system, a target area is divided into multiple sub-areas. Each sub-area is monitored by one unmanned aerial vehicle (also known as a drone) and several sensors/actuators located on the ground. The drone is equipped with a microcontroller and a microcomputer for local data processing in addition to its HD camera and a transmit/receive block. The microcontroller is responsible for combining the sensed data received from the sensors on the ground and from the drone's camera at each time interval into a single task. The task is then executed by the microcomputer to detect any critical situations. If a critical situation is detected, the drone sends an alert message to actuators located on the ground without waiting for a human decision. In addition, a drone can request neighboring drones to share its workload if it becomes overloaded. Using a Markov chain, the authors model the proposed system to evaluate its performance. The results show that the proposed system can provide an average response time delay of 1 ms and a job loss probability of $10^{-7}$, indicating the system's high reliability achieved by increasing the drones' processor capacity. Therefore, this work meets both the ultra-low latency ($C_1$) and ultra-high reliability ($C_2$) criteria.

*4) Industrial Automation*

Mekikis *et al.* [104] propose a testbed to investigate the capabilities of NFV and SDN-based networks for supporting Tactile Internet industrial applications. The testbed relies on a well-defined Smart End-to-end Massive IoT Interoperability, Connectivity and Security framework [107] which consists of three main layers, denoted as backend/cloud, field and network layers. In the testbed, the backend/cloud layer includes computing and storage nodes, called the local cloud, on which VNF, database, and NFV management and orchestration modules are deployed. The field layer is a virtualized industrial IoT gateway that connects sensors and actuators in a field. The virtualized industrial IoT gateway can also act as a MEC node on which VNFs are hosted to reduce the delay due to its proximity to the sensors/actuators. Finally, the network layer, comprised of s SDN controllers and SDN switches, is responsible for managing the traffic of each application traversing between the backend/cloud and the field layers. This layer allows the creation of network slices to support multiple applications by reserving traffic bandwidth for each application with respect to its QoS requirement through the SDN switches. The authors evaluate the performance of the proposed testbed in terms of bandwidth reservation configuration, VNF migrations and round-trip delay. In the testbed, two applications are considered; each of which requires one VNF. The experimental results show that the bandwidth reservation can be reconfigured successfully among the applications, and that there is no service disruption during the migration while both applications are running. In addition, a round-trip delay on order of 1 ms can be achieved by hosting the VNFs at the virtualized industrial IoT gateway. This work thus meets the ultra-low latency criterion ($C_1$). However, a performance evaluation for the case of multiple VNFs in a service chain needs to be further investigated. We note that this work also meets the ultra-high reliability criterion ($C_2$) because it shows that there is zero downtime during the migration, which indicates the reliability of the system.

*5) Smart Cities*

Smart cities are another field of applications that will benefit from the emerging Tactile Internet. To ensure users' satisfaction, Wei *et al.* [105] propose a quality of experience (QoE)-driven Tactile Internet architecture for smart cities. The architecture is designed to not only achieve low latency and high reliability, but to also improve users' QoE by exploiting audio, visual and haptic information. The architecture contains five layers: sensing, transmission, storage, computing and application. The sensing layer consists of distributed sensors and actuators that are responsible for collecting audio, visual and haptic information. The transmission layer relies on a generalized frequency division multiplexing-based system that can achieve low latency due to its flexibility in reducing the transmitted signal length. The storage layer is based on a distributed storage approach in which data is stored in distributed edge datacenters to reduce the time for users to receive their required data. A blockchain technique is also adopted to increase safety and privacy. The computing layer adopts osmotic computing, in which an application is first decomposed into microservices and then these microservices are dynamically distributed across cloud and edge datacenters depending on their characteristics. Finally, the application layer is responsible for QoE management. In this layer, users' QoE can be perceived by analyzing user preference and QoS parameters (e.g., throughput, jitter and packet loss rate). The user preference can be determined through historical data associated with a user's behavior and context information. Based on the obtained QoE, intelligent decision-making is performed to promote the QoE, such as making adjustments in the QoS parameters. The authors also introduce a QoE management framework based on a broad learning system to predict the user's QoE. Simulations were conducted to evaluate the performance of the proposed broad learning system. These simulations show that the proposed system outperforms deep learning and decision tree systems in terms of both accuracy and training time. Even though the authors claim that their proposed architecture can achieve low latency and high reliability due to the selected techniques applied in each layer, its delay and reliability performance is not explicitly evaluated. Therefore, this work does not cover either of the criteria.

*6) General E-health Systems*

To support general health Tactile Internet applications, EPON technology is a viable solution to provide a low latency and reliable network architecture due to its high capacity and flexibility. Toward this end, Wong *et al.* [106] propose an EPON-based local area network architecture as a networking infrastructure to serve the health Tactile Internet applications in a hospital campus. It should be noted that unlike the OCLDN [99] discussed in Section III.A.4 that leverages on NG-EPON, this architecture relies on an IEEE 802.3av 10 Gb/s EPON technology. It mainly consists of multiple ONUs and a single OLT. The ONTs are distributed all around the campus to interface with wireless APs in body area networks. Each of the ONTs is also connected to the single OLT, located in a CO (e.g., the main hospital building), via fiber links. The OLT acts as a gateway to/from wide area networks. It is co-located with a tactile control server that is responsible for accommodating caching prediction, resource allocation prediction for the control/steering of Tactile Internet traffic, human action prediction and other related predictions. The authors introduce a predictive resource allocation algorithm that dynamically adjusts the number of active wavelengths based on traffic prediction. The algorithm increases the number of active wavelengths at high traffic loads to maintain the required delay and reduces them at light traffic loads to improve the energy efficiency. The detail of the proposed algorithm will be discussed later in Section V.B. Wong *et al.* [106] do not evaluate the performance of their proposed architecture separately from the resource allocation framework, and so it does not cover either of the criteria.

*C. Protocols*

In this subsection, we review protocols proposed to date for the Tactile Internet. These protocols include a migration protocol over MEC [108], a handoff protocol [109] and a haptic handshake scheme for orchestrating heterogeneous tactile devices [110]. The reader should note that no protocol has been proposed so far by the IEEE P1918.1 standard, although interfaces have been identified [3]. Note also that the protocols for haptic communications are not discussed here, as they are all extensively reviewed in reference [21], except for reference [110], which has been published after reference [21].

Braun *et al.* [108] introduce an application-level migration protocol called Agile Cloud Migration (ACM) for Tactile Internet applications. This protocol aims at providing fast migrations with zero downtime over MECs to ensure that applications (e.g., a control server for autonomous driving) are always close to users. The key approach of this protocol is to transfer only the state of the application during migrations, thereby optimizing the amount of data transmitted. However, this protocol requires that applications be designed such that their engine is isolated from their state. The authors present both their architecture and the migration procedure. The architecture consists of three types of entities. The first type is a centralized controller that is responsible for initiating and orchestrating migrations. The second type is composed of clients that act as client users. The last entity type are engines in which an application server resides. A prototype was developed and implemented to evaluate the performance of the ACM protocol. This prototype was developed based on a browser-based multiplayer gaming application, where the game server can be hosted on the edge cloud and Amazon Web Services platforms. Several migrations between different cities were carried out to assess its performance. The results demonstrate that the ACM protocol has less migration overhead than KVM and Docker migrations, resulting in faster migration time, and reveal that the ACM protocol has no downtime during the migrations. The authors do not discuss the communication delay aspect, and so this work does not cover the ultra-low latency criterion ($C_1$). However, the ACM protocol can provide zero downtime, which guarantees the system availability during migration. Therefore, it meets the ultra-high reliability criterion ($C_2$).

Du *et al.* [109] propose a handoff protocol for vehicular to infrastructure (V2I) communication. They assume that vehicles can be connected to both LTE and IEEE 802.11p networks. Their goal is to select the network to which vehicles should be connected and how to seamlessly switch connections between those two networks to ensure that the vehicles experience a minimum delay. Being able to utilize both of these networks makes it possible to best use each of their advantages and to minimize their disadvantages. The LTE network provides better delay performance than the IEEE 802.11p network when V2I distance is large and vehicle density is high, thanks to its broader coverage and higher capacity. However, it is worse than the IEEE 802.11p network when the V2I distance is smaller and vehicle density is lower, due to its handshaking protocol during the channel access procedure. The authors first conducted extensive simulations to develop relationship models between the delay and vehicles' positions over both networks. Based on these models, they derived a position-based switching point. Finally, they propose how to define an alert zone, in which vehicles should initiate a handoff procedure. The authors evaluate the proposed protocol based on simulations. Their results indicate that the proposed protocol can seamlessly switch connections between these two networks. Their protocol also ensures that vehicles are always connected to the network that provides the minimum delay at a given time. However, this work cannot offer delay that is on the order of 1 ms, and so it does not meet the ultra-low latency criterion ($C_1$). For the reliability aspect, the proposed protocol can provide zero downtime during handoffs, guaranteeing vehicular connectivity. Thus, this work meets the ultra-high reliability criterion ($C_2$).

Iiyoshi *et al.* [110] focus on a haptic handshake scheme for orchestration between heterogeneous tactile devices participating in a Tactile Internet interaction. In general, these devices often have different specifications and requirements in terms of sensing and display, data compression compatibility, and application requirements, among others. Before a Tactile Internet application connection is set up, every node must be aware of the capabilities and requirements of other nodes so that an adequate amount of information can be transmitted in an efficient manner. For instance, a tactile node with both tactile and kinesthetic sensing capabilities should only transmit the tactile information to receiving tactile nodes that have appropriate tactile display capabilities. In fact, this mandates the need for designing a comprehensive handshaking mechanism, through which the participating tactile nodes advertise their capabilities and requirements, thereby reaching a consensus on the relevant parameters for realizing proper interaction. To tackle this issue, Iiyoshi *et al.* [110] designed a haptic handshake mechanism to coordinate between the (often) heterogeneous tactile devices. They also developed a messaging format, referred to as Tactile Internet message, with which to exchange metadata during haptic handshake. The authors experimentally validated their proposed protocol on a WebRTC-based platform and real-world haptic devices, achieving a mean handshake latency of 47.25 ms. This work focuses on evaluating the performance of the proposed mechanism in a connection setup phase rather than that in an operation phase. Therefore, it does not cover any of the criteria.

### D. Intelligent Prediction Schemes

As AI techniques evolve, their integration into future mobile communication networks in general and the Tactile Internet in particular is inevitable. According to [6], AI is a prominent solution towards getting around the excessive delay caused by the limitations mainly risen by the speed of light. In particular, the AI engines can be cached and run in close proximity of the Tactile Internet users and help them experience a seamless, immersive experience. The research on the application of AI in the Tactile Internet has just started to receive attention (e.g., [49], [50], [61], [111]).

The authors of [49] and [50] have investigated the feasibility of using multi-layer perceptron (MLP) artificial neural networks in performing multi-sample ahead-of-time forecasting of force/torque samples in teleoperation systems, with the main objective of decoupling haptic feedback from the impact of excessive latencies. More specifically, after investigating the potential and limits of teleoperation coexistent with conventional H2H and M2M traffics in Fi-Wi enhanced LTE-A HetNets, the authors of [49] have leveraged the AI-enhanced edge servers collocated with ONU-APs/BSs to run their artificial neural network (ANN)-based force sample predictors within a 1 ms waiting threshold to realize an increased togetherness between the human operators and the corresponding teleoperator robots. According to the results, for non-local teleoperation, the proposed ANN-based predictor decreases the mean squared error from $0.9\times10^{-3}$ to $0.65\times10^{-3}$, translating into an improvement of 27.8%. For local teleoperation, it is able to keep the mean squared error close to zero (between $0.006\times10^{-3}$ and $0.007\times10^{-3}$) for low to medium background traffic loads. While this paper meets both ultra-low latency ($C_1$) and ultra-high reliability ($C_2$) criteria for the local teleoperation, it only meets the ultra-low latency criterion ($C_1$) for the non-local teleoperation.

More recently, a hidden Markov model (HMM) is used in [111] to encapsulate a set of expert force/torque profiles and corresponding parameters during an offline training process, followed by exploiting a Gaussian mixture regression to reproduce a generalized version of the force/torque profile during the prediction of the lost haptic feedback information in remote needle insertion application. HMM is used to account for the variability of the spatial and temporal data. Given the training sequence and model parameters, the Baum-Welch algorithm is used to estimate the optimum parameters of the developed HMM model. Next, the encoded HMM parameters were stored in a library to be used for the online retrieval of force/torque profiles. Finally, Gaussian mixture regression is used to retrieve the generalized version of the force/torque profile and corresponding parameters encoded during the training using HMM. Predictions can then be made by querying the temporal information for the corresponding spatial information estimated via regression. According to the results, the prediction times are kept below 1 ms. Further, the prediction accuracy measured by the root mean squared error does not exceed $7.5\times10^{-3}$. As a result, while this paper meets the ultra-low latency criterion ($C_1$), it does not meet the ultra-high reliability criterion ($C_2$).

In a recent experimental work [61] (which is the extension of [112] and [113]), the authors have leveraged on the emulation of the remote environment to alleviate the detrimental impact of the physical distance between master and slave from the perceived latency. More specifically, AI-embedded cloudlets comprising H2M servers are deployed to predict the haptic feedback coming from the salve and then deliver the predicted feedback to the human operator. The feasibility of the developed approach is examined in a VR-based teleoperation setup, where the human operator (i.e., master) interacts with the virtual environment via a pair of VR gloves, which helps a human touch a virtual ball (i.e., slave) on a computer running the VR application. The forecasting and delivery of haptic feedback from slave to master are realized via the so-called event-based haptic sample forecast module, which relies on a two-stage AI model. The first stage comprises a set of ANN-based binary classifiers (deployed per finger), which are responsible for detecting the instants at which a touch event is triggered by the human operator. The second stage is responsible for forecasting and delivering the haptic feedback to the human operator using a reinforcement learning scheme. The results show that the proposed ANN-based classifier can detect the touch event with an accuracy of ~99%, while the reinforcement learning-based module can achieve a prediction accuracy of 87% to 100% for 4 and 1 different material options, respectively. As a result, this work meets both the ultra-low latency ($C_1$) and the ultra-high reliability ($C_2$) criteria.

*E. Summary, Insights and Lessons Learned*

In this subsection, we first provide a short summary of architectures, protocols and intelligent prediction schemes proposed for the Tactile Internet. Then, we discuss insights and lessons learned from our literature review.

*1) Summary*

While multiple works have introduced application-agnostic and application-specific architectures for the Tactile Internet, a few have proposed protocols as architectural building blocks as well as intelligent prediction schemes. Table III provides a summary of the key contributions of the architectures, protocols and intelligent prediction schemes reviewed in this section.

A standard architecture has been proposed by IEEE P1918.1 [3]. It aims at interoperability and defines functional entities and interfaces. However, it does not meet any of the criteria in itself because the implementation of the functional entities is outside the scope of standardization. Four of the applications-agnostic architectures harness MEC in a cellular environment [94]-[97], one of them harnesses Wi-Fi in a cellular environment [98], and the sixth relies on EPON [99]. The first application agnostic architecture [94] that harnesses MEC in a cellular environment relies on a multi-level cloud system for computation offloading. The second [95] builds on the first by leveraging SDN in addition to cloud computing. The third [96] relies on a DNA network for traffic offloading and the fourth [97] proposes a new low latency approach for service function chaining. The only architecture proposed to date for harnessing Wi-Fi in cellular environment aims at enhancing LTE-A Hetnets with Wi-Fi access for high capacity and traffic offloading purposes [98]. On the other hand, the architecture based on NG-EPONs allows optical network units to transmit over multiple wavelength channels at the same time to improve the network capacity [99].

Two of the proposed application-specific architectures focus on real-time tele-operations [50][100], two on virtual reality [101] [102], one on real-time video surveillance [103], one on industrial automation [104] and one on smart city applications [105], and one on general health applications [106]. One of the architectures that deals with tele-operations [50] combines the Wi-Fi and EPON-enhanced cellular network architecture presented in reference [98] with AI-enhanced MEC. Another relies on sub-GHz technology for the transmission of tactile feedback and robotic control [100]. As far as virtual reality is concerned, one of the proposals builds on MEC and mmWave communications for a multiplayer VR game [101]. The second [102] builds on fog computing for a remote VR phobia treatment application. For video surveillance, a proposed architecture [103] aims at a system with very stringent requirements (1 ms response time, $10^{-7}$ failure rate). The architecture for industrial automation [104] aims at providing a testbed for evaluating the use of NFV and SDN networks for industrial applications. The architecture for smart city applications [105] not only aims at achieving low latency and high reliability, but also at enabling high QoE. Finally, the sixth one relies on EPONs and aims at health applications in a campus hospital [106].

A migration protocol over MEC [108] and a handoff protocol [109] are the two protocols proposed so far for Tactile Internet, besides the protocols proposed for the haptic aspects. The migration protocol aims at transferring only the application state over MEC, with a goal of zero downtime [108]. The second protocol produces a handoff between vehicles and infrastructure, with the assumption that the vehicles can be connected to both LTE and 802.11p networks [109]. A recent haptic communication protocol worth mentioning is the haptic handshake scheme proposed in reference [110]. It aims at orchestration between heterogeneous tactile devices.

As for the intelligent prediction schemes, MLP-ANN is exploited in [49] and [50] to predict the delayed and/or lost samples in the feedback path of a teleoperation system. It is shown that the proposed edge sample forecaster is capable of making accurate predictions within the 1 ms deadline. Moreover, after training a HMM model with a set of expert force/torque profiles, the authors of [111] made predictions using a Gaussian mixture regression during a remote needle insertion in robotic telesurgery applications. Further, the authors of [61] have designed a two-stage AI model, where a set of ANN-based binary classifiers are used along with reinforcement learning to forecast the haptic feedback.

TABLE III
The main contributions and criteria for works proposing architectures, protocols and intelligent prediction schemes for the Tactile Internet. In the criteria columns, Y indicates that the criterion is met, P indicates that the criterion is partially met, and N indicates that the criterion is not met

| Scope | References | Main Contributions | $C_1$ | $C_2$ |
|---|---|---|---|---|
| Application-agnostic Architectures | Holland et al. [3] | Propose a generic Tactile Internet architecture which aims at defining functional entities and interfaces between the entities to ensure interoperability. | N | N |
| | Ateya et al. [94] | Introduce a multi-level cloud system in cellular networks to provide offloading capabilities. | Y | N |
| | Ateya et al. [95] | Propose a core network that leverages SDNs and a multi-level cloud system. | Y | N |
| | Shafigh et al. [96] | Propose a radio access network that integrates DNA networks with a CRAN network. | N | N |
| | Xiang et al. [97] | Propose a new design approach to implement low-latency service function chains in NFV-based MEC systems for Tactile Internet services. | Y | N |
| | Beyranvand et al. [98] | Propose a Fi-Wi-enhanced LTE-A HetNets network that utilizes EPON and Gigabit WLAN technologies in LTE-A HetNets networks. | Y | Y |
| | Neaime et al. [99] | Propose a high-speed, programable and scalable optical access network that relies on a next generation EPON and a mini datacenter. | N | N |
| Application-specific Architectures | Maier et al. [50] | Introduce a Fi-Wi-enhanced LTE-A HetNets network with AI-enhanced MEC to support teleoperation. | Y | N |
| | Holland et al. [100] | Propose a system architecture that utilizes sub-GHz technology and mobile networks to support teleoperation. | N | N |
| | Elbamby et al. [101] | Propose a system architecture based on MEC and mmWave technologies for multiplayer VR interactive gaming in an indoor VR gaming arcade. | N | N |
| | Jebbar et al. [102] | Introduce a fog-based architecture for a remote VR phobia treatment application. | N | N |
| | Grasso et al. [103] | Propose a video surveillance system that relies on unmanned aerial vehicles with computation capability. | Y | Y |
| | Mekikis et al. [104] | Introduce an NFV and SDN-based testbed for Tactile Internet industrial applications. | Y | Y |
| | Wei et al. [105] | Propose a QoE-driven Tactile Internet architecture for smart cities. | N | N |
| | Wong et al. [106] | Introduce an EPON-based local area network for general health applications in a hospital campus. | N | N |
| Protocols | Braun et al. [108] | Propose an application-level migration protocol aiming at ensuring fast migration and zero migration downtime. | N | Y |
| | Du et al. [109] | Propose a handoff protocol to seamlessly switch connections between LTE and IEEE 802.11p networks. | N | Y |
| | Iiyoshi et al. [110] | Propose a haptic handshake scheme to orchestrate heterogeneous tactile devices involved in Tactile Internet interaction. | N | N |
| Intelligent Prediction | Maier et al. [49][50] | Propose an MLP ANN-based scheme to forecast delayed and/or lost force/torque feedback for 1- and 6- DoF teleoperation systems in Fi-Wi enhanced LTE-A HetNets with AI-enhanced MEC. | Y | Y |
| | Boabang et al. [111] | Introduce an HMM-based haptic feedback prediction for the remote needle insertion application. | Y | N |
| | Mondal et al. [61] [112] [113] | Propose a two-stage AI model based on ANN-based binary reinforcement learning to forecast haptic feedback in a VR-based teleoperation system. | Y | Y |

*2) Insights and Lessons Learned*

A first insight is that the number of papers published so far on the architectural aspects of the Tactile Internet is relatively low: a grand total of 15 papers from 2014 to early 2020, 7 on application-agnostic architectures and 8 on application-specific architectures. In comparison, the number of papers published on fog computing through 2017 (i.e., 2013-2017) was 32, 19 on application-agnostic architectures and 16 on application-specific architectures [73]. A lesson we can learn from this is that the architectural aspects of the Tactile Internet are still being investigated. Much more research is needed, especially on the application-agnostic aspects.

Another insight is that the most popular techniques used to date for meeting the requirements are offloading and running components at the edge. Offloading consists of offloading either computation [94][95][101][103] or traffic [96][98]. This is currently the case for both application-agnostic architectures [94][95][96][98] and application-specific ones [101][103]. In references [97][50][102][104], the technique consists of running components at the edge. However, the outcome so far is not compelling, as none of these proposals meets both criteria. A lesson we can learn here is that research in this area needs to go beyond offloading and running components at the edge of the network. Some of the potential avenues will be sketched in the research directions section.

A third insight is that most of the available works do not provide experimental validation, even though experimental validation is a sine qua condition for any sound architecture design. Instead, they rely on simulations and do not even include proof-of-concept prototypes. The three notable exceptions are references [97], [102] and [104], which include prototypes and measurements made on the prototypes. The lesson we can infer from this observation is that researchers should aim more and more at comprehensive architectures that include experimental validation.

An insight from the protocol perspective is that work remains scarce, aside from the protocols for haptic communications which are well discussed in reference [21]. To date, this area only contains two proposals, one targeting the Tactile Internet at large [108] and the other Tactile Internet for a specific application, vehicular networks [109]. Clearly, more research is required in this area.

An important insight learned from the intelligent prediction schemes is the feasibility of using ANN-based classifiers and predictors in the context of haptic-enabled VR, teleoperation, and telesurgery applications. Note, however, that the research in this avenue is still in its infancy and much further efforts are needed to design more sophisticated prediction modules with the capability of being generalized to a larger group of applications and user behaviors.

## IV. Radio Resource Allocation Algorithms for the Tactile Internet

Significant research efforts have focused on radio resource allocation algorithms to achieve ultra-low latency and ultra-high reliability in cellular RANs for the Tactile Internet. These algorithms can be classified as: (i) uplink transmission, (ii) downlink transmission, and (iii) joint uplink and downlink transmission algorithms. Furthermore, some works have leveraged on the network slicing concept to design efficient radio resource allocation schemes by creating multiple logical radio networks to guarantee the unique requirements of Tactile Internet applications. In this section, we first review the radio resource allocation algorithms designed for uplink only, for downlink only, and then those for joint uplink and downlink. Next, we review the network slicing-based radio resource allocation algorithms. Finally, we provide some insights and draw the important lessons that can be learned from this section.

### A. Uplink Transmission

In cellular wireless networks, a key challenge is to address the problem of multiple access, especially in the uplink direction, where the user equipment (UE) needs to enter a scheduling and grant procedure. Typically, this procedure starts when a UE sends a grant request to the BS and then waits for the scheduling grants (i.e., radio resources) before transmitting its data. In the downlink direction, however, the BS transmits the downlink data immediately after the data reaches the buffer. Clearly, the contribution of the uplink latency in total end-to-end latency in RANs is much larger than that of the downlink latency, especially given that the contention in the uplink direction may lead to failures and retransmissions. This has spurred a great deal of interest in designing low-latency radio resource allocation mechanisms for the uplink direction. To support low-latency Tactile Internet applications in 5G cellular networks, several algorithmic works have proposed novel radio resource allocation algorithms for the uplink transmission in order to meet the ultra-low latency as well as the ultra-high reliability requirements.

The uplink radio resource allocation algorithms for the Tactile Internet can be either grant-based [114]-[116] or grant-free [117]-[121] algorithms. In grant-based schemes, the UE is required to enter a scheduling and grant procedure for uplink transmission. As a result, the main objective of these algorithms is to optimally allocate the radio resources upon the reception of a grant request from each UE to improve the latency and the reliability. On the other hand, the grant-free algorithms aim to reduce the uplink delay by eliminating the scheduling delay in the scheduling and grant procedure. An effective way to achieve this is to pre-reserve the resources for the UEs such that each UE can send packets immediately whenever it has packets to send. However, reserving dedicated resources for each UE may lead to low resource usage efficiency. Therefore, the grant-free algorithms mainly focus on how to efficiently utilize the resources while achieving low uplink delay. In the following, we discuss the grant-based algorithms, followed by the grant-free algorithms in more detail.

A grant-based radio resource allocation algorithm is proposed in She *et al*. [114], where the authors aim to optimize the uplink transmission in the context of mMTC. They consider a single BS with multiple antennas serving multiple single-antenna MTC devices. In the uplink transmission procedure, the device first sends a grant request to the BS, and then the BS assigns multiple subchannels to the device. Next, the device selects one of the assigned subchannels, whose instantaneous channel gain is greater than a given threshold for the uplink transmission. The threshold is derived such that the packet can be transmitted successfully; otherwise, there is a packet loss. The authors optimize the number of subchannels, the sub-channel bandwidth, and the threshold for each device accordingly by formulating an optimization problem with the objective of minimizing the total required bandwidth while ensuring the reliability. To solve the problem, it is first decomposed into multiple single-device problems. An exact linear search algorithm is then proposed to solve the single-device problem for each device separately and then merge the results. The evaluation results show that the proposed solution leads to a smaller total required bandwidth compared to the bandwidth reservation scheme, where the bandwidth is reserved for each device. The authors do not consider the end-to-end delay. Instead, they consider an uplink delay threshold of 0.3 ms. In addition, they consider an uplink packet loss probability threshold of $10^{-7}$. Therefore, the ultra-low latency ($C_1$) and ultra-high reliability ($C_2$) criteria are partially met in this work.

Similarly, Zhou *et al*. [115] propose an access control and resource allocation algorithm for uplink transmission in MTCs. They focus on a scenario with a single BS serving multiple MTC devices. This algorithm consists of two stages: access control and resource allocation. For the first stage, the authors design a contract-based incentive mechanism to reduce the peak-time access demands by motivating delay-tolerant devices to postpone their access demands in exchange for higher access probabilities. For the second stage, they propose long-term cross-layer online resource allocation algorithm to find the optimal sensing rate control in the application layer as well as the optimal power allocation and channel selection in the physical layer. This resource allocation algorithm aims to maximize the time-averaged satisfaction of all devices while satisfying the long-term constraints related to power consumption, transmission delay, and backlog queue stability. To solve the problem, it is transformed into a series of two short-term sub-problems: (i) sensing rate control and (ii) joint power allocation and channel selection using Lyapunov optimization. The former is a convex problem, which can be solved using the well-known KKT approach, whereas the latter is NP-hard and is solved using a pricing-based stable matching approach. The simulation results indicate that the proposed access control can effectively flatten the peak-time access demand when a high number of delay-tolerant and sensitive devices attempt to access the network simultaneously. Moreover, the proposed resource allocation algorithm has better delay and energy efficiency performance over time than the existing algorithm, which maximizes the transmission rate without considering the long-term constraints or the sensing rate control. However, the delay requirement considered in this paper is on the order of a second, and so it does not meet our ultra-low latency criterion ($C_1$). Also, it does not cover the ultra-high reliability criterion ($C_2$) either.

Recently, we are witnessing the growth of scaled-up research towards adopting the concept of small-cell networks to improve the coverage, latency, and throughput of cellular networks, making this a promising solution for Tactile Internet applications with stringent QoS requirements [122]. In the context of networked microgrids, where a small end-to-end delay of control messages is needed, Elsayed *et al.* [116] study the joint problem of resource allocation and user association for uplink transmission. The authors consider the coexistence of both critical user devices (e.g., microgrid controllers) and non-critical users (e.g., conventional user equipment) over a small-cell wireless network. This work leverages deep Q-networks (DQNs), which are deep neural networks used for estimating the quality value (Q-value) function of Q-learning algorithms. To balance the trade-off between allocating more resource blocks (RBs) and associating the critical user devices to BSs with higher channel quality, they propose a distributed deep Q-learning-based algorithm, which is referred to as a delay minimizing deep Q-network (DM-DQN). The results indicate that the DM-DQN scheme achieves a 41% reduction of the delay for critical users, even under heavy traffic loads. Their results also show that the proposed DM-DQN scheme not only converges faster than the Q-learning scheme, but it is also capable of keeping the average packet delay of critical users below 15 ms. Given that the smallest achievable packet delay is around 5 ms and the packet loss rate is not even considered, this work does not meet either of our criteria.

Unlike [114], [115] and [116], Ma *et al.* [117] propose a grant-free multiple access scheme, in which tactile users do not need to enter a scheduling and grant procedure for uplink transmission to improve the uplink delay. The proposed scheme also adopts frequency diversity, where the tactile users transmit duplicated data via multiple subcarriers to improve the reliability. The authors assume that tactile users have a sporadic traffic pattern, meaning that most of the time they do not have packets to transmit, while non-tactile users have a non-sporadic traffic pattern. In their proposed scheme, tactile users are assigned a dedicated group of subcarriers, which last for a relatively long time without entering the scheduling and granting procedure, thereby reducing the delay. The tactile users also transmit the same data via all subcarriers in the same group to reduce the transmission error, given that subcarriers in each group have independent fading. On the other hand, non-tactile users have access to a group of subcarriers for uplink transmission through a scheduling and grant procedure. A subcarrier index modulation technique is also adopted to improve the spectral efficiency. The authors also propose a low-complexity maximum likelihood detection algorithm, which exploits the traffic characteristics of both types of users. They analyze the BER expression in an additive white Gaussian noise (AWGN) channel and conduct simulations over an LTE Extended Vehicular A model channel to evaluate the proposed scheme in a multipath fading environment. As a baseline, the authors have considered the conventional orthogonal scheme, where the subcarrier index modulation technique is not considered. The simulation results indicate that the proposed scheme outperforms the orthogonal scheme for tactile users in terms of the BER. For instance, for SNR = 10 dB, the BER of the proposed scheme is $10^{-4}$, while that of the conventional orthogonal scheme is $10^{-2}$ in the LTE Extended Vehicular A model channels. We note that the authors claim that the uplink delay of tactile users is equivalate to a few transmission time intervals or time slots, since the delay in the scheduling and grant procedure can be eliminated. However, they do not show explicitly that the proposed scheme can achieve the uplink delay on the order of 1 ms. Thus, this work does not meet the ultra-low latency criterion ($C_1$). Also, it does not meet the ultra-high reliability criterion ($C_2$) because the proposed scheme is not shown to ensure a reliability of $10^{-7}$.

Likewise, Condoluci *et al.* [118] introduce a soft resource reservation scheme for uplink transmission which extends the legacy LTE scheduling and grant scheme in order to meet Tactile Internet latency requirements. The proposed scheme can be divided into two steps. The first step operates almost like the legacy scheme, where the haptic device sends a grant request to the BS for the soft reservation, and then waits for a grant response from the BS before transmitting the uplink packets. In the second step, for the incoming packets, the haptic device sends a grant request to the BS followed by an immediate transmission of the packets into the resources assigned in the first grant response, without waiting for a new grant response. Given that the resources were already reserved via the first grant request, the latency is significantly reduced. However, if the BS does not receive the grant request from the haptic device, the BS will realize that this device does not have any packets to transmit at this time, and so the BS can release the resources previously reserved for this device to serve other devices in order to avoid resource underutilization. Evaluated in a simulative environment with multiple haptic devices under a single BS, the proposed strategy is shown to outperform the legacy procedures in terms of latency. However, the results show that an end-to-end delay remains on the order of several milliseconds, with a minimum obtained value of 4.5 ms. Therefore, this work does not meet the ultra-low latency criterion ($C_1$). In addition, the authors do not discuss any reliability-related metric and thus this work does not cover the ultra-high reliability criterion ($C_2$).

As discussed earlier, grant-free multiple access schemes may exhibit poor resource usage efficiency. This happens especially for Tactile Internet applications with bursty traffic [118], as the bandwidth generally needs to be pre-reserved for tactile devices. To efficiently handle the burstiness of tactile traffic for uplink transmission, Hou *et al.* [119] propose a bandwidth allocation algorithm that optimally allocates the bandwidth in each next time slot using traffic prediction, with the objective of minimizing the bandwidth usage while meeting the given latency and reliability constraints. The authors consider tactile users communicating with a multi-antenna BS. They also model the burstiness of tactile traffic, using a switched Poisson process. In the proposed algorithm, the BS first classifies tactile users into high- and low-traffic users based on their history. Two classification methods are considered: (i) the Neyman-Pearson-based method, and (ii) a *k*-means-based unsupervised learning method. Next, the BS allocates a dedicated bandwidth for high-traffic users, as opposed to low-traffic users, who are only allowed to use bandwidth sharing. In addition, the authors formulate the bandwidth optimization problem with respect to the two types of users and then propose a three-step method to solve this problem. It is worth mentioning that an error correction mechanism is also proposed to readjust the required bandwidth when users are misclassified. The simulation results indicate that both the Neyman-Pearson and the *k*-means methods require lower bandwidth than the baseline method, where a dedicated bandwidth is allocated to all users without any classification. It is also shown that the required bandwidth in both algorithms decreases as the severity of burstiness increases. The authors do not consider an end-to-end delay, and instead ensure an uplink latency threshold of 0.5 ms. Therefore, this work partially meets the ultra-low latency criterion ($C_1$). As for reliability, the authors consider a packet loss probability of only $10^{-5}$, and so this work does not meet the ultra-high reliability criterion ($C_2$).

The authors of Feng *et al.* [120] introduce a grant-free, predictive semi-persistent scheduling (SPS) scheme to handle tactile traffic for uplink transmission. The proposed scheme not only schedules the transmission resources more efficiently compared to the existing adaptive SPS schemes, it also consumes a minimal physical downlink control channel resource, making it particularly beneficial for the transmission of other latency-tolerant data transmissions. To be more specific, the authors leverage deadband coding to reduce the incoming traffic load and then propose a novel predictive SPS scheme that exploits the transmission history to effectively reduce the uplink latency of LTE systems to below 5 ms by conducting multi-timeslot-ahead prediction. The simulation results demonstrate that the proposed predictive SPS scheme can outperform conventional scheduling and existing SPS schemes in terms of uplink latency. This scheme achieves an uplink latency of close to 4 ms under a wide range of traffic loads, and so, this work does not meet the ultra-low latency criterion ($C_1$). The authors do not consider the reliability aspect and therefore the ultra-high reliability criterion ($C_2$) is not covered.

NOMA is an emerging technology to improve the spectral efficiency, which is the main challenge of typical orthogonal multiple access (OMA) systems. The basic idea of NOMA is to serve multiple users in the same radio resource (e.g., a subcarrier) while relying on a highly complex mechanism in a receiver to separate the non-orthogonal signals of different users. To reap the benefits of NOMA to achieve high spectral efficiency, a grant-free NOMA access scheme for the uplink transmission is proposed by Ye *et al.* [121]. We note that the grant-free access mechanism in NOMA can lead to reliability degradation due to unexpected inter-user interference. To optimize the reliability, the authors develop a variational optimization problem. Due to the high complexity of the problem, they propose a neural network model based on a deep variational auto-encoder to parameterize the problem, which is generally intractable. The proposed model contains both encoding and decoding networks. The encoding network, which has one input and four hidden layers, models the random user activation and the symbol spreading in each time interval. Meanwhile, the decoding network, with four hidden layers, one reconstructed symbol output layer, and one activity estimation output layer, estimates the active users and the amount of transmitted data. A multi-loss-based network training algorithm is introduced, which utilizes historic data on the probability of user activation to train the proposed model offline. The simulation results indicate that the proposed scheme outperforms the conventional grant-free NOMA scheme, with a 5 dB gain at the same symbol error rate (SER). In addition, an SER of $10^{-3}$ can be achieved, even for higher-order modulations, indicating robustness in a larger-sized model. In this work, although the authors claim that the proposed scheme can achieve a low latency, the end-to-end packet delay is not evaluated. Therefore, it does not cover the ultra-low latency criterion ($C_1$). Also, the best achievable SER of the proposed scheme reported in this work is $10^{-3}$, and so it does not meet the ultra-high reliability criterion ($C_2$).

### B. Downlink Transmission

As discussed above, downlink delay is relatively smaller than uplink delay in RANs. However, to achieve ultra-low end-to-end delay on the order of 1 ms, the downlink delay may not be negligible, especially since the downlink queuing delay could grow longer as the incoming aggregated traffic to the BS increases. In addition, the heterogeneity of the service requirements of different packets may be leveraged to further reduce the latency experienced by the delay-sensitive packets with tighter deadlines than those of the delay-tolerant packets. The above discussion clearly mandates the need to design proper allocation mechanisms in the downlink direction, with the main objective of ensuring a given latency budget for tactile packets. In the context of the Tactile Internet, only two algorithmic works address the radio resource allocation problem in the downlink direction.

Khorov *et al.* [123] study the radio resource scheduling problem for downlink transmission to handle low-latency communication (LLC) traffic generated by Tactile Internet and Industrial Internet applications. While considering an LTE system architecture, the authors leverage low-rate modulation and coding schemes and very short transmission time intervals (TTIs) to decrease the airtime scheduling and signal processing delay. They evaluate two different schedulers. The first scheduler, referred to as the LLC – maximum rate (LLC-MR), is designed to maximize the goodput of users with LLC traffic across all TTIs before packet expiration. The second, referred to as the LLC – proportional fair (LLC-PF) scheduler, creates proportional fairness among LLC users across all TTIs before packet expiration times. Utilizing the objectives of both schedulers, the authors model the problem via an integer linear programing (ILP) formulation with the main focus on ensuring that the packet deadlines can be met. A novel dynamic programming-based algorithm is proposed to solve this ILP problem. The proposed algorithm is evaluated in a simulative environment with one eNodeB and multiple UEs. Simulation results show that the proposed algorithm outperforms the existing algorithms in terms of goodput and the number of satisfied users. The simulation environment considers a QoS threshold of 10 ms and a packet loss ratio of at least 5%. Therefore, the ultra-low latency ($C_1$) and ultra-high reliability ($C_2$) criteria are not met in this work.

The concept of using LTE in an unlicensed band (LTE-U) for the Tactile Internet is introduced by Su *et al.* [124]. Since the LTE-U operates at the same frequency band as Wi-Fi, inefficient resource allocation over LTE-U and Wi-Fi networks can lead to a poor QoS for both tactile and Wi-Fi users. In [124], the authors study a power and time slot allocation problem over LTE-U and Wi-Fi networks for downlink transmission. They consider multiple small cell-BSs and Wi-Fi APs, which are distributed under the coverage of a macro BS to serve tactile and Wi-Fi users. Note that tactile users can be served through both licensed and unlicensed bands, whereas Wi-Fi users are served only through the unlicensed band. The authors aim to optimally allocate the transmission power in both licensed and unlicensed bands for tactile users and to allocate the time slots in the unlicensed bands for both users such that the system utility is maximized. The developed optimization problem considers channel interference and traffic conditions. The problem is then divided into two sub-problems: (i) power and (ii) time allocation. The power allocation sub-problem is solved first and then its solution is used to solve the subsequent time allocation sub-problem. The simulation results indicate that the proposed algorithm can improve the throughput and system utility compared to the fixed transmission power scheme with a fixed proportion of time slots equally allocated to both types of users. However, we note that the proposed formulation does not account for the delay or reliability constraints. Therefore, this work does not cover either of our criteria.

*C. Joint Uplink and Downlink Transmission*

Several radio resource allocation algorithms have aimed to jointly tackle both uplink and downlink transmission in order to achieve the desired end-to-end delay and reliability for the Tactile Internet. She *et al.* [125] and Gholipoor *et al.* [126] focus on radio resource allocation for general Tactile Internet applications. Aijaz [127] introduces a radio resource allocation algorithm for haptic communications in LTE-A RAN and Singh *et al.* [128] and Budhiraja *et al.* [129] study the radio resource allocation problem over cellular networks to enable device-to-device communications for the Tactile Internet. She *et al.* [130] and She *et al.* [131] focus on radio resource allocation for vehicle communications. Finally, Avranas *et al.* [132] investigate the performance of the incremental redundancy hybrid automatic repeat request scheme for URLLC.

She *et al.* [125] focus on radio resource allocation in a RAN architecture for supporting general Tactile Internet applications. They propose a cross-layer transmission optimization framework, where the queuing and transmission delays are considered for the end-to-end delay and different packet error/loss probabilities are taken into account for the reliability. Their objective is to minimize the transmission power required to satisfactorily meet the ultra-high reliability and ultra-low latency requirements. To ensure the desired reliability, the authors propose a proactive packet dropping mechanism that allows some packets in deep fading to be dropped before transmission. Starting from a single-user scenario, they then extend their framework to a multi-user scenario to find the optimal resource allocation policy and packet dropping policy such that the required maximal transmission power is minimized. To solve the single-user scenario, they propose a two-step method that relies on the exhaustive search method. In the multi-user scenario, the allocation of bandwidth is also added to the decision variables of the optimization problem, which is a mixed integer programming problem. To reduce the complexity of the developed problem, a heuristic is proposed based on the steepest descent method. To validate the optimality of their proposed heuristic, the authors compare its performance with the exhaustive search method. The results show that the proposed algorithm can obtain optimal solutions. The authors consider an end-to-end delay of 1 ms as a delay bound, and so this works meets the ultra-low latency criterion ($C_1$). By imposing a packet loss probability of $10^{-7}$, this paper also meets the ultra-high reliability criterion ($C_2$).

Sparse code multiple access (SCMA) is a new emerging coding-domain NOMA scheme designed to improve the spectral efficiency of an orthogonal frequency division multiple access (OFDMA). It allows multiple user signals to be transmitted in the same subcarrier via code domain multiplexing (i.e., codebooks). In [126], Gholipoor *et al.* focus on codebooks and power allocation in SCMA-based cellular networks to support general Tactile Internet applications. The authors consider a single-cell wireless network with multiple users and teleoperators. Each user is assumed to generate both traditional data and tactile data. The objective is to maximize the uplink sum rate of the traditional data when assigning codebooks and power to users and teleoperators for both uplink and downlink transmissions. The problem is formulated as an optimization problem with the given transmit power and end-to-end delay constraints. The delay model includes the queueing delay at the source node and at the BS. The problem is divided into two sub-problems. The first sub-problem focuses solely on the codebook assignment, whereas the second sub-problem deals with the power allocation. Both sub-problems are solved iteratively until the difference between the allocated powers is minimal over two consecutive iterations. The simulation results indicate that the proposed algorithm can satisfy the 1 ms end-to-end delay constraint of haptic devices by allocating more resources to them. Therefore, this work meets the ultra-low latency criterion ($C_1$). However, the authors do not discuss the reliability aspects and therefore this paper does not cover the ultra-high reliability criterion ($C_2$).

Unlike traditional multimedia communications, haptic communications require the following three specific requirements: (i) bounded delay, (ii) minimum guaranteed data rate, and (iii) joint and symmetric resource allocation. The last one indicates that the transmission rates in both uplink and downlink directions need to be symmetric to account for the inherent bidirectional communications aspect of haptic traffic. However, it should be noted that symmetric transmission rates in both directions may not be necessary if perceptual coding techniques (i.e., perception-based data reduction) are considered. Perceptual coding aims to reduce the data transmission rate by sending a haptic sample only if humans can notice the changes in haptic perception compared to the previously transmitted sample. Considering the aforementioned requirements, Aijaz [127] study radio resource allocation for haptic communications. Specifically, the authors introduce two radio resource allocation problems with a focus on power and RB allocation for both symmetric design and perceptual coding cases. In the former, the objective is to minimize the data rate difference, whereas in the latter case, the objective is to maximize the overall network utility. These problems also take into account the specific power and RB allocation constraints of uplink and downlink MAC schemes in LTE-A. Due to the high complexity of the developed problems, the authors propose their so-called canonical duality theory-based greedy heuristic to solve the problem with the systematic case. As for the perceptual coding case, a Hungarian method-based solution is proposed. The simulation results show that the proposed solutions outperform the conventional round robin-based and best channel quality indicator-based resource allocation algorithms in terms of the objectives for both systematic and perceptual coding. In this work, the authors considered a 1 ms end-to-end delay bound and a delay bound violation probability of $5 \times 10^{-2}$ as the reliability metric. Thus, this work only meets the ultra-low latency criterion ($C_1$).

Diversity is known as a viable solution toward increasing the reliability of wireless communications in cellular networks. Recently, spatial diversity in the form of using device-to-device (D2D) communications has attracted more research attention. Using D2D communications can decrease the average end-to-end delay by reducing the conventional two-hop cellular communication to only one hop via a direct wireless link. In general, D2D transmissions can be realized using either time division duplex (TDD) or frequency division duplex (FDD) systems. We note that a more efficient resource usage can be achieved in TDD compared to current FDD deployments. However, TDD systems are subject to higher requirements in terms of switching between uplink and downlink. The time synchronization for inter-operator D2D support is also more challenging. Given that TDD systems increase the latency, they may not be well suited for Tactile Internet URLLC applications. On the other hand, in current LTE FDD, only the uplink spectrum is allowed for D2D resource usage, which may result in an inefficient resource usage for URLLC purposes.

In [128], Singh et al. study the problem of resource allocation for cellular D2D networks in FDD systems. The authors adopt a 5G numerology in which a 0.125 ms TTI is considered to possibly achieve the 1 ms end-to-end latency. In addition, the authors aim to maximize the availability in multi-user, multi-cell networks, which is defined as the minimum data rate offered in a network's coverage area. The allocation between D2D and cellular traffic is optimized such that by granting resources to the worst-performing user equipment to improve the rates achieved with 99.999% availability throughout the network. According to the simulation results, the proposed D2D resource allocation algorithm is instrumental in characterizing a complexity-gain trade-off. Even though the authors claim that the end-to-end delay of 1 ms can be achieved by the proposed D2D resource allocation algorithm with a short TTI, they only target the rate availability of 99.999% which can be translated to a reliability of $10^{-5}$. Therefore, this approach only meets the ultra-low latency criterion ($C_1$).

Similar to [128], Budhiraja et al. [129] also introduce a two-hop D2D communication scheme for 5G to support the Tactile Internet. The authors consider a heterogeneous cellular network consisting of a macro-base station and several cellular users. They adapt NOMA for two-hop D2D communications to maximize the throughput for cellular users and propose a sub-channel allocation for the first- and second-hop transmission. The transmission power of D2D users is optimized using successive convex approximation to control the channel interference. Simulation results show improvement in the latency and throughput compared to the conventional NOMA and OMA approaches. While the proposed solution optimizes the system throughput, there is no guarantee that the delay requirement is met for tactile users. Thus, this work does not meet the ultra-low latency criterion ($C_1$). Similarly, although interference reduction is considered, there is no guarantee of the reliability of each transmission. Therefore, this work does not meet the ultra-high reliability criterion ($C_2$).

Considering vehicular communication systems with a collision avoidance application, She et al. [130] aim at optimizing the bandwidth allocation among tactile users. With BSs relaying information, each user exchanges safety information with its neighboring users. Seeking the best bandwidth allocation among users while minimizing the average transmission power, the authors also aim to ensure the end-to-end delay and reliability requirements. The considered delay model takes both queueing and transmission delays into account. Also, the reliability aspect is considered via the packet error rate, queuing bound violation, and packet dropping rate during deep channel fading periods. They consider a TDD with an OFDM system in a block fading channel. The problem is formulated via a convex optimization problem, which is then solved using the interior optimization method. Their simulation results indicate that for different packet arrival rates, the maximum end-to-end delay bound of 1 ms can be satisfied. Furthermore, they show that to ensure a high reliability on the order of $10^{-7}$, increasing the transmission power remains more efficient than increasing the number of transmitters. Therefore, the proposed solution meets both criteria.

In the context of a vehicular safety application, She *et al.* [131] address the problem of allocating resources for tactile users while maximizing the energy efficiency as well as meeting the QoS constraints. In this work, the authors consider the end-to-end delay bound, where the end-to-end delay can be broken down into air interface and queuing delays. They also consider a delay-bound violation probability that includes the queueing-delay violation probability, packet error and/or drop probability, utilizing a TDD OFDM system. To maximize the energy efficiency, the authors aim to optimally allocate the transmission power and bandwidth to each user by considering the queue state information (QSI) and channel state information (CSI) using a QSI- and CSI-dependent resource allocation policy. The analytical and simulation results of a highway vehicle scenario show that the proposed policy leads to a near-optimal energy consumption. This work focuses on ensuring an end-to-end delay of 1 ms and a packet loss probability of $10^{-7}$. Consequently, it meets both criteria.

The incremental redundancy (IR) hybrid automatic repeat request (HARQ) is known as a standard technique to improve wireless transmission reliability. In [132], the authors investigated how to characterize the fundamental energy-latency trade-off and optimize the IR-HARQ for URLLC purposes. To be more specific, the authors investigate whether it is more beneficial to perform a one-shot transmission without any HARQ or to split the packet into sub-codewords and use IR-HARQ by pursuing the following objectives: (i) to obtain the HARQ mechanism that minimizes the average energy consumption for a given packet error probability and latency constraint, and (ii) to find the optimal number of transmission rounds for different feedback delay models. A dynamic programming-based algorithm is proposed for energy efficient IR-HARQ optimization in terms of the number of retransmissions, the block length, and the amount of power per round. Furthermore, the impact of feedback latency on the energy performance of IR-HARQs is investigated and an expression for the solution of the average delay minimization problem is derived mathematically. It is shown by means of analysis as well as numerical results that a properly optimized IR-HARQ scheme can be beneficial in terms of energy if the feedback delay is reasonable compared to the packet size. In addition, the results indicate that outage probability on the order of $10^{-7}$ can be achieved by properly adjusting the number of retransmissions. However, to maintain such an outage probability, the proposed scheme is not shown explicitly to guarantee the 1 ms latency. Therefore, this work only meets the ultra-high reliability criterion ($C_2$).

*D. Network Slicing*

Network slicing is expected to be a key concept for enabling different vertical applications with diverse and rather orthogonal requirements over a shared infrastructure [133]. Three algorithmic works adopt the network slicing concept in radio resource allocation to achieve the stringent requirements of the Tactile Internet.

A. Aijaz [134] leverages the network slicing concept to create multiple logical radio networks over a shared LTE-A RAN to support haptic communications. To be more specific, the author studies the problem of radio resource slicing along with radio resource customization for haptic communications. The author first proposes a radio resource slicing strategy that aims to optimally allocate RBs for different vertical slices over the network while considering the dynamics and the utility requirements of the different slices. This work also adopts the OFDM numerology from [135] to achieve a TTI of ~100 to 200 microseconds, which is a necessary condition to be able to meet latency 1~2 ms requirement. After modeling the slicing problem as a Markov decision process, an optimal radio resource slicing strategy is derived based on the application of reinforced learning without any requirement for a priori knowledge about wireless environments. To deal with the relatively slow performance of the reinforcement leaning-based method in practice, the efficiency of the radio resource slicing strategy has been improved through a post-decision state learning approach, which requires partially-known system dynamics. Furthermore, considering the same key haptic communication requirements as those discussed in [127], the author then develops a radio resource allocation formulation to optimally allocate transmitting power and RBs in the slice to each haptic user. The formulation also accounts for the specific power and RB allocation constraints of uplink and downlink MAC schemes in LTE-A networks. To reduce the complexity, the original optimization problem is decomposed into a binary integer programming problem, which is then solved using a low-complexity greedy heuristic algorithm. The simulation results indicate that the proposed slicing strategy outperforms the existing solution based on a fixed slicing period in terms of overall radio resource utilization. In addition, this proposed radio resource allocation algorithm for haptic communications is superior to conventional round robin and best channel quality indicator-based algorithms in terms of satisfying the haptic communication requirements (e.g., a symmetrical data rate in uplink and downlink and an end-to-end delay bound of 1 ms). Therefore, this work meets the ultra-low latency criterion ($C_1$). However, this work considers a 1 ms delay bound violation probability of $5 \times 10^{-2}$ as the reliability metric, and so it does not meet the ultra-high reliability criterion ($C_2$).

Considering the flexible cloud-based radio access network, as discussed in Section III.A.2, Shafigh *et al.* [96] propose a two-layer/slice traffic-aware resource allocation framework to serve tactile users. The main concept of this framework is that it will offload low-priority users from a CRAN to a DNA if the CRAN resources are not sufficient to satisfy the tactile user requirements. Here, the authors assume that tactile users require a bounded delay, while low-priority users prefer a minimum data rate. Additionally, tactile users can only be served by a CRAN due to its higher reliability and security. This framework applies a dynamic network slicing concept. Two dynamic network slices, one for high-throughput and one for low-latency, are created for low-priority users and tactile users, respectively. A two-step matching game is utilized to assign resources so that both types of users meet their requirements while optimizing resource utilization ratios. Due to the instability of the proposed matching game, an algorithm based on two-sided exchange stable matching and Nash stability concepts is proposed to solve this problem. The simulation results show that increasing the number of dynamic access points in the DNA can improve the user satisfaction of both types of users, especially low-priority users. In addition, the proposed algorithm requires less than 10 ms to allocate resources to tactile users. However, the delay bound value for the tactile users in the simulations is not specified, and therefore this framework does not cover the ultra-low latency criterion ($C_1$). Also, it does not consider the reliability aspect, and so it does not cover the ultra-high reliability criterion ($C_2$).

Another radio resource allocation algorithm relying on the network slicing concept is proposed by Ksentini *et al.* [136], where the authors focus on providing a MAC scheduling solution in order to sustain URLLC requirements in the context of a slicing-ready 5G network. To be more specific, they aim to dynamically manage the radio resources of a sliced 5G network to meet the stringent latency and reliability requirements of URLLC by means of a proposed two-level MAC scheduling framework. This proposed framework aims to handle the uplink and downlink transmissions of the network slices of different characteristics over a shared RAN. It is comprised of two-level scheduling steps. The first level of UE scheduling is slice-specific, which is carried out by a slice-specific scheduler. The second scheduling level is responsible for inter-slice resource partitioning, which shares the channel among slices and enforces the scheduling decisions of the first level using the slice-dedicated bandwidth policy communicated by the so-called slice orchestrator. The results demonstrate that the proposed scheme ensures an end-to-end delay of less than 16 ms for both short- and large-sized packets. Note, however, that there is a need to carefully dimension the slice-dedicated bandwidth value and use an admission control scheme to limit the number of UEs per URLLC slice. Since the achievable average end-to-end delay obtained by the proposed framework is still greater than 3 ms, this work does not meet the ultra-low latency criterion ($C_1$). Furthermore, as for the reliability, the authors claim that the reliability can be ensured using a robust modulation and coding scheme. We note, however, that increased reliability may come at the expense of an increased delay. Given that this work does not show that it can achieve a reliability of $10^{-7}$, it does not meet the ultra-high reliability criterion ($C_2$).

*E. Summary, Insights and Lessons Learned*

To conclude the discussions above, we first present a short summary of all works focusing on radio resource allocation. We then derive some insights and identify the important lessons learned.

*1) Summary*

Radio resource allocation has been studied extensively for the Tactile Internet, as it has a significant impact on the overall delay and reliability in RANs. These works focus on the radio resource allocation problem for either uplink, downlink or joint uplink and downlink transmission. Alternatively, the network slicing concept has been adopted in some works to guarantee the stringent requirements of delay-sensitive Tactile Internet services coexistent with conventional delay-tolerant services. For completeness, Table IV summarizes the main contributions of the radio resource allocation algorithms reviewed in this section.

The resource allocation algorithms for the uplink direction can only be either (i) grant-based [114]-[116] or (ii) grant-free [117]-[121]. In grant-based schemes, the BS is responsible for arbitrating the channel access in the uplink direction, and the allocation is carried out only after it receives the users' grant requests. In contrast, grant-free schemes aim to eliminate the scheduling and grant procedures, thus accelerating the channel access process in the uplink direction. The proposed algorithms in this group are based on dedicated reservation, soft reservation, predictive reservation, semi-persistent scheduling, or deep learning. It is worth mentioning that although both grant-based and grant-free algorithms have shown an equal potential to meet the ultra-low latency criterion, none of the proposed grant-free algorithms was able to meet the ultra-high reliability criterion. This mandates the need to address the reliability aspect of the grant-free algorithms.

Our survey indicates that only two papers [123][124] have studied the problem of resource allocation in the downlink-only direction. The main idea in these works is to adopt a queue-scheduling mechanism at a given BS to ensure the timely delivery of delay-sensitive Tactile Internet packets.

Eight papers [125]-[132] have considered joint uplink and downlink transmission, which stands as a very important research direction, especially for haptic communications with its inherent bidirectional exchange of haptic information. The works in this category mainly consider communication for general Tactile Internet applications, haptic communication, FDD-based D2D communication, NOMA-based D2D communication and vehicular communication.

There are three network slicing-based radio resource allocation algorithms proposed for the Tactile Internet [96][134][136]. These works are mainly based on reinforced learning, game theory, and two-level MAC scheduling.

*2) Insights and Lessons Learned*

One overall insight from our literature review of radio resource allocation algorithms for the Tactile Internet is that most of the studies in this context focus on resource allocation for OFDMA-based cellular systems. These studies mainly aim at allocating orthogonal resources to the users to avoid interference while considering delay and reliability either as the main objective functions to be optimized or as the constraints to be satisfied. We note, however, that the main drawback of these systems is their spectral efficiency. For this reason, OFDMA-based cellular systems may not be appropriate to support the Tactile Internet, especially for large numbers of users. To cope with this drawback of OFDMA-based systems, NOMA is an emerging concept that enables non-orthogonal resource allocation among multiple users and utilizes sophisticated inter-user interference cancellation at the receiver side to separate the signals from different users. Not only is NOMA capable of supporting a larger number of users than OFDMA, it can also achieve smaller transmission delay, as it does not necessarily require a scheduling and grant procedure for uplink transmission [137]. Despite its potential, only three studies [121][126][129] consider NOMA-based cellular systems for the Tactile Internet. A lesson learned here is that more effort is needed on designing radio resource allocation for NOMA-based cellular system to unleash the potential of NOMA in realizing the Tactile Internet.

Yet another insight is that the studies that meet both our criteria (i.e., [114][130][131]) rely on short packet transmission. These studies are essential for haptic communications, because packets generated by haptic devices are generally very small in size (e.g., 8, 24, and 48 Bytes for 1-, 3- and 6-DoF devices, respectively [3]). Nevertheless, we note that some Tactile Internet applications may require larger packet sizes (e.g., 1.5 kB for video traffic in virtual reality applications [3]). Therefore, the above-mentioned studies may not be able to meet our criteria for large-size packet transmission. This is more critical for the ultra-high reliability criterion, as packet error rate can easily grow as packet size increases. The lesson we can infer from this insight is that further in-depth studies are needed to design proper radio resource allocation schemes for large-size packet transmission in the context of the Tactile Internet.

A third insight is that only a few works (e.g., [96][124]) have tapped into the significant potential of low-cost data-centric Ethernet technologies (e.g., Wi-Fi) in conjunction with the coverage-centric cellular mobile networks to increase the overall network capacity. This leads to the conclusion that there is a need to complement cellular networks with already widely deployed Wi-Fi access networks as well as other technology networks. This will open the possibility to tap into the unlicensed frequency band, thus helping reach the envisioned 1 ms latency and $10^{-7}$ reliability at a lower cost.

**TABLE IV**
**The main contributions and evaluated criteria for the algorithmic works on radio resource allocation for the Tactile Internet. In the criteria columns, Y indicates that the criterion is met, P indicates that the criterion is partially met, and N indicates that the criterion is not met.**

| Scope | | Ref. | Main Contributions | Criteria | |
|---|---|---|---|---|---|
| | | | | $C_1$ | $C_2$ |
| Radio Resource Allocation | Uplink | She et al. [114] | Introduce a radio resource allocation algorithm utilizing the frequency diversity technique to ensure the reliability while optimizing the total required bandwidth for mMTCs under Tactile Internet requirements. | P | P |
| | | Zhou et al. [115] | Propose access control and long-term radio resource allocation algorithms to guarantee long-term constraints related to power consumption, transmission delay, and backlog queue stability for MTCs. | N | N |
| | | Elsayed et al. [116] | Propose a joint user association and resource allocation algorithm in multi small-cell networks based on a deep Q-network technique to minimize the uplink delay. | N | N |
| | | Ma et al. [117] | Propose a fast and reliable grant-free multiple access scheme using frequency diversity and subcarrier index modulation techniques. | N | N |
| | | Condoluci et al. [118] | Introduce a soft resource reservation scheme for uplink transmission to provide fast access for tactile users without a degradation of spectral efficiency. | N | N |
| | | Hou et al. [119] | Propose a burstiness-aware bandwidth reservation algorithm based on traffic prediction to optimize the total required bandwidth while supporting the bursty tactile traffic. | P | N |
| | | Feng et al. [120] | Introduce a predictive semi-persistent scheduling scheme exploiting the traffic history to minimize the uplink delay in LTE systems. | N | N |
| | | Ye et al. [121] | Introduce a deep learning-aided grant-free multiple access scheme for NOMA-based systems to achieve low latency uplink transmission. | N | N |
| | Downlink | Khorov et al. [123] | Propose a radio resource scheduling algorithm maximizing the number of users whose requirements are satisfied in downlink transmission. | N | N |
| | | Su et al. [124]. | Introduce a resource allocation framework for downlink transmission over coexistent LTE-U and Wi-Fi networks to maximize the system utilization while serving both tactile and Wi-Fi users. | N | N |
| | Joint uplink and downlink | She et al. [125] | Propose a cross-layer resource allocation algorithm for 5G cellular RANs to optimize the transmission power under Tactile Internet latency and reliability constraints. | Y | Y |
| | | Gholipoor et al. [126] | Introduce a cross-layer resource allocation algorithm in the SCMA-based cellular system to maximize the overall uplink throughput while ensuring the ultra-low delay constraint. | Y | N |
| | | Aijaz [127] | Proposes radio resource allocation algorithms with a focus on the key requirements of haptic communications. | Y | N |
| | | Singh et al. [128] | Propose a flexible resource allocation framework for FDD-based cellular networks with D2D communications to achieve ultra-low latency and ultra-reliable communication. | Y | N |
| | | Budhiraja et al. [129] | Introduce a resource allocation algorithm for a NOMA-based system with two-hop D2D communication to improve the throughput and delay. | N | N |
| | | She et al. [130] | Propose a bandwidth allocation method minimizing the average transmission power for vehicular communication systems under Tactile Internet requirements. | Y | Y |
| | | She et al. [131] | Propose radio resource allocation maximizing the energy efficiency under Tactile Internet latency and reliability constraints for vehicular safety applications. | Y | Y |
| | | Avranas et al. [132] | Introduce an energy efficiency optimization framework for the IR-HARQ scheme to achieve low latency and high reliability. | N | Y |
| | Network Slicing | A. Aijaz [134] | Proposes a dynamic radio resource slicing strategy based on the reinforced learning method for haptic communications with partial or no advance knowledge about the wireless environment. | Y | N |
| | | Shafigh et al. [96] | Propose a two-layer/slice traffic-aware resource allocation framework over a flexible cloud-based radio access network based on two-sided exchange-stable and Nash stability concepts. | N | N |
| | | Ksentini et al. [136] | Propose a two-level MAC scheduling framework for radio resource slicing to guarantee the Tactile Internet requirements. | N | N |

## V. Non-radio Resource Allocation Algorithms on Lower Layers for the Tactile Internet

In this section, we compile the research on non-radio resource allocation algorithms that mainly focus on lower layers (i.e., PHY and MAC layers) for the Tactile Internet. These algorithms include bandwidth allocation and scheduling in WLAN, WBAN, and PON for the Tactile Internet. They also cover algorithms whose main focus is on the physical layer (PHY layer) and which go beyond shortening the air interface to realize URLLC. This section is concluded by presenting the insights and important lessons learned from this review.

### A. WLAN and WBAN

A number of researchers have focused on introducing WLAN [138]-[141] and WBAN [142][143] technologies as potentially alternative wireless access networks to serve Tactile Internet applications due to their low cost, low power consumption, and wide deployment. Some mechanisms have been developed to enhance the WLAN and WBAN technologies to meet the stringent Tactile Internet requirements, as conventional mechanisms cannot fulfill these requirements.

The conventional IEEE 802.11 MAC protocol in WLANs is a contention-based protocol based on the distributed coordination function (DCF), in which packet collision is avoided via random back-offs. However, packet collision may still occur, which results in an unbounded latency. Hence, DCF can only provide a soft QoS. In contrast, the hybrid coordination function (HCF)-controlled channel access (HCCA) scheme deploys a centralized QoS-aware mechanism, whereby a hybrid coordinator assigns resources to the associated users based on the traffic characteristics reported in a traffic specification frame (TSPEC). The TSPEC contains information such as the maximum service interval, mean data rate, and nominal MAC service data unit size. Feng *et al.* [138] have conducted a feasibility analysis of WLAN infrastructure in general, and of the widely deployed IEEE 802.11 HCCA MAC protocol in particular, for supporting Tactile Internet low-latency applications. They consider a scenario where multiple hubs transmit data gathered from several haptic sensors wirelessly to an AP. In this study, the data arrival rate at each hub follows a Poisson process. After developing an analytical M/G/1 queueing model of the average uplink delay of the HCCA, the authors of [138] conducted a global sensitivity analysis, which helps to better investigate the impact of various network parameters on the uplink latency. They also proposed a novel strategic parameter selection algorithm to reduce the uplink latency by adjusting the value of the transmission opportunity and the service interval duration in each iteration. The results indicate that the proposed strategic parameter selection algorithm can achieve an average uplink latency of <0.09 ms. We note, however, that this work focuses only on the uplink delay rather than the overall end-to-end delay. Therefore, it partially meets the ultra-low latency criterion ($C_1$). The authors did not discuss or evaluate the reliability aspect of their proposed design approach, and so it does not cover the ultra-high reliability criterion ($C_2$).

Similar to [138], Engelhardt *et al.* [139] aim to enhance the existing IEEE 802.11 MAC protocol for the Tactile Internet. More precisely, the authors investigate the feasibility of achieving a 1 ms end-to-end latency in three-hop wireless networks by means of a proposed tactile coordination function (TCF) to complement the widely deployed HCF of the enhanced distributed channel access (EDCA) mechanism of the IEEE 802.11 MAC protocol. To be more specific, a new access category (AC) is designed to prioritize haptic data stream with a 1 ms latency requirement above the ACs for voice and video with 10 and 100 ms latency requirements, respectively. Typically, the user priority is determined by the transport layer port. It is then classified as one of the following five categories sorted from high to low priority: haptic data, voice, video, background information, and best effort. On the MAC layer, the EDCA mechanism then implements the prioritization by scheduling the higher priority queues first and tuning the inter-frame spaces and transmission opportunity of the frames such that they are also prioritized in the contention phase between stations. These steps reduce the average end-to-end latency and jitter significantly, while being backward-compatible with the existing IEEE 802.11 deployments. The simulation results indicate that current IEEE 802.11n technology utilizing TCF instead of HCF can already support up to three wireless hops while keeping the average end-to-end latency below 1 ms and jitter below 0.5 ms. Therefore, this work can meet the ultra-low latency criterion ($C_1$), but only for 3 hops. However, this work indicates that the proposed TCF scheme can achieve a 0.1% loss rate for 5 hops. Thus, it does not meet the ultra-high reliability criterion ($C_2$).

To improve the uplink delay in WLANs, Lv *et al.* [140] propose a novel MAC scheme referred to as request-based poll access (RPA). The authors of [140] consider a single AP communicating wirelessly with multiple hubs. The proposed scheme is based on a slotted approach, where the time is divided into multiple beacon intervals. Each of the intervals is further divided into several sub-cycles. Each sub-cycle consists of three periods: a broadcasting period (BP), a slotted request uploading period (RUP), and a polling period (PP). During each sub-cycle, the AP first broadcasts a beacon message in the BP to assign uploading request slots in the RUP to all hubs. After the hubs receive the beacon message, they send uploading requests, according to their buffer status, to their assigned uploading request slots during the RUP. Finally, during the PP, all the hubs are polled by the AP sequentially according to the order of the assigned slots in the RUP. The authors compare the proposed scheme with two conventional MAC schemes, the DCF and the point coordination function (PCF), via simulations. The results indicate that under light loads, the DCF provides the smallest average uplink delay. As load increases, the uplink delay of both the DCF and PCF schemes increases significantly, as opposed to that of the proposed scheme, which grows at a slower pace and eventually provides the smallest uplink for medium to high loads. The results indicate that the proposed scheme can achieve an average uplink delay of 200 $\mu$s. Given that this work only considers the uplink direction, it partially meets the ultra-low latency criterion ($C_1$). This work does not cover the ultra-high reliability criterion ($C_2$).

We note that the RPA scheme proposed in [140] has some drawbacks. These occur mainly because during the PP, the AP polls all hubs sequentially according to the order of the assigned slots in the RUP without considering the buffer status of each hub. This process can lead to an extra queuing delay for the hubs that have a high number of packets in their buffer and for those whose packets have been queued for a long time. To tackle this issue, Lv *et al.* [141] propose the so-called dynamic polling sequence arrangement scheme to further improve the delay and jitter. The authors introduce two dynamic polling sequence arrangement schemes, namely RPA-sum and RPA-packet. In the RPA-sum scheme, the AP will give the higher priority to the hub that has the higher value of the sum of the packet queuing time during the PP. In the RPA-packet scheme, the AP prioritizes the hubs with the larger number of packets in their buffers. The authors compare the two proposed schemes with the original RPA scheme proposed in [140] via simulations. The results demonstrate that the proposed schemes are superior to the conventional RPA scheme in terms of average uplink delay and jitter. Furthermore, it is shown that the proposed schemes can achieve an uplink delay on the order of 200 $\mu$s. Given that this work only considers the uplink delay without mentioning other delay components, this work partially meets the ultra-low latency criterion ($C_1$). Similar to [138] and [140], the reliability aspect has not been discussed in this work and therefore it does not cover the ultra-high reliability criterion ($C_2$).

Unlike [138]-[141], Ruan *et al.* [142] investigate the feasibility of Smart Body Area Networks (SmartBANs) to support tele-health Tactile Internet applications. The authors study the delay performance of exhaustive transmission mechanisms based on a SmartBAN MAC frame in downlink transmission. They consider a scenario where a hub is connected to multiple on-body actuators and broadcasts the incoming packets to all actuators. They also assume that the downlink packet arrival follows a Poisson process. In this study, both conventional exhaustive and fixed-length exhaustive transmission schemes are considered. In a conventional exhaustive transmission scheme, the duration of the MAC frame dynamically varies according to the number of packets in a buffer, while in a fixed-length exhaustive transmission scheme, the duration of the MAC frame is fixed. To evaluate the downlink delay performance, the authors develop an M/D/1 queuing model to analyze the conventional exhaustive transmission scheme. They adopt an embedded Markov Chain model to analyze the downlink delay performance. Simulations are also carried out to validate the analytical models. The simulation and the analytical results are shown to closely match each other. The results also indicate that the fixed-length transmission scheme outperforms the conventional exhaustive transmission scheme in terms of both downlink delay and energy efficiency. Lastly, the analysis shows that increasing the MAC frame duration in the fixed-length exhaustive transmission scheme can further improve the downlink delay. This work provides some guidance for choosing the most suitable MAC frame duration such that downlink delay on the order of 1 ms is achieved. We note, however, that an end-to-end delay is not considered in this work. Therefore, it only partially meets the ultra-low latency criterion ($C_1$). In addition, it does not cover the reliability criterion ($C_2$).

Nanonetworks, also known as body area nanonetworks (BANNs), are an emerging technology to enable e-health systems at the nanoscale. By allowing haptic communications over nanonetworks, nanodevices can be manipulated inside a human body via a human operator for disease treatment in real time. Typically, the nanonetwork technology operates at the terahertz band, which may incur high path loss due to molecular absorption. To satisfy the high throughput and low latency required for haptic communications, a resource allocation problem for nanonetworks is studied in Feng *et al.* [143], in which the authors consider a scenario where multiple in-body nanodevices transmit haptic information to a human operator. Each nanodevice is assumed to have the capability of harvesting energy to power itself. The authors aim to select the optimal sub-frequency bands, transmission power, energy harvesting, and packet drop rate for each device such that the average data rate at each time slot is maximized, while satisfying the energy consumption and latency constraints. The packet drop rate is considered here to ensure the stability of nanodevices, as each of them has finite power and communication blocks. The authors adopt a Brownian motion model to describe the mobility of devices in each time slot and derive a time-variant THz channel model. Based on this channel model, they formulate a stochastic resource allocation optimization problem. Utilizing the Lyapunov optimization theory, the problem is converted to potentially three solvable subproblems: (i) energy harvesting, (ii) packet drop rate and (iii) joint allocation of transmission power and frequency. Note that the third sub-problem is nonlinear and thus the authors solve it via a heuristic algorithm that is based on the swarm intelligence technique. Via simulations, the performance of the proposed heuristic was evaluated and compared with the performances of normal genetic and traditional particle swarm optimization algorithms. The results show that the proposed heuristic scheme provides a higher average data rate, while requiring less convergence time. We note that only the unlink latency bound of the Tactile Internet latency requirement is considered. Thus, it partially meets the ultra-low latency criterion ($C_1$). In addition, the packet drop rate, which reflects the reliability, is considered as a constraint in the problem. However, this work does not explicitly mention the maximum packet drop rate value that the proposed algorithm aims to satisfy in the simulations, and so it does not cover the ultra-high reliability criterion ($C_2$).

### B. Passive Optical Network

Due to their high bandwidth and flexibility, PONs are promising candidates with which to realize low-latency Tactile Internet applications, especially for fixed users. Several algorithmic works [99][106],[144]-[147] have studied the problem of scheduling and bandwidth allocation for PONs. In addition, [99] and [106] have suggested their PON-based network architectures for the Tactile Internet, as discussed in Section III.A.4 and Section III.B.6, respectively. However, it should be noted that the PON-based network architectures considered in [99][106],[144]-[147] share a similar design concept. These network architectures mainly consist of multiple ONUs integrated with wireless APs to interact with wireless access networks. All ONU/APs are also connected to a single OLT via fiber links. In addition, the OLT is co-located with a tactile control server at a CO. Considering this specific architecture, the main objective of these works is to introduce an efficient bandwidth allocation mechanism to ensure ultra-low latency and ultra-high reliability communication between the ONU/APs and the tactile control server.

Neaime *et al*. [99] introduce a dynamic wavelength and bandwidth allocation algorithm to guarantee the latency requirement of tactile services. The authors assume that the distribution of the incoming traffic is not known in advance. In their proposed algorithm, each ONU/AP conducts a classification of the incoming traffic based on the QoS requirements, and then sends a bandwidth request via REPORT messages to the CO. The proposed algorithm consists of two main modules. The first module is for grant sizing; it determines the total grant size of each class of every ONU/AP (aggregate time slots). It allocates the grant size equal to the requested bandwidth for the tactile service class while allocating the grant sizes equally among the ONU/APs for the non-tactile service classes. The second module is a grant scheduler, which is responsible for scheduling the grants (i.e., time slots) over multiple uplink channels. It first schedules the grants of the tactile service class in dedicated wavelengths in order to meet the strict delay requirement. Next, a modified water-filling algorithm considering traffic priorities is proposed to schedule the grants of the given non-tactile service classes in order to maximize the network throughput. The proposed algorithm is compared to two existing approaches; the first assigns wavelengths in a round-robin fashion, and the second assigns available wavelengths to non-tactile traffic once the tactile queue is empty. The simulation results indicate that the proposed algorithm provides the best performance in terms of throughput and packet drop rate. It can also achieve an average end-to-end delay on the order of 1 ms. Furthermore, the delay performance is shown to slightly decrease by increasing the number of wavelengths. Therefore, this work meets the ultra-low latency criterion ($C_1$). In addition, the results indicate that for the normalized traffic load of less than 0.7, the proposed algorithm can achieve a zero packet drop rate, reflecting the reliability of the system. Thus, it also meets the ultra-high reliability criterion ($C_2$).

We note that the work presented in [99] relies on a water-filling-based scheme along with an advanced intra-ONU scheduling mechanism. However, this requires the ONUs to have the capability of transmitting over multiple wavelengths concurrently, which can be accomplished by deploying a bank of costly fixed transceivers or tunable lasers. To fulfill the stringent low-latency requirements of tactile services in a cost-efficient manner, Valkanis *et al.* [144] study the problem of intra-ONU scheduling and present their so-called double per-priority queue dynamic wavelength and bandwidth allocation algorithm for PONs, with the objective of preserving the stringent QoS demands of Tactile Internet applications without degrading the other non-tactile applications served by the same access network.. To be more specific, for each class of service (CoS) (e.g., tactile and non-tactile packets) the authors deploy two different queues at each ONU, namely, high priority (HP) and low priority (LP) queues. The transmission priority of an incoming packet depends not only on the type of the queue where the packet is buffered, but also on the CoS of the packet. In the proposed scheme, following a specific packet transfer procedure designed to avoid monopolizing the most demanding traffic type, a packet with a lower priority CoS buffered in an HP queue takes precedence over another packet with a higher priority CoS buffered in an LP queue. The performance of the proposed algorithm was evaluated for the coexistence of five classes of service over a 10G-PON with the ONU buffer size set to 10 Mbytes. The results indicate the proposed scheme can keep the average end-to-end delay below the given threshold for each CoS, most notably that of the Tactile Internet, i.e., 1 ms. In addition to the average latency, it is shown that the proposed scheme can keep the maximum end-to-end delay below the 1 ms for a wide range of loads. Therefore, this work meets the ultra-low latency criterion ($C_1$). From a reliability point of view, the authors have evaluated the performance of their proposed scheme in terms of packet loss and delay rate (PLDR), which is defined as the summation of the number of packets lost due to buffer overflow and the number of delayed packets not meeting the latency constraint of the corresponding CoS. According to the results, the proposed scheme can achieve a close-to-zero PDLR for loads <0.8. Consequently, this work meets the ultra-high reliability criterion ($C_2$).

To improve the latency in PONs, proactive bandwidth allocation via traffic prediction may be a viable approach. Toward this end, Wong *et al.* [106] propose a predictive Tactile Internet-capable dynamic wavelength and bandwidth allocation algorithm to support tele-health applications. The proposed algorithm aims to dynamically adjust the number of active wavelengths according to the predicted traffic load in the network to maintain the required end-to-end delay. In this work, both non-tactile (i.e., healthcare monitoring data) and tactile traffic are considered. In addition, non-tactile traffic at each ONU/AP is assumed to be periodic, whereas the non-bursty and bursty tactile traffic are characterized by Poisson and Pareto traffic models, respectively. In the proposed algorithm, each ONU/AP first predicts its traffic load and reports it to the CO during each transmission polling cycle via REPORT message. Bayesian and maximum-likelihood sequence estimation methods are used to estimate the non-bursty and bursty tactile traffic, respectively. Meanwhile, the arithmetic averaging method is adopted to estimate the non-tactile traffic. Based on the predicted traffic load, the CO calculates the required bandwidth of each ONU/AP while ensuring the end-to-end delay requirements to dynamically adjust the number of active wavelengths. To evaluate the proposed algorithm, the authors conduct simulations under intra- and inter-building configurations. The simulation results indicate that the proposed algorithm can keep the average end-to-end delay on the order of 100 µs when the ratio of tactile traffic to non-tactile traffic is below 0.5. For this reason, this work meets the ultra-low latency criterion ($C_1$) under LAN coverage. However, it does not discuss the reliability aspect, and so it does not cover the ultra-high reliability criterion ($C_2$).

To further improve the decision making on proactively allocating bandwidths to the the delay-sensitive traffic in the upstream direction, some works have recently leveraged machine learning to perform traffic prediction. Towards this end, Ren *et al.* [145] propose an ANN-based predictive dynamic bandwidth allocation (DBA) algorithm, which is able to achieve near-optimal bandwidth allocation decisions supervised by the underlying ANN. To be more specific, an ANN is deployed at the CO to first learn the relation between upstream latency performance and different bandwidth allocation strategies as well as network features, and to then allocate the bandwidth in the upstream direction to ONU/APs. This step, in turn, is proven to improve the upstream latency. The simulation results indicate that the proposed ANN-DBA can adaptively allocate the bandwidth, thus improving the latency performance over the conventional DBA schemes. It can also achieve upstream delay on the order of 100 µs. Since this work focuses only on the upstream direction, it only partially meets the ultra-low latency criterion ($C_1$). However, the evaluation metrics related to the reliability aspect are not discussed. Therefore, it does not cover the ultra-high reliability criterion ($C_2$).

In [146], Ren *et al*. also introduce a machine learning-based predictive DBA algorithm to address the upstream bandwidth contention and latency bottleneck of PONs. Their proposed algorithm predicts the ON and OFF status of bursty H2M traffic arriving at each ONU/AP using an ANN deployed at the CO, thereby estimating the uplink bandwidth demand of each ONU/AP. As such, the CO allocates the required bandwidth to the forthcoming packet bursts without the need to have them wait until the following transmission cycle. The simulation results show that the trained ANN is able to achieve a > 90% accuracy in predicting the ON and OFF status of ONU/APs, thus improving the accuracy of the bandwidth demand estimation significantly. The proposed ANN prediction is then exploited at the CO to facilitate the bandwidth allocation decisions through the proposed machine learning-based predictive DBA, called the MLP-DBA. The simulation results indicate that using ANN-enabled prediction, the proposed MLP-DBA successfully improves upstream latency performance and reduces the packet drop ratio as compared to the conventional limited-service DBA and existing predictive DBA algorithms. More specially, it can guarantee the 1 ms upstream delay, and zero packet drop rate can be achieved when the normalized aggregated traffic load is smaller than 0.6. However, this work only partially meets both criteria since it focuses only on the upstream direction without considering the downstream flow.

To conduct a better bandwidth allocation, it is necessary to understand the unique characteristics of the haptic traffic. In [147], Ren *et al*. set out to achieve an end-to-end latency of 1 to 10 ms over converged fiber and wireless networks with edge intelligence placed at the CO and at the interface of ONU/APs. They focus on non-local H2M communications, as such applications are more susceptible to the network domain latency performance of converged access networks. First, the authors aim to understand the unique characteristics of human control and haptic feedback traffic by means of developing interacting H2M applications to experimentally investigate the H2M traffic and the impact of its aggregation over optical access networks. After identifying a high correlation between real-time control and feedback traffic, the authors exploit such correlation to design an artificial intelligence-based interactive bandwidth allocation (AIBA) algorithm for converged fiber and wireless access networks. The AIBA algorithm aims to reduce the latency for converged H2M traffic delivery over access networks by predicting and pre-allocating the bandwidth for feedback based on the control traffic forwarded at the CO, thereby expediting haptic feedback delivery by up to 60% compared to a baseline scheme. According to the simulation results, the average end-to-end latency is kept below 105, 120, and 350 microseconds for 16, 32, and 64 ONU/APs, respectively. For this reason, this work meets the ultra-low latency criterion ($C_1$). However, it does not cover the ultra-high reliability criterion ($C_2$), as that aspect is not investigated.

### C. PHY Layer

As discussed in several papers (e.g., [3][17]), an important step towards achieving 1 ms latency in LTE cellular networks is to reduce the air interface by shortening the TTI in the PHY layer to allow for an accelerated user scheduling. Shortening the TTI can also improve the reliability, as the number of retransmission opportunities within the latency bound can increase. However, this improvement comes at the expense of increased subcarrier spacing. Therefore, tackling other aspects of the PHY layer rather than just shortening the TTI remains a key challenge to achieve ultra-low latency and ultra-high reliability. Several algorithmic works have tried to improve the latency and reliability from a PHY layer point of view. Tarneberg *et al.* [148] and Li *et al.* [149] focus on multi-input multi-output (MIMO) systems for the Tactile Internet, and Joshi *et al*. [150] investigate mmWave communications for haptic-enabled VR/AR applications. Ma *et al*. [151] propose a novel modulation technique and Mountaser *et al.* [152] introduce a reliable and low-latency fronthaul network for Tactile Internet applications. Finally, Szabo *et al.* [153] and Gabriel *et al.* [154] leverage network coding to achieve low latency and high reliable transmission.

Tarneberg *et al.* [148] give insights into finding the proper number of antennas required in the massive MIMO system to satisfy Tactile Internet requirements. The authors consider a bilateral teleoperation scenario, where each robotic joint is operated with one antenna and manipulated independently over wireless links through a massive MIMO BS. They then characterize the trade-offs between the required number of antennas and the number of simultaneously served UEs (or joints) under ultra-high reliable and ultra-low latency conditions. The authors first formulate an optimization problem to find the maximum number of simultaneously served UEs that satisfies a given end-to-end latency requirement. This problem is formulated for an OFDM system. Next, the authors conduct simulations to study the relationship between the required number of antennas and the number of simultaneously served UEs for a given BER requirement. In the simulations, both maximum-ratio combining (MR) and zero-forcing (ZF) are considered as the precoding scheme, while conventional coding is utilized as the channel coding scheme. The simulation results indicate that the ZF scheme outperforms the MR scheme in terms of the required number of antennas. In addition, channel coding does not show any significant performance improvement for the ZF scheme, as opposed to the MR scheme, where the performance improves considerably. In this work, the authors consider the end-to-end delay requirement of 1 ms, and therefore this work meets the ultra-low latency criterion ($C_1$). However, for the reliability, the authors consider a BER of only $10^{-5}$. As a result, it does not meet the ultra-high reliability criterion ($C_2$).

For efficient MIMO system operation, the MIMO transceivers should be provided with CSI, which can be obtained using uplink channel training (UCT) and downlink channel training (DCT) strategies. To achieve low latency for short-packet communications, overhead generated by both strategies cannot become negligible, as it can be a dominant part of the overall delay in comparison to the transmission delay of a short packet size. In [149], Li *et al.* suggest the use of UCT for estimating the downlink CSI for downlink transmission to support low-latency short-packet communications in a multiple-input single-output (MISO) system. This method works because the overheads generated by the UCT are independent of the number of transmitter antennas, which may result in smaller latency compared to DCT. However, unlike the DCT strategy, the achievable data rate using the UCT strategy for estimating the downlink CSI is highly influenced by the channel reciprocity between uplink and downlink. In order to select proper training strategies for low-latency short-packet communications, the authors derive a close-form expression of the minimal channel reciprocity coefficient (CRC) to indicate the minimal level of channel reciprocity, which allows the UCT strategy to achieve a higher data rate than the DCT strategy for the downlink transmission. In the derived expression, the authors consider a MISO system with a TDD mode. They also assume independent quasi-static Rayleigh fading channels with a finite block length. In addition, simulations are conducted to verify the accuracy of the derived close-form expressions. Both analytical and simulation results show that the data rate achieved by the UCT increases as the channel reciprocity increases. The minimal CRC is also shown to be feasible, and it decreases as the number of antennas increases, which indicates the advantage of the use of UCT in TDD MISO systems. In this work, although the authors claim that the UCT can achieve lower latency than the DCT, their delay performance is not explicitly evaluated. Therefore, it does not cover the ultra-low latency criterion ($C_1$). As for the reliability, the results indicate that the minimal CRC remains feasible, with the error probability of $10^{-7}$ for the downlink transmission. Thus, it partially meets the ultra-high reliability criterion ($C_2$).

To realize high data rate Tactile Internet applications such as haptic-enabled VR/AR, a promising solution is to move towards higher frequency bands, e.g., mmWave communications. We note, however, that the use of narrow beamwidth directional antennas, which is inevitable for reducing the high path loss at mmWave, may result in frequent link outages due to antenna misalignment. To support Tactile Internet applications, a hybrid radio access architecture is considered in [150], where sub-6 GHz access is used for the transmission of haptic information, while mmWave access is used for the high data-rate transmission of audiovisual information. The authors of [150] investigate the trade-offs between the high gain of narrow beamwidth antennas and the susceptibility to misalignment in mmWave links. Further, a beamwidth-adaptation scheme is proposed to stabilize the link throughput performance and alleviate the impact of random antenna misalignment. More specifically, to show the existence of optimum transmitter-receiver beamwidths, a novel link capacity optimization framework is developed, which jointly considers the beam-misalignment and beam-searching overhead. After showing that the narrow beam links are not suitable to support Tactile Internet applications due to their high susceptibility to beam misalignment, the authors propose a beamwidth-adaptation scheme that is mis-alignment aware. This adaptation scheme stabilizes the quality of the mmWave links and increases the average link capacity. The authors only focus on the throughput performance of the proposed scheme. Therefore, it does not cover either of criteria.

To enable tactile IoT applications, Ma *et al.* [151] consider a frequency-modulated differential chaos shift keying ultra-wideband system, which is known to be low-complex, low-power, and robust against multipath fading. The authors propose a new energy-based receiver, called an enhanced energy-based detector (EED), to mitigate the impact of narrow-band interference (NBI) while maintaining low power consumption and system complexity at the receiver side. In their proposed EED, the received signal is first processed by a square law device and then processed by a variance calculator. Finally, the original data bit is decided based on the energy obtained. The authors derive an analytical BER expression of the proposed EED over AWGN and IEEE 802.15.4a multipath fading channels, and conduct simulations to validate the analytical expressions. The results show that the analytical and simulated BERs closely match each other. The proposed EED is shown to have a performance gain of up to 5 dB and 2 dB at the same BER over AWGN and multipath fading channels, respectively, in comparison with the conventional energy-based detector. Finally, this EED is also more robust against the presence of NBIs. However, the delay performance of the proposed system is not evaluated. Thus, this paper does not cover the ultra-low latency criterion ($C_1$). In addition, although this approach offers a promising performance in terms of the BER, this paper does not explicitly show that it achieves a packet error rate on the order of $10^{-7}$. Therefore, this work cannot meet the ultra-high reliability criterion ($C_2$).

Mountaser *et al.* [152] introduce an Ethernet-based fronthaul with multi-path transmission for CRANs to improve the reliability and latency of downlink transmission. In CRANs, the fronthaul is an interface connecting a central unit performing baseband functionalities and multiple remote radio heads. The authors consider a scenario where packets are delivered from the central unit to the remote radio heads through the Ethernet-based multi-path fronthaul. In the fronthaul, each path from the central unit is connected to a switch that can transmit the packets to any of the remote radio heads. Two multiple-path fronthaul solutions are proposed: (*i*) multi-path with duplication (MPD) and (*ii*) multi-path with erasure coding (MPC). In the MPD scheme, the central unit duplicates each incoming packet and sends it to all the paths, whereas in the MPC scheme, the central unit encodes each incoming packet using erasure coding before sending it onwards. The authors derive the analytical expressions of the average latency as well as the reliability-latency trade-off to evaluate the performance of their proposed solutions. They developed and ran simulations to validate the analytical expressions. The analytical and the simulation results closely match each other. These analyses and simulations show that both MPD and MPC schemes improve the average downlink delay significantly compared to a single path fronthaul solution. Moreover, at a given error probability, the MPC scheme achieves the smallest average downlink delay when compared to the MPD scheme and the single path fronthaul solution. The authors show that the proposed solutions can achieve a packet error probability of $10^{-7}$ within an average delay of 1 ms for the downlink transmission. Therefore, this work partially meets both criteria.

To cope with the inflexible nature of the store-and-forward scheme in current packet switched networks, Szabo *et al.* [153] explore the feasibility of network coding enabled by SDN. To be more specific, they focused on a special type of network coding referred to as random linear network coding (RLNC), which has two new functionalities: (i) recoding and (ii) a sliding window. Unlike the end-to-end-based coding strategies (e.g., Reeed Solomon, Raptor, etc.), the recoding functionality of the RLNC replaces the store-and-forward approach with the so-called compute-and-forward, in which each node is able to evaluate the current situation for the next hop communication and adapt its coding scheme accordingly. The authors first analyzed the performance of block codes (i.e., end-to-end coding and hop-by-hop coding) in comparison with RLNC in terms of the number of packet re-transmissions and latency in a single-path scenario. Next, they verified their theoretical findings by creating a fully-fledged implementation of the compute-and-forward router for different networking scenarios. The expected results and the measurements match well, indicating that the number of packets increases linearly for the hop-by-hop coding and RLNC, whereas it increases exponentially for the end-to-end approach, as link loss probability increases. They also showed that RLNC has a lower latency compared to the hop-by-hop coding and end-to-end schemes over a wide range of parameters. Overall, the latency values remained greater than 50 ms, and so the work does not meet the ultra-low latency criterion ($C_1$). We note that the authors focus only on the latency performance of the proposed strategy. Thus, it does not cover the ultra-high reliability ($C_2$).

Gabriel *et al.* [154] propose using multi-path transmissions with network coding to achieve reliability in wireless networks. The authors consider a scenario where one source can send traffic to one destination through multiple wireless channels. The traffic can be encoded by forward error correction coding before transmitting. The authors also assume that all channels have time-varying capacities. The transmission error considered here occurs if the source sends traffic through channels that do not have enough capacity at a given time. Formulating the problem as an optimization problem, the authors search for the transmission rate allocation on all channels that minimize the probability of transmission error. To solve the problem, two available channels are considered, LTE and Wi-Fi. Their distribution of the channel capacity is modelled based on real world data of their throughputs. Using simulation, the authors evaluate the performance of the proposed multipath rate allocation strategy with different forward error correction coding over two transmission paths. It should be emphasized that the transmission error considered here occurs if the source sends traffic through channels that do not have enough capacity at a given time. The simulation results show that the proposed strategy using forward error correction coding can significantly improve the error probability compared to the case where traffic is allocated equally to both channels, with or without forward error correction coding. However, this advantage comes with an increase of 5 ms in the delay to achieve significant reliability, and thus it does not meet the ultra-low latency criterion ($C_1$). Moreover, the error probability remains greater than 0.1, and therefore it does not meet the ultra-high reliability criterion ($C_2$).

### D. Summary Insights and Lessons Learned

This subsection briefly summarizes non-radio resource allocation algorithms tackling the lower layer aspects of the Tactile Internet and discusses insights and potential lessons learned.

*1) Summary*

In this section, we review the algorithmic works that study resource allocation in WLANs/WBANs, and dynamic wavelength and bandwidth allocation in PONs for the Tactile Internet. We also discuss the algorithmic works that mainly focus on the PHY layer. For completeness, Table V provides a summary of the key contributions of the papers reviewed in this section.

Six works have proposed novel resource allocation algorithms for WLANs and WBANs in the context of the Tactile Internet. Four of them focus on WLANs, aiming to either propose an efficient mechanism to enhance the existing IEEE 802.11 MAC protocols (i.e., HCCA, EDCA) [138][139] or to design novel MAC schemes [140][141] to meet the strict Tactile Internet requirements. As for the other two works targeting WBANs, the enhanced SmartBAN MAC protocol is proposed in [142], and an efficient resource allocation framework is proposed in [143] for haptic communications over WBANN.

The existing dynamic wavelength and bandwidth allocation algorithms in PONs for the Tactile Internet are either non-predictive [99][144] or predictive [106][145][146][147]. The non-predictive algorithms mainly aim to optimally allocate the bandwidths based on the requested bandwidths reported from each ONU in each polling cycle while prioritizing the Tactile Internet service to ensure its requirements compared to other services. On the other hand, the predictive algorithms allow the bandwidths to be proactively allocated based on traffic prediction to further reduce latency. The predictive algorithms rely on either Bayesian and maximum-likelihood sequence estimation methods [106] or machine learning-based methods [145][146][147] for traffic prediction.

Unlike the above-mentioned works that take a system point of view, the PHY layer algorithms proposed for the Tactile Internet include an antenna selection method [148], a channel training strategy [149], beamwidth adaptation [150], receiver design [151], multi-path transmission [152], and network coding-based approaches [153][154]. Clearly, these algorithms cannot be evenly compared, as they deal with different problems in their efforts to address different PHY layer issues. Most notably, multi-path transmission using packet duplication and erasure coding has shown promising results in CRAN networking infrastructure.

*2) Insights and Lessons learned*

Despite initial efforts on realizing the Tactile Internet over WLAN and WBAN technologies, none of the studies have considered joint resource allocation in both uplink and downlink directions. This aspect is particularly crucial for haptic communications due to their bidirectional exchange of control commands in the forward path (i.e., from the master domain to the slave domain) as well as haptic feedback in the reverse path (i.e., from slave domain to the master domain) [127]. Particularly, shortly after the transmission of a control command to a slave device via the uplink direction, the master device expects to receive a haptic feedback being fed back from the slave side via the downlink direction. Therefore, allocating the resources in each direction independently from each other may not be an efficient approach to ensure the timely real-time exchange of control commends and haptic feedback for a frictionless, immersive haptic communication. The lesson we can learn from this insight is that the joint resource allocation in both uplink and downlink directions in WLAN and WBAN should be further investigated to efficiently meet the stringent Tactile Internet requirements. This could push forward WLAN and WBAN technologies as potential alternative wireless access technologies for the Tactile Internet.

Another insight on WLAN and WBAN resource allocation is that none of the studies has addressed the reliability aspect that is also critical for Tactile Internet applications, as discussed in Section II.C. We note that although most of the proposed algorithms rely on contention-free mechanisms [138][140][141] to avoid packet collision, packet loss can still occur due to erroneous transmission caused by excessive path loss, given that packet dropping occurs mainly due to buffer overflow [139][143]. The lesson we can infer from this is that the reliability aspect of the WLAN and WBAN technologies

TABLE V
The main contributions and evaluation of the algorithmic works on non-radio resource allocation on lower layers for the Tactile Internet. In the criteria columns, Y indicates that the criterion is met, P indicates that the criterion is partially met, and N indicates that the criterion is not met

| Scope | | Ref. | Main Contributions | Criteria | |
|---|---|---|---|---|---|
| | | | | $C_1$ | $C_2$ |
| Non-radio Resource Allocation on Lower Layers (PHY and MAC Layers) | WBANs/WLANs | Feng et al. [138] | Analyze the delay performance of the IEEE 802.11 HCCA MAC protocol and propose a novel HCCA parameter selection algorithm to achieve the Tactile Internet latency requirements. | P | N |
| | | Engelhardt et al. [139] | Propose the tactile coordination function to enhance the IEEE 802.11 MAC protocol for multi-hop transmission in order to reduce delay and jitter. | Y | N |
| | | Lv et al. [140] | Propose a novel request-based poll access MAC scheme in WLAN to reduce uplink latency. | P | N |
| | | Lv et al. [141] | Propose two dynamic polling sequence arrangement mechanisms considering queuing times and buffer status to enhance the request-based poll access scheme [140] in terms of delay and jitter. | P | N |
| | | Ruan et al. [142] | Introduce an enhanced fixed-length exhaustive transmission mechanism in the SmartBAN MAC protocol to achieve low latency downlink transmission. | P | N |
| | | Feng et al. [143] | Propose a resource allocation algorithm for haptic communications over BANNs with the objective of satisfying high throughput and low latency. | P | N |
| | Passive Optical Networks | Neaime et al. [99] | Propose a dynamic resource allocation algorithm for NG-EPONs without a priori knowledge of traffic distribution to ensure that the requirements of tactile services are met. | Y | Y |
| | | Valkanis et al. [144] | Present a double per priority queue dynamic resource allocation algorithm for coexistent tactile and non-tactile users in EPONs. | Y | Y |
| | | Wong et al. [106] | Propose a predictive dynamic resource allocation algorithm for EPONs to achieve the delay requirement of tactile services under varying loads. | Y | N |
| | | Ren et al. [145] | Propose an ANN-based predictive dynamic bandwidth allocation algorithm for EPONs to reduce the upstream delay. | P | N |
| | | Ren et al. [146] | Propose a machine learning-based predictive dynamic bandwidth algorithm with a focus on handling the burstiness of H2M traffic for uplink transmission. | P | P |
| | | Ren et al. [147] | Introduce an AI-based interactive bandwidth allocation algorithm with a focus on capturing the unique characteristics of haptic communications. | Y | N |
| | PHY Layer | Tarneberg et al. [148] | Introduce an effective framework for selecting the number of antennas in MIMO systems to achieve Tactile Internet requirements | Y | N |
| | | Li et al. [149] | Present an effective way of choosing the channel training strategies for CSI estimation in MISO systems to achieve low latency and reliable communications. | N | Y |
| | | Joshi et al. [150] | Introduce a mmWave beamwidth-adaptation scheme to improve the throughput performance for haptic enabled VR/AR applications. | N | N |
| | | Ma et al. [151] | Propose an energy-efficient and low-complexity receiver in the frequency-modulated differential chaos shift keying ultra-wideband system to manage the presence of narrowband interference for tactile IoT applications. | N | N |
| | | Mountaser et al. [152] | Introduce an Ethernet-based fronthaul network with multi-path transmissions using packet duplication and erasure coding in a CRAN to achieve low-latency and reliable downlink transmission. | P | P |
| | | Szabo et al. [153] | Introduce random linear network coding enabled by SDN to achieve low latency communications in wired access networks. | N | N |
| | | Gabriel et al. [154] | Propose a transmission rate allocation algorithm for multi-path transmission with network coding over multiple wireless channels to improve reliability. | N | N |

must be further investigated in order to fulfill the Tactile Internet reliability requirement.

In the PHY layer, we have seen in [152] that the application of erasure coding on the MAC frames transported by the fronthaul network can reduce the transport overhead significantly. Further, multi-path transmission can be used along with erasure coding to allow the central unit to split the original MAC frames into smaller blocks and send them over multiple paths, thereby helping to achieve improvements over the conventional single-path transmission fronthaul solution. Despite all these efforts that pushed the error probability down to $10^{-5}$, further investigations are needed to achieve the ultra-high reliability of the Tactile Internet, i.e., an error probability of $10^{-7}$.

Finally, to cope with the emerging haptic-enabled VR systems' need for ultra-high data rates, one viable solution is to leverage on the mmWave frequency bands via mmWave communications. Investigating the feasibility of using mmWave communications to realize haptic-enabled VR applications has begun to attract real interest (e.g., [150]). Despite all the benefits that mmWave communications may offer in terms of ultra-high data rates and security, such systems still suffer from a major disadvantage. Specifically, the narrow beamwidth makes the mmWave system susceptible to link misalignment, which may deteriorate the reliability-latency performance. This reinforces the lesson that it is necessary to investigate the feasibility of smart beamwidth adaptation techniques designed to stabilize the link performance of mobile end-users. Advances along these lines could not only accelerate the transmission of multi-modal packets, they could also increase the reliability of wireless communication links and significantly reduce the packet error rate.

## VI. Non-radio Resource Allocation Algorithms beyond Lower layers for the Tactile Internet

This section focuses on non-radio resource allocation algorithms that go beyond the lower layers. We first review the algorithmic works on the allocation of edge resources including computation, storage, and communication resources for computation task offloading. Next, we discuss the recent works on co-design algorithms for the Tactile Internet. We then present the algorithmic works that address the routing and switching aspects. Finally, after summarizing our review, we discuss some insights and important lessons learned.

### A. Edge Resource Allocation

Typically, mobile users have limited computation resources, and thus local processing of computation-intensive, low-latency tasks at the mobile devices may not meet the stringent requirements of the Tactile Internet. A recent viable approach is the so-called computation task offloading, where mobile users can offload their computation-intensive tasks to computing and storage resources—variously referred to as cloudlets, micro datacenters, or fog nodes—which are placed at the Internet's edge in proximity to wireless end devices in order to achieve low end-to-end latency, low jitter, and scalability. Several papers have studied the feasibility of leveraging on computation task offloading in the Tactile Internet context. Xiao et al. [155] and Aazam et al. [156] focus on computation distribution for task offloading over fog computing networks. Tang et al. [157] tackle joint computation, storage, and communication allocation over mobile edge devices, while Ning et al. [158] consider joint computation offloading and radio resource assignment for vehicular applications. In addition, Xu et al. [159] investigate storage allocation for content caching in MEC. Each paper is explained in greater detail below.

The physical proximity and reliability of edge servers combined with the huge computational capabilities of the remote cloud suggests that the true potential of computation offloading can be unleashed through so-called *cooperative computation offloading*. Towards this end, Xiao et al. [155] focus on an energy-efficient computation distribution over fog computing-supported networks to enable Tactile Internet applications. The proposed approach allows a given fog node to offload a portion of its workload to other fog nodes or to the remote cloud to balance the workload in the network, thereby improving the response time of end-users as well as the power consumption of fog nodes. In this work, the objective is to optimally distribute the workloads over the fog nodes to minimize the overall response times, while satisfying the energy consumption of each node. The problem is formulated as an optimization problem. Global information, such as the computational capacities of fog nodes and the round-trip transmission latency among fog and cloud nodes is required to solve this optimization problem. However, some or all of this information may not be accessible due to privacy issues. To address this issue, the authors propose two distributed optimization algorithms to solve the computation distribution problem so that the optimization is performed at each fog node using its local data/information, allowing for a trade-off between the response time and power consumption. The authors conduct simulations based on a self-driving buses use case to evaluate the performance of the proposed solutions. The simulation results show that the proposed solutions can significantly improve the response time compared to a case where computation is done without any cooperation between fog nodes. We note that the developed optimization problem aims to minimize the response time rather than ensuring a given response time as a constraint. Given that the response time may be compromised for power consumption, the proposed scheme may lead to the violation of the given latency requirement. In addition, the achievable response time remains greater than 0.3 seconds. Thus, this work does not meet the ultra-low latency criterion ($C_1$). Further, the reliability aspect is not considered, and therefore the ultra-high reliability criterion ($C_2$) is not covered.

Aazam et al. [156] emphasize the importance of the QoE from a user perspective in the Tactile Internet era. They discuss the importance of fog resource utilization to offer high QoE, and propose a statistical metric, the Net Promotor Score-ratio, which represents the QoE of a specific customer in comparison to the QoE of all customers in a system when Tactile Internet services are used. The proposed metric can be used to provide a dynamic fog resource allocation such that more resources are allocated to a customer when the customer's QoE is poor. They investigate how the existence of such dynamic allocation reduces customer quitting and increases the service provider's profit. The simulation results indicate that the proposed model works efficiently to improve the QoE by allowing the dynamic selection of appropriate fog nodes for Tactile Internet services to reduce the overall latency. However, no solution has been provided to guarantee ultra-low latency and ultra-high reliability. Therefore, this work does not meet any of our criteria.

Tang *et al.* [157] study resource sharing among mobile edge devices through device-to-device communications. In particular, the objective is the coordination of the so-called 3C resources: communication, computation and caching resources. Each task is initiated at an edge node, which receives a set of contents as inputs either from the Internet or from the caches of other resources. The task is then executed by a computational edge resource, when then either uploads contents to the Internet or stores the contents in edge caches as the output. The content download/upload and task execution are assigned to the edge resources such that the energy consumption is minimized while the delay bound (which includes download, upload, and computation delay) is satisfied. The problem is modeled as a non-convex optimization problem, which is transformed into an equivalent ILP problem. The resultant ILP problem is then solved using the proposed heuristic approach, where the main idea is to iteratively solve multiple modified versions of the ILP problem where the delay constraints are relaxed and some tasks are prevented from being allocated to certain devices by standard optimizers until the delay constraints of the obtained solution are satisfied. The simulation results show that the proposed 3C resource sharing algorithm can lead to a 27% energy savings compared to the existing 1C/2C resource sharing approach. We note that the authors consider a delay bound on the order of seconds and do not consider the reliability aspect. Therefore, this work does not meet any of criteria.

Despite all the potential benefits that can be offered by providing cloud computing capabilities at the edge of access networks, the inefficient allocation of radio and computing resources for transmission of the computation tasks to computing entities may lead to the degradation of the overall latency and network utilization. To achieve low-latency and to utilize radio resources efficiently, Ning *et al.* [158] study joint computation offloading decisions, power allocation, and channel assignment in 5G networks. The authors consider a vehicular application scenario where vehicle users can offload their computation tasks to either a micro-cell or to nearby roadside units. The users communicate with the micro-cell and with roadside units through NOMA-based and vehicle-to-vehicle (V2V) technologies, respectively. The authors also assume that due to the limited number of roadside units, offloading the tasks to roadside units presents a higher task forwarding delay than offloading to a micro-cell. An optimization problem is then developed to determine the optimal task offloading decision, power allocation, and channel assignment such that the total network utility is maximized while satisfying the task response delay. To solve the problem, it is divided into the following two sub-problems: (i) task allocation and (ii) joint power and channel allocation. First, the expected task forwarding delay to the nearest roadside unit is derived to determine the task offloading decision. Next, a heuristic algorithm is proposed to solve the subsequent sub-problem in an iterative manner. The simulation results indicate that the proposed scheme outperforms the two baseline schemes in terms of network utility, which suggests that a larger number of tasks can be offloaded. Also, the results demonstrate that the proposed algorithm outperforms the OFDM-based solutions. We note that this work does not explicitly specify the maximum tolerable delay value, and no constraint related to the reliability aspect is discussed. Consequently, it does not meet any of the criteria.

In addition to computing resources, edge computing can provide storage resources, which can be used as cache nodes in which to store contents that are frequently demanded by end-users. Deployment of caching at the edge servers gives way to the so-called *mobile edge caching*[1], which helps to alleviate the mobile traffic in the backhaul and thus reduce the content delivery latency. In [159], Xu *et al.* propose an energy-efficient hybrid edge caching scheme for the Tactile Internet in 5G. They consider a three-tier heterogeneous caching system. Their system model consists of a server tier, a macro-base station tier, and a small base station tier. The server tier consists of central cloud servers and storage units with their high processing and storage capabilities. The server tier also owns a complete file library. Macro and small BSs offer edge caching capabilities to mobile users including smart devices and vehicles. Mobile users request files from the file library. A Zipf distribution for file popularity is assumed. When a user requests a file, the local cache on the mobile device is checked first. In the case of a cache-miss, the two following caches are checked: (i) the caches of the small BSs covering the user and its neighbors, and (ii) the cache of the macro BS covering the user. The authors cover the caching problem by using two separate optimization problems designed to: (i) minimize the average end-to-end latency to access the requested files and (ii) maximize the overall energy efficiency for transmission of the user requests. The authors propose a hybrid heuristic-based edge caching scheme to solve the two problems. The simulation results indicate that the proposed scheme can improve the overall energy efficiency and the average end-to-end latency in comparison to the existing schemes. It can also achieve an average end-to-end latency of 1 ms or even less for users to access their requested content files. Therefore, this work meets the ultra-low latency criterion ($C_1$). This work does not consider reliability, and so it does not meet the ultra-high reliability criterion ($C_2$).

### B. Co-design

The Tactile Internet sits at the crossroads of communication, computation, and robotization. The idea of leveraging on these crucial aspects of the Tactile Internet to achieve immersive, reliable tactile experiences gives rise to the so-called co-design approaches, in which the interdependencies between communications, robotics, and cloud computing are taken into consideration in order to increase the user's QoE. Only a few studies have taken a co-design approach in the context of the Tactile Internet. Xu *et al.*

---

[1]For a comprehensive survey on mobile edge caching and its role in 5G cellular networks, the interested reader is referred to [160].

[161], Liu *et al.* [162] and Chang *et al.* [163] focus on the co-design of communication and control schemes for Tactile Internet applications. Ashraf *et al.* [164] investigate the joint design of energy management and data rate control for energy harvesting-enabled tactile IoT nodes. Garcia-Saavedra *et al.* [165] study the problem of joint deployment of radio split functions and service functions in centralized RANs for Tactile Internet services. These works are presented in greater detail below.

While necessary, the design of reliable low-latency networking infrastructure and/or algorithms is not sufficient to realize an immersive teleoperation experience. In fact, efficient control schemes are needed to enable a smooth, glitch-free teleoperation, especially in the presence of communication-induced artifacts such as latency and jitter. Towards this end, Xu *et al.* [161] investigate the potential of the dynamic switching of control schemes in a bilateral teleoperation system, comprising a master and slave domain. In this study, the objective is to dynamically select the appropriate control scheme maximizing the QoE while considering the time-varying communication latency. The authors have considered two control schemes for a virtual teleoperation environment of a one-dimensional spring-damper system. The first control scheme is referred to as a time-domain passivity approach (TDPA), which is highly sensitive to delay, but offers support for highly dynamic interactions between the operator and the remote environment. The second control scheme is model-mediated teleoperation (MMT), which can tolerate large delays, and therefore is not suitable for highly dynamic environments. Both schemes are used with a perceptual deadband-based data reduction technique to down-sample the transmission of haptic packets. The authors propose a dynamic switching scheme that selects between the two control schemes based on the delay evaluation. To validate their proposed solution, the authors ask participants to interact with the virtual teleoperation environment of the one-dimensional spring-damper system to measure users' experience. The experimental results show that the QoE of each approach is observed to degrade as delay increases, but it remains more stable for the MMT with the perceptual deadband. Moreover, the MMT with the perceptual deadband starts to provide better QoE than the TDPA with the perceptual deadband when the delay is greater than 50 ms, which indicates a switching point of the proposed solution that allows it to maintain high QoE. This work covers end-to-end delay values ranging from 0 to 200 ms. However, ultra-low delay values are shown to require larger average packet rates, which can go up to 50 packets per second. As a result, this paper meets the ultra-low latency criterion ($C_1$). However, no reliability metric is considered, and so the ultra-high reliability criterion ($C_2$) is not covered by this paper.

In [162], Liu *et al.* tackle the joint control and radio resource allocation problem for bilateral teleoperation. More specifically, after presenting a quantitative measure of QoE vs. communication latency for the TPDA and MMT schemes under study, the authors propose a new uplink scheduling algorithm for haptic communication over 5G networks to maximize the total QoE of teleoperation sessions by allowing flexible switching between the two control schemes, depending on the value of the measured round trip delay (which is varied from <1 ms up to 200 ms). The authors develop an exponentially complex optimization problem and then present a heuristic solution by changing the original problem into a max-min problem for real-time implementation. The simulation results indicate that the proposed scheme can outperform the proportional fair scheduling method in terms of throughput and delay, meaning that it offers an improved QoE. Although this work indicates that the proposed scheme can achieve an uplink delay of 1 ms, the reliability aspect is not considered here. Therefore, it only partially meets the ultra-low latency criterion ($C_1$).

In [163], Chang *et al.* focus on reducing the wireless resource consumption in real-time wireless control by means of dynamic QoS allocation from a control-communication co-design perspective. They first develop a communication-control co-design model that develop includes both critical communication and control parameters. To be more specific, they consider a typical model-based linear wireless control, in which it is assumed that the wireless communication between the sensor and the remote controller is non-ideal, whereas the communication between the remote controller and the plant is ideal. Next, the authors obtain a criterion to evaluate the control performance for the above-mentioned system model. After obtaining the control cost as the criterion, an optimal control method is developed to obtain the parameters used in cost control. They then propose a dynamic QoS allocation method to reduce the wireless energy consumption with minor control performance loss. This is done by designing the appropriate threshold to allocate the extreme high QoS and low QoS throughout the control process, where the high QoS is allocated to critical control periods and low QoS is given to the non-critical ones. According to the simulation results, the energy consumption of the proposed method shows only a 5% increase compared to the low QoS, and an 80% decrease compared to the high QoS method. In this work, the authors consider the end-to-end delay of 1 ms and packet loss of $10^{-5}$. Therefore, this work meets the ultra-low latency criterion ($C_1$), but it does not meet the ultra-high reliability criterion ($C_2$).

Energy efficiency remains a key challenge in managing energy harvesting-enabled tactile IoT nodes for supporting Tactile Internet applications, as these nodes still have limited energy storage. Inefficient energy management and transmission rate control can lead to energy starvation, where these nodes might not have enough energy to operate in emergency and critical situations. In order to achieve an improved reliability performance, Ashraf *et al.* [164] study the combined energy management and rate control problem for energy harvesting-enabled tactile IoT nodes. The problem is presented as a queue control problem. In each node, its energy and data buffers are first modeled as energy and data queues, respectively, where both linear and non-linear models of energy and data queues are considered. Next, control methods based on predictive model control and non-linear control theory are proposed to design controllers for linear and non-linear cases, respectively. The objective is to control the state of both energy and data queues close to predefined reference values. This can ensure the availability of the node and keep the queuing delays experienced by packets in the data queue as small as possible. The simulation results show that the proposed methods for both linear and non-linear cases can indeed regulate the state of energy and data queues over time, successfully maintaining their state close to the predefined reference values. This indicates that the device always has enough energy to operate, thus ensuring the reliability of the system. Therefore, this work meets the ultra-high reliability criterion ($C_2$). The authors also claim that maintaining the state of the data queue close to the predefined reference value can ensure a small queuing delay. However, they do not evaluate the packet delay achieved by the proposed methods. Therefore, this work does not cover the ultra-low latency criterion ($C_1$).

Centralized radio access networks (CRANs), also referred to as cloud-RANs, are a viable solution towards achieving cost-efficient access networks, where various base station functions are transferred from remote radio heads to a central unit. Although C-RANs may offer a number of benefits such as reducing cost and improving spectrum efficiency, they impose stringent latency and bandwidth requirements on the fronthaul, operating as the interface connecting the central unit and remote radio heads. To cope with these issues, the so-called virtual RAN (vRAN), which relies on leveraging softwarization to enable a flexible selection of centralization degree of each BS given the available network resources, has recently been presented. In [165], the authors investigate the problem of the joint optimization of deployment of MEC services and radio split functions, with the main objective of maximizing the MEC service performance while minimizing the vRAN cost. The stringent latency requirement of Tactile Internet services is reflected within the constraints of the developed optimization problem, which is solved using Benders' decomposition method. Both analysis and simulation results reveal that the joint and flexible MEC-vRAN design achieves significant cost savings of up to 2.5 times compared to that of non-optimized fully centralized RAN (i.e., CRAN) or fully decentralized RAN (i.e., DRAN) systems. It can also guarantee the delay bound of the Tactile Internet services on the order of 1 ms, and therefore this work meets the ultra-low latency criterion ($C_1$). Since the reliability is not considered in this work, it does not cover the ultra-high reliability criterion ($C_2$).

*C. Routing*

Due to its stringent QoS requirements, the Tactile Internet requires novel routing schemes that consider the coexistence between haptic traffic and conventional triple-play traffic. To date, three studies have aimed to design novel routing algorithms for the Tactile Internet. While Lumbantoruan *et al.* [166] and Farhoudi *et al.* [167] focus on one-to-one routing, Ren *et al.* [168] consider broadcast routing.

In wired networks, the most significant portion of the end-to-end latency is traced back to the queueing delay, which mainly occurs due to node and/or link congestion. Other latency contributors such as transmission and processing delays are often negligible or only account for a relatively small percentage compared to the queueing delay. To address the problem of congestion in wired networks, one viable solution is a process known as *traffic engineering*, which aims to balance the traffic load and optimize the packet routing process by means of designing suitable routing and traffic management strategies. Toward this end, Lumbantoruan *et al.* [166] propose a multi-plane routing (MPR)-based algorithm to obtain reliable and low-latency paths for Tactile Internet traffic coexistent with best-effort traffic. Specifically, an optimization framework is developed with the objective of finding a suitable routing strategy for multiple commodity flows consisting of both tactile and best-effort packets. This routing strategy maximizes the total flow in the network while ensuring that the QoS requirements of tactile packets are met and the routing cost is minimized. The proposed scheme uses MPR as the underlying routing protocol for implementing multipath routing for the purpose of load balancing by means of prioritizing the tactile packets, thus satisfying as many potential future demands as possible. The MPR builds on multiple logical routing planes, which represent instances of the open shortest path first (OSPF) routing algorithm, thereby ensuring the availability of multiple routing paths for each source-destination pair. The results indicate that the proposed algorithm can achieve an average latency of 1 ms for tactile traffic over an IP access network. Therefore, this work meets the ultra-low latency criterion ($C_1$). The results indicate that the proposed algorithm outperforms the OSPF and equal cost multi-path routing algorithms in terms of packet delivery ratio and throughput. As for the reliability, the proposed algorithm was shown to achieve a packet delivery ratio of 100% under low to medium loads. Therefore, it also meets the ultra-high reliability criterion ($C_2$).

In [167], Farhoudi *et al.* propose a traffic routing policy to handle both Tactile Internet and traditional traffic. The proposed policy is based on MPR, which is extended to handle the haptic traffic. Their solution considers the deployment of two queues at each router, one for priority traffic (i.e., haptic) and the other for non-priority traffic. To further enhance the reliability, MPR enables the forwarding of duplicate haptic packets on two of the best available routing planes. The solution operates an offline stage followed by an online stage. The offline stage defines the different routing planes, whereas the online stage determines the best routing plane for a given session. The online stage is modeled as an ILP problem to find the routing plane for each session while satisfying the QoS requirements in terms of delay, jitter, and packet loss. An online policy is then proposed to solve the developed ILP problem. The operations are evaluated on campus and metropolitan networks. The proposed solution is shown to outperform the OSPF routing algorithm and offers near-optimal solutions that can achieve an average delay of 0.88 ms, a jitter value of below 20 microseconds, as well as a completely eliminated packet loss. Based on these results, this approach meets both the ultra-low latency ($C_1$) and ultra-high reliability ($C_2$) criteria.

Ren *et al.* [168] leverage on the construction of minimum spanning trees for message broadcasts in the Tactile Internet. They introduce a framework that constructs such trees at a minimum total cost while satisfying a given end-to-end delay bound. The proposed framework employs a dynamic algorithm operating on a given network graph. This dynamic algorithm includes a preprocessing technique, which discards the infeasible and sub-optimal edges from the graph. Edge scores are then derived based on their costs. The algorithm iterates over the given set of edges in a non-increasing score order. A given edge is eliminated only if its elimination does not affect the delay constraint; otherwise, it is added to the spanning tree. The edge elimination process continues until all the edges can ensure a connected graph among all the nodes. This framework is shown to outperform the existing heuristic approaches, namely, Prim's algorithm and the competitive decision algorithm, in terms of average cost and stability. This work considers different delay bounds ranging from 6 to 40 ms, but not the 1 ms delay bound. Therefore, it does not meet the ultra-low latency criterion ($C_1$). Moreover, this paper focuses on the overall cost, and there is no assessment of its reliability. Therefore, this work does not cover the ultra-high reliability criterion ($C_2$).

### D. Switching

Two algorithmic works focus on scheduling policies in switches and/or routers to efficiently steer the Tactile Internet traffic. Shin *et al.* [169] introduce a low-latency scheduling algorithm in cut-through switching-based networks, and Deng *et al.* [170] investigate delay-constrained input-queued switching using virtual output queueing for the Tactile Internet.

Today's store-and-forward switching-based network architecture leads to an increased transmission delay, which is not scalable with the number of forwarding hops, as per-hop packet transmission delays are accumulated at every router. Cut-through switching is a promising technique to significantly decrease the transmission delay of store-and-forward switching. In [169], Shin *et al.* aim to reduce per-router transmission delay using the cut-through switching paradigm, which divides a packet into multiple flits and concurrently forwards them from the ingress link to the egress link, as opposed to the conventional store-and-forward switching structure, where per-hop transmission delay accumulates at every router along the path, thus lacking scalability with the number of forwarding hops. Leveraging on a mechanism of discarding the packets that are expected to violate the delay requirement, the authors propose their so-called flit-based earliest-deadline-first and flit-based shortest-processing-time-first scheduling algorithms to achieve low end-to-end latency with high reliability, where the reliability is measured by the ratio of packets satisfying the delay requirement. The simulation results state that the proposed scheduling algorithms significantly outperform the existing algorithm and conventional first-in first-out scheduling algorithms with store-and-forward and cut-through switching in terms of both packet delivery rate and latency. However, the minimum average end-to-end delay remains greater than 4.5 ms, and the achievable packet delivery rate cannot guarantee a reliability of $10^{-7}$. Therefore, this work does not meet any of the criteria.

In [170], Deng *et al.* study the delay-constrained input-queued switch using virtual output queueing, where each packet has a deadline and it will expire if it is not delivered before its deadline. More specifically, the capacity region is first characterized in terms of the timely throughput of all input-output pairs. After that, a throughput-optimal scheduling policy that can support any feasible timely throughput requirements in the capacity region is designed to be suitable for inelastic applications with stringent minimum timely throughput requirements. The authors then aim to design a scheduling policy to maximize the network utility with respect to the achieved timely throughput, which is important for elastic applications with no stringent minimum timely throughput but large utility requirements. To tackle these problems, the authors leverage on the Markov decision process to characterize the capacity region and design their feasibility-optimal scheduling policy. Given that their Markov decision process-based approach suffers from the curse of dimensionality, they characterized the capacity region with only a polynomial number of linear constraints by means of combinatorial matrix theory. By leveraging the Lyapunov-drift theorem, the authors showed that the problem of minimizing Lyapunov drift is a maximum-weight *T*-disjoint-matching problem, and then solved it optimally by means of their proposed polynomial-time, bipartite-graph edge coloring algorithm. The simulation results show that the proposed scheduling policy outperforms the existing scheduling policies that are not designed for delay-constrained traffic in terms of timely throughput with different deadline setting. However, this work does not show the performance of the proposed scheduling policy when the packet deadline is on order of 1 ms, and so it does not meet the ultra-low latency criterion ($C_1$). Also, no evaluation metrics related to the reliability aspect are considered. Therefore this work does not cover the ultra-high reliability criterion ($C_2$).

### E. Summary Insights and Lessons Learned

In this subsection, we briefly summarize the works on non-radio resource allocation algorithms that go beyond lower layers. We then identify some key insights and discuss the important lessons learned.

*1) Summary*

This section covers edge resource allocation, co-design, routing and switching algorithms for the Tactile Internet. Table VI summarizes the key contributions of the works reviewed in this section.

A number of studies have focused on edge resource allocation for the Tactile Internet. Their main objective is to efficiently utilize resources (i.e., computation, storage, and communication) at the network edge to help end-users with limited resources meet the strict Tactile Internet requirements. These studies have tackled the computation resource allocation problem over fog systems [155][156], resource sharing problem among mobile edge devices [157], joint computation and radio resource allocation problems [158], and content edge caching problems [159] for the Tactile Internet.

Given the multi-disciplinary nature of the Tactile Internet, co-design approaches are an interesting group of algorithms with which to tackle the latency and reliability challenges by taking into account the crucial aspects of communication, computation, and control. Despite its paramount importance, this area of research has not received much attention. While four works [161]-[164] have leveraged the interdependence between the communication and control domains of telerobotic systems to increase the QoE, one work [165] has taken a communication-computation approach.

The works that consider the routing aspect mostly aim to address the reliability. The proposed routing algorithms are based on MPR [166][167] and minimum spanning tree [168] approaches. The MPR protocol outperforms intra-domain routing protocols (e.g., OSPF). In addition, the dynamic edge-elimination-based algorithm [168] can efficiently construct a rooted delay-constrained minimum spanning tree framework for message broadcasts while ensuring the given end-to-end delay constraints.

TABLE VI
The main contributions and evaluated criteria for the algorithmic works on non-radio resource allocation beyond lower layers for the Tactile Internet. In the criteria columns, Y indicates that the criterion is met, P indicates that the criterion is partially met, and N indicates that the criterion is not met.

| Scope | | Ref. | Main Contributions | Criteria | |
|---|---|---|---|---|---|
| | | | | $C_1$ | $C_2$ |
| Non-radio Resource Allocation beyond Lower Layers | Edge Resource Allocation | Xiao et al. [155] | Present an energy-efficient computation distribution algorithm based on a distributed approach over a fog computing network to minimize the overall delay under the power constraints of fog nodes. | N | N |
| | | Aazam et al. [156] | Introduce a QoE-driven dynamic resource allocation framework based on the Net Promotor Score-ratio method in a fog computing network to maximize the QoE of tactile users. | N | N |
| | | Tang et al. [157] | Propose a 3C resource sharing framework among mobile edge devices through D2D communications to minimize energy consumption under a delay constraint. | N | N |
| | | Ning et al. [158] | Propose a joint computation offloading decision, power allocation, and channel assignment algorithm for vehicular applications over 5G cellular networks to maximize the number of offloaded tasks. | N | N |
| | | Xu et al. [159] | Present an energy-efficient hybrid edge caching scheme that jointly optimizes the energy efficiency and content access delay. | Y | N |
| | Co-design | Xu et al. [161] | Propose a control scheme selection with the objective of optimizing the QoE in teleoperation systems. | Y | N |
| | | Liu et al. [162] | Present a joint control and radio resource allocation framework for bilateral teleoperation to maximize the QoE. | P | N |
| | | Chang et al. [163] | Propose a real-time wireless control scheme to enable dynamic QoS allocation to ensure URLLCs. | Y | N |
| | | Ashraf et al. [164] | Introduce a joint energy management and rate control framework for harvesting enabled energy-constrained tactile sensing nodes to prolong their lifetime and maintain a small queuing delay. | N | Y |
| | | Garcia-Saavedra et al. [165] | Present an optimization framework for the joint deployment of MEC services and radio split functions in CRANs to jointly optimize the QoS of the MEC services as well as the routing and computation costs. | Y | N |
| | Routing | Lumbantoruan et al. [166] | Present an MPR-based algorithm to obtain reliable and low-latency paths for the Tactile Internet in wired access networks. | Y | Y |
| | | Farhoudi et al. [167] | Propose an MPR-based routing and scheduling algorithm for the Tactile Internet in wired access networks with the objective of reducing delay, jitter, and packet loss. | Y | Y |
| | | Ren et al. [168] | Propose a minimum spanning tree-based routing algorithm for wired access networks to minimize the cost while ensuring low latency. | N | N |
| | Switching | Shin et al. [169] | Introduce a low-latency scheduling for cut-through switches with the objective of reducing per-router transmission delay. | N | N |
| | | Deng et al. [170] | Present an efficient scheduling policy for delay-constrained input-queued switching policy using virtual output queueing with the objective of maximizing the throughput. | N | N |

Finally, two papers have proposed scheduling policies in switches and/or routers to efficiently handle tactile traffic. Reference [169] focuses on cut-through switching-based networks, while Reference [170] investigates delay-constrained input-queued switching using virtual output queueing for the Tactile Internet.

*2) Insights and Lessons Learned*

An insight on edge resource allocation is that most of the studies do not consider user mobility. These studies may not be suitable for the Tactile Internet applications that require (almost) constant mobility (e.g., autonomous driving) because computation resources need to be allocated according to the user's movement in order to meet the stringent delay and reliability requirements. For instance, in autonomous driving, the application migration and edge resource allocation to run an application server has to be made according to the movement of vehicular users. If not, the stringent delay and reliability requirements between the application server and the vehicular users are not likely to be met. We note that although reference [158] considers the node mobility on the computation resource allocation, it does not satisfy either of our criteria. The lesson we can learn from this insight is that more research efforts are needed on mobility management while allocating resources at the edge for the Tactile Internet.

Among our surveyed papers, advanced co-design solutions that address the Tactile Internet challenges (i.e., [161]) rely on the use of data reduction techniques (e.g., perceptual coding) in conjunction with different stability-ensuring control schemes (e.g., TPDA or MMT) in the presence of end-to-end latency between the human operator and telerobot. Further, it has been shown in [161], [162], and [163] that each control scheme is associated with a certain amount of delay tolerance, so that the selection of the proper control scheme for a given end-to-end delay may be as effective as reducing the latency itself. Another insight is that communication delay and packet loss can be modeled into the underlying control system via a communication-control co-design, where the communication energy consumption is reduced using dynamic QoS allocation throughout the control process [163]. Therefore, a lesson we can learn is that further investigation is needed to design mechanisms to reduce the latency while adaptively allocating the proper control schemes.

From a routing perspective, the first insight is that the MPR scheme can ensure that the QoS requirements of haptic packets are met while maximizing the total network flow as well as minimizing the routing cost (e.g., [166][167]). It is also shown that shortest-path algorithms cannot handle haptic traffic. An optimized MPR algorithm has achieved significant improvements over equal-cost multi-path algorithm in terms of throughput, delay, and packet delivery ratio [167]. We note, however, that optimization approaches to find the best routing plane in an MPR scheme may add extra computational burdens, especially for large-sized graphs. This mandates the need for designing efficient heuristics to solve achieve (sub)optimal optimized MPR solutions to realize ultra-reliable, low-latency routing for haptic traffic.

Another important insight from a routing perspective is that none of the existing studies has tackled the problem of traffic routing for wireless multi-hop networks. This is especially important for the cases where establishing a direct communication link between the BS/AP and the mobile users becomes infeasible. This might be also necessary in case of excessive network congestion in the core network, which may result in an unacceptable end-to-end latency. In these cases, a traffic routing mechanism is needed to establish an efficient routing path using D2D-enabled wireless multi-hop networks in order to meet the stringent Tactile Internet requirements. The lesson we can learn from this insight is that more research efforts on this aspect are still needed to realize the Tactile Internet over D2D-enabled multi-hop wireless access network.

## VII. RESEARCH DIRECTIONS

This section provides examples of some potential research directions. Exhaustiveness is impossible due to the richness and novelty of the topic.

### A. Architectures

As indicated in the various lessons learned, there is a need for architectures that go beyond offloading and running components on MEC servers, since the results of those approaches are rather mixed. An example of a new direction would be to run components closer to end-users. Another example could involve using parallel processing of the traffic that goes through VNFs in 5G settings. A third example is to try "forcing" the streams generated by multi-stream transport protocols through multiple paths in networks. Each of these directions are elaborated below.

**Running components closer to end-users:** Generally, several devices are much closer to end-users than MEC servers. Furthermore, these devices are getting more and more powerful. Some examples are smart phones, set top boxes and even laptops. Running applications (or at least some of the components of applications) on these devices will certainly help in meeting stringent latency requirements. However, there are still very few Tactile Internet architectures that exploit the proximity and the capacity of these devices. To the best of our knowledge, only reference [96] (already discussed in this survey) exploits end users' cellular phones as Aps for offloading cellular traffic on Wi-Fi. However, it does not really use the processing capabilities of these devices. There is indeed a need to research architectures that integrate devices (in the vicinity of end-users) in a processing layer.

A potential starting point for these architectures is the concept of mobile ad hoc clouds [171][172]. A mobile ad hoc cloud allows computation by leveraging the resources of a group of mobile devices in the same vicinity. Local stationary devices such as set top boxes can also be included. Mobile ad hoc clouds can indeed constitute the processing layer that integrates the devices in the vicinity of end-users. However, integrating mobile ad hoc clouds in Tactile Internet architectures entails multiple challenges. A first challenge is that several issues related to the very concept of mobile ad hoc clouds (e.g., incentives, resource management) are still not yet resolved in a satisfactory manner [173]. Yet another challenge is that this new processing layer will need to interact with the two other processing layers generally considered in the Tactile Internet, meaning cloud and edge layers.

**Parallel processing of traffic that goes through VNFs in 5G:** In 5G, VNF chains installed on commodity servers process the traffic. Reference [174] shows that it is not always possible to meet stringent deadlines in such settings with a sequential approach. It therefore proposes a parallel approach that splits the traffic and sends it through duplicate VNFs installed on servers selected from pools installed along the paths.

However, in order to make this parallel approach a reality, innovative SDN-based architectures are required to gear the traffic through the different duplicate VNFs in a scalable and highly available manner. The very high-level system view sketched by reference [174] (which focuses on the algorithmic aspects) could serve as the basis for the design of such architectures. It comprises an application plane (routing application and server provisioning entities), a control plane (an SDN controller and NFV management and orchestration entities) and an infrastructure plane (with NFV infrastructure and SDN routers/switches). The input comes from the server provisioning entity and consists of the VNF chain, the input traffic size, and the user defined deadline. The routing application entity decides how to distribute the traffic among the parallel servers that duplicate the execution of the same VNF. It also interacts with the SDN controller entity to program the switches.

**"Forcing" streams generated by multi-stream transport protocols through multiple paths in networks:** Innovative architectures that gear the streams generated by multi-stream transport protocols through multi-paths in networks could also rely on SDNs. Examples of such protocols are QUIC [175] and SCTP [176]. Reference [177] discusses the details of multipath routing. There are now multi-path versions of both QUIC [178] and SCTP [179]. However, these still assume that the node (generally a client) has several network interfaces, and each of these interfaces maps the traffic onto different paths.

Unfortunately, nodes do not always have several network interfaces. Furthermore, one-to-one mapping between network interfaces and paths is not always optimal from a latency reduction perspective. General architectures that map multi-streams onto multi-paths in a flexible and dynamic manner will be more appropriate for latency reduction. SDNs seem to be the ideal paradigm on which to base such architectures. Reference [180] presents early efforts in that direction and may serve as a basis for further research.

*B. Protocols*

While to the best of our knowledge, only two papers expressly address the challenge of new protocols for the Tactile Internet, few other works address the general issue of latency reduction in protocols. One example is reference [181]. It aims at preventing congestion collapses of the backhaul traffic and at ensuring ultra-high data rates of up to 400 Gb/s. It is an improved version of TCP and consequently runs in the kernel.

However, protocols that run in the kernel are not suitable for low latency applications due to potential cross–talk between the various applications that use the protocols simultaneously [182]. Protocols such as QUIC implemented in the user space are therefore more promising for achieving the ultra-low latency requirement. QUIC aims at offering better performance than TCP in terms of latency. However, the improvements it brings are at best incremental, according to the concrete measurements made to date [183][184]. Thus, another research direction would be to investigate how to make QUIC quicker.

Designing new transport protocols (to be implemented in the user space) with ultra-low latency as the key design goal is another promising avenue. Yet another example would be the design of application-specific transport protocols (to be implemented in the user space), with the specific requirements of an application, including ultra-low latency, incorporated as part of the objective. Reference [182] provides an example of an application-specific transport protocol implemented in user space. It targets vehicular networks and its design goals include low (but not ultra-low) latency. By the same token, one may well envision a transport protocol for remote robotic surgery with ultra-low latency as one of its key design goals.

*C. Intelligent Prediction*

Machine learning is an integral part of artificial intelligence, and can be defined as a mechanism that gives a computer the ability to learn and act without the need to be programmed [185]. Investigating the application of machine learning for prediction purposes in the context of the Tactile Internet is a very important research direction, especially in relation to tactile artificial agents, such as those mentioned in this paper, as part of the remote robotic surgery use case (see Section II.B). These tactile intelligent agents are pertinent to all tactile teleoperations.

In robotic surgery, unlike other areas such as tele-phobia treatment, substantial work has already been done on the use of machine learning, although not for predicting packet loss. The work completed to date on the use of machine learning for robotic surgery is generally focused on fully automating specific procedures and covers all four phases of surgery: access to the body cavity, tissue dissection, tissue destruction and tissue reconstruction. For instance, [186] focuses on access to the body cavity. It uses Gaussian mixture regression to enable the autonomous execution of ultrasound scanning. Reference [187] deals with the tensioning required to cut circular patterns, a fundamental procedure in robotic surgery. Reinforcement learning is a key to this aspect. On the other hand, reference [188] deals with tissue temperature dynamics during laser exposure. Correctly monitoring these dynamics is essential to being able to predict the effects of the laser-tissue interaction during surgery. Finally, reference [189] focuses on in-situ microscopic surgery using Gaussian mixture regression to perform autonomous tissue surface scanning and reconstruction.

Substantial research is still needed to bring the work done in this context of (local) robotic surgery to the world of remote robotic surgery. The main reason this additional work is needed is that the approaches proposed thus far do not account for the eventual packet losses in the network, as their key assumption is that the surgical procedure is executed locally. For example, being able to assess prediction accuracy for remote robotic surgery will require evaluation techniques that evaluate accuracy under various network conditions. Many approaches that are considered accurate today may very well become inaccurate when network conditions are considered.

Typically, in a predictive robotic telesurgery system, a forecast sample is generated and delivered to the operator rather than waiting for a delayed action. To ensure the stability and tracking performance of such systems, the prediction error must be kept very small, preferably smaller than the so-called just-noticeable difference (JND) of humans. According to the well-known Weber's law, JND$^2$ is defined as the minimum change in the magnitude of a stimulus that can be detected by humans [190]. Recall from Section III.D that the authors of [49] and [50] conducted a trace-driven analysis of Tactile Internet traffic. We note, however, that the available traces studied in [49] and [50] are restricted to only 6-DoF teleoperation and 1-DoF telesurgery systems, respectively, thus posing limitations to the reported results due to being application- and/or device-specific, especially given that the emerging advanced robotic teleoperation systems with relatively larger numbers of DoF may rely on the exchange of a wider variety of haptic feedback samples than simply 6-DoF force-torque samples. Thus, reference [49] may be considered as a starting point from which to suggest methods of investigating the haptic traffic characteristics by means of a more comprehensive trace-driven study. Hence, in order to validate the findings of [49], it would be worthwhile to explore a more diverse set of traces gathered from various teleoperation systems. Further investigations are obviously needed to generalize the proposed ANN-based forecaster in references [49] and [50] to other deployment scenarios apart from the studied 1- and 6-DoF teleoperation use cases. Towards this end, the applicability of more sophisticated machine learning models, such as deep neural networks and/or long-short-term-memory neural networks, may be further investigated for detecting the patterns within highly complicated multi-dimensional data.

Moreover, given that using a first-order HMM in [111] is shown to achieve promising results, exploiting the temporal correlation within haptic samples by considering a larger number of past observations rather than just the preceding sample appears to be promising. As such, it would be worthwhile to investigate the impact of higher order HMMs on reducing the prediction error.

---

$^2$ The JND threshold for force perception with a hand and an arm is $7 \pm 1\%$ and for velocity around $8 \pm 4\%$ [20].

Note also that even though achieving a very low latency on the order of 1 ms and providing the human operator with accurate forecasts may be necessary to realize immersive Tactile Internet experiences, these steps are not sufficient to guarantee a desired QoE. This suggests the need to conduct subjective and objective studies to allow for both qualitative and quantitative measuring of how closely a human operator is coupled with the involved experience enhanced by machine learning-based prediction.

*D. Algorithms*

The algorithmic aspects of the Tactile Internet have received much more attention from the research community than the architectural and protocol aspects, accounting for approximately seventy-five percent of the total work done in this area. However, as highlighted in the lessons learned discussed at the end of each algorithmic section, there are still some research gaps that need to be filled before the Tactile Internet can become a reality. In the following five subsections, we discuss possible research directions on the algorithmic aspect of the Tactile Internet, mainly derived from the lessons learned from our literature review. These include the research directions on radio resource allocation, WLAN and WBAN resource allocation, edge resource allocation and routing for the Tactile Internet.

*1) Radio Resource Allocation*

As discussed in a number of recent papers (i.e., [6][11][14]), 5G is anticipated to underpin the Tactile Internet, especially at the wireless edge. To realize this, an efficient radio resource allocation framework is required to ensure the stringent latency and reliability requirements of Tactile Internet applications, especially given its significant impact on the overall latency and reliability in RANs. Radio resource allocation has been studied in a number of research works, among which some promising performances has been reported, e.g., [125][127][130][131][134]. Despite the benefits achieved by these works, some issues remain to be addressed before 5G wireless networks can be relied upon to effectively support the Tactile Internet with its full capacity. These issues are discussed in the following paragraphs.

**Mobility management:** Mobility management in the context of radio resource allocation has not received much attention, even though mobility is at the core of mobile networks. Allocating radio resources to mobile devices in a Tactile Internet is particularly challenging due to the stringent latency and reliability requirements. In the existing studies on radio resource allocation, the node mobility has been covered as part of a targeted vehicular networking application (i.e., [130][131]). We note, however, that the proposed mechanisms were not designed to handle node mobility, and therefore they lack the ability to efficiently manage the mobility of nodes in a Tactile Internet scenario. These algorithms should ensure that the end users are allocated the radio resources they need as they move, while still guaranteeing the Tactile Internet requirements. Toward this end, a starting point could be the work in [191], which targets mobility management in CRANs. As discussed in [96][152][165], the CRAN has proven to be a potential technology for the Tactile Internet in 5G RANs. The strategy in [191] could be extended to meet the Tactile Internet requirements.

**NOMA-based cellular systems:** In order to support massive connectivity along with the ultra-low latency of the Tactile Internet, NOMA has proven to outperform OFDMA due to its higher spectral efficiency and lower latency [22][137]. However, only a few radio resource allocation studies focus on the NOMA-based cellular system. Allocating radio resources in NOMA-based cellular systems is still challenging due to the presence of inter-user interference, which can lead to poor reliability performance. In this context, two notable studies are [121] and [126], which rely on deep learning and a cross layer approach, respectively. We note, however, that none of these works meets the ultra-high reliability requirement. Therefore, there are still no efficient radio resource allocation algorithms for the Tactile Internet in NOMA-based cellular systems to exploit NOMA's specific advantages over OFDMA. References [121] and [126] can serve as a basis for further research. The main goal here is to support the massive number of connected devices while guaranteeing both stringent Tactile Internet requirements in NOMA-based cellular systems.

**Large-size packet transmission**: As identified in the lesson learned, there is a lack of efficient resource allocation algorithms to handle the transmission of large-size packets to enable multi-sensory Tactile Internet applications such as haptic-enabled virtual reality. Achieving the reliable transmission of a large-size packet is not straightforward, especially given that the probability of erroneous transmission increases as the packet size increases. In order to achieve the ultra-low latency and high reliability for the transmission of large-size packets, a viable staring point would be to investigate the feasibility of the existing radio resource algorithms proposed for small-size packet transmission (i.e., [125][127][130][131][134]) with larger packet sizes and then to enhance them using advanced error control schemes.

**Mobile network technology in coordination with other low-cost technologies:** To support the massive number of connected devices, one of the ultimate goals of 5G, mobile network operators must coordinate their mobile network with other low-cost technology networks to increase network capacity. Supporting Tactile Internet applications coexistent with other applications over these coordinated networks requires efficient resource allocation frameworks to ensure the stringent requirements of the Tactile Internet. Examples of the works tackling this aspect are references [96] and [124], which propose resource allocation frameworks over mobile networks in coordination with a DNA and Wi-Fi networks, respectively. Another possible example of this research direction would be to introduce resource allocation frameworks over mobile networks in coordination with WiMAX in order to maximize the number of connected devices, while guaranteeing the stringent requirements of tactile devices. A preliminary study on the feasibility of using WiMAX for the Tactile Internet is presented in [41].

*2) Resource Allocation in WLAN/WBAN*

Thanks to their low cost, low power consumption, and wide deployment, WLAN and WBAN technologies have attracted considerable attention as an alternative wireless access technology for the Tactile Internet. As reported in [109], WLAN technologies can also achieve lower latency than LTE for short-range transmission. This makes WLAN technologies a suitable candidate, especially for some Tactile Internet applications that do not require high user mobility all of the time (e.g., e-health applications). In order to guarantee the Tactile Internet requirements, some resource allocation algorithms have been proposed for WLAN/WBAN technologies to reduce the latency, with promising results. It is worth noting that there are still some research directions that could enhance WLAN and WBAN technologies to efficiently support Tactile Internet applications.

**Joint resource allocation in both uplink and downlink directions:** As discussed in [127], the joint allocation of resources in both uplinks and downlinks is virtually essential for haptic communications to ensure the timely real-time exchange of control commends and haptic feedbacks. However, the studies on resource allocation for WLANs and WBANs have focused on either uplink- or downlink-only directions. Given that considering only one direction may neglect other contributors to overall latency, these studies may not be sufficient to fulfill the latency requirements of haptic communications. Therefore, one promising research direction is the joint design of resource allocation in both uplink and downlink directions for WLANs/WBANs. Toward this end, reference [140] can be used as a starting point, as it proposes a novel MAC scheme for the uplink direction in WLANs relying on an efficient polling-based mechanism to achieve low latency. This scheme could be extended to assign resources for both uplink and downlink transmission simultaneously while ensuring the achievable sum of uplink and downlink delays within 1 ms.

**Ultra-reliable resource allocation:** As highlighted in the lesson learned, there is still a lack of efficient mechanism to meet the stringent Tactile Internet reliability requirements in WLANs/WBANs. To achieve low latency and high reliability, most of the studies (e.g., [138][140][141]) rely on contention-free mechanisms to eliminate packet collision. However, packet loss can still occur due to the packet dropping caused by buffer overflow [139][143]. Therefore, an interesting research avenue is to design efficient buffer management mechanisms to be used in conjunction with the underlying channel access algorithm in order to meet the ultra-high reliability requirement without compromising the delay for the tactile traffic.

*3) Edge Resource Allocation*

Edge computing is one of the key enablers for the Tactile Internet. It pushes computation and storage resources to the edge of the network so that Tactile Internet applications can run close to end-users, or end-users can offload tasks to the more powerful computation units in their vicinity, thereby reducing the overall delay. However, due to the stringent latency and reliability requirements, resource allocation at the network edge remains a challenge for the Tactile Internet in case of the high user mobility. As indicated in the lessons learned, none of the studies has addressed mobility management while allocating resources at the edge for the Tactile Internet. To efficiently tackle this issue, appropriate task migration schemes are needed. These schemes will have to derive migration and resource allocation decisions for application tasks, i.e., application components, according to the movements of tactile devices, while meeting the Tactile Internet requirements. Reference [192] can be considered as a starting point. The authors propose a strategy that allows application tasks to migrate in fog systems according to the movement of devices. This approach could be extended to further consider Tactile Internet requirements.

*4) Routing*

D2D communication over wireless multi-hop networks is another potential solution for realizing Tactile Internet requirements [139]. It can avoid the need to transmit tactile data through congested core networks or through direct communication in wireless networks over long-distances, thereby improving the latency and reliability performance. However, there is still a lack of routing schemes capable of efficiently handling tactile traffic coexistent with conventional traffic in wireless multi-hop networks. Towards this end, two interesting research avenues are suggested: (i) prioritized and (ii) reservation-based D2D routing. In prioritized routing, QoS assurance methods can be used to guarantee a certain performance level (in terms of latency and reliability) for high-priority flows and yet provide enough flexibility to establish other low-priority requests on a given network topology with a limited capacity. In contrast, reservation-based routing mechanisms allow a predefined set of routes/resources to be reserved for flows with more stringent latency and reliability requirements.

## VIII. CONCLUSIONS

The Tactile Internet will completely revolutionize the way humans communicate once it becomes a reality. The main challenges are ultra-low latency and ultra-highly reliable communications. However, the Tactile Internet is still in its early stages. Research on all of its aspects is still needed to overcome its unique challenges. This paper provides a comprehensive survey of the solutions proposed to date in the literature on this topic. It first discusses illustrative use cases in three different application domains (remote robotic surgery, autonomous driving and remote phobia treatment) to define a set of evaluation criteria for the Tactile Internet. Based on this evaluation criteria, the solutions covering architectures, protocols, intelligent predication and algorithms were then critically reviewed. In addition, insights and lessoned from our comprehensive literature review were derived. Finally, we discuss the remaining challenges and promising research directions in this topic based on our lessons learned.

ACKNOWLEDGMENT

This work is partially funded by the Canadian Natural Sciences and Engineering Research Council (NSERC) and by Ericsson through a CRD grant, as well as by the Canada Research Chair program and Zayed University (United Arab Emirates) through the Research Incentive Fund (RIF) program.

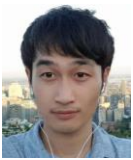
**Nattakorn Promwongsa** received his Master's degree in Telecommunications from Asian Institute of Technology, Pathumthani, Thailand in 2016 and obtained his Bachelor's degree in Electrical Engineering from Thammasat University, Pathumthani, Thailand in 2014. He is currently pursuing his Ph.D. degree in Information System Engineering at Concordia University, Montreal, Canada. His research interests include Tactile Internet, 5G and wireless communications.

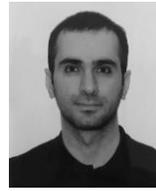
**Amin Ebrahimzadeh** received the B.Sc. and M.Sc. degrees in electrical engineering from the University of Tabriz, Iran, in 2009 and 2011, respectively, and the Ph.D. degree (Hons.) in telecommunications from the Institut National de la Recherche Scientifique (INRS), Montréal, QC, Canada, in 2019. From 2011 to 2015, he was with the Sahand University of Technology, Tabriz, Iran. He is currently a Horizon Post-Doctoral Fellow with Concordia University, Montréal, QC, Canada. His research interests include 6G, Tactile Internet, FiWi networks, and multi-access edge computing. He was a recipient of the doctoral research scholarship from the B2X program of Fonds de Recherche du Québec-Nature et Technologies (FRQNT).

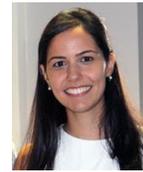
**Diala Naboulsi** received the M.S. degree in computer science from INSA Lyon, France, and the M.Eng. degree in telecommunications from Lebanese University, Lebanon, in 2012, and the Ph.D. degree in computer science from INSA Lyon in 2015. She was a Research Associate, a Postdoctoral Researcher, and a Course Lecturer with Concordia University, Canada. She was also a Visiting Researcher at Ericsson, Canada. She was a visiting Ph.D. student at the Politecnico di Torino, Italy, in 2013. She is currently a Research Professional with ÉTS, Canada, and a Course Lecturer with McGill University, Canada. Her research interests are in mobile networks, virtualized networks, and wireless networks.

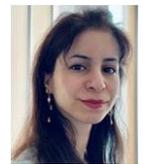
**Somayeh Kianpisheh** received the B.S. degree in software computer engineering from the University of Tehran, Iran, in 2004, and the M.S. and Ph.D. degrees in computer engineering from Tarbiat Modares University, Iran, in 2010 and 2016, respectively. Since 2018, she has been a Post-Doctoral Fellow with Concordia University, Canada. She has published several papers in the international journals/conferences and performed reviews for several international journals/conferences. Her research interests include distributed systems particularly resource allocation and performance modeling in 5G, fog/cloud systems, and content driven networks.

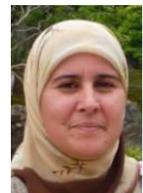
**Fatna Belqasmi** received the M.Sc. and Ph.D. degrees in electrical and computer engineering from Concordia University, Montreal, QC, Canada. She was a Research Associate with Concordia University and a Researcher with Ericsson Canada. She was a part of the IST Ambient Network Project (a research project sponsored by the European Commission within the Sixth Framework Programme—FP6). She was a Research and Development Engineer with Maroc Telecom, Morocco. She is currently an Associate Professor with Zayed University, Abu Dhabi, United Arab Emirates. Her research interests include next-generation networks, service engineering, distributed systems, and networking technologies for emerging economies.

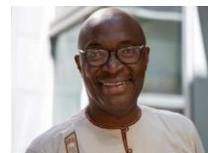
**Roch H. Glitho** (M'88–SM'97) received the M.Sc. degree in business economics from the University of Grenoble, France, the M.Sc. degree in pure mathematics and the M.Sc. degree in computer science from the University of Geneva, Switzerland, and the Ph.D. (Tech.Dr.) degree in informatics from the Royal Institute of Technology, Stockholm, Sweden. He is currently a Full Professor at Concordia University where he holds a Canada Research Chair. He also holds the Ericsson/ENCQOR-5G Senior Industrial Research Chair in Cloud and Edge Computing for 5G and Beyond. He has worked in industry and has held several senior technical positions (e.g., senior specialist, principal engineer, and expert) with Ericsson, Sweden and Canada. He has also served as an IEEE Distinguished Lecturer, and the Editor-in- Chief of the IEEE Communications Magazine and the IEEE Communications Surveys & Tutorials.

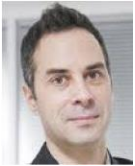

**Noel Crespi** received the master's degrees from the University of Orsay (Paris 11) and the University of Kent, U.K., the Diplome d'Ingenieur degree from Telecom ParisTech, and the Ph.D. and Habilitation degrees from Paris VI University (Paris-Sorbonne). Since 1993, he has been with CLIP, Bouygues Telecom, and then with Orange Labs in 1995. He took leading roles in the creation of new services with the successful conception and launch of Orange prepaid service, and in standardization (from rapporteurship of IN standard to coordination of all mobile standards activities for Orange). In 1999, he joined Nortel Networks as a Telephony Program Manager, architecting core network products for the EMEA region. In 2002, he joined the Institut Mines-Telecom and is currently a Professor and the Program Director, leading the Service Architecture Laboratory. He coordinates the standardization activities for the Institut MinesTelecom, ITU-T, ETSI, and 3GPP. He is an Adjunct Professor with KAIST, an Affiliate Professor with Concordia University, and a Guest Researcher with the University of Goettingen. He is also the Scientific Director of the French-Korean Laboratory ILLUMINE. His current research interests are in softwarization, data analysis, and Internet of Things/services.

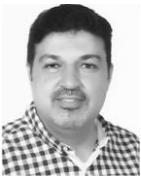

**Omar Alfandi** received the M.Sc. degree in telecommunication engineering from the University of Technology Kaiserslautern, Germany, in 2005, and the Ph.D. (Dr.rer.nat.) degree in computer engineering from the Georg-AugustUniversity of Goettingen, Germany, in 2009. From 2009 to 2011, he enjoyed a Postdoctoral Fellowship at the Telematics Research Group and he founded the Research and Education Sensor Lab, where he is currently as a Lab Advisor. He carried his Ph.D. Research as part of an Industry, Academia, and Research centers collaboration European Union (EU) project. He was a Package Leader of EU DAIDALOS II in the 6th Framework Project. In 2015, he was appointed as the Assistant Dean for Abu Dhabi Campus. He is currently an Associate Professor with the College of Technological Innovation, Zayed University. He is also the Co-Founder and the Co-Director of the Sensors and Mobile Applications Research and Education Lab, CTI. He published numerous articles on Authentication Framework for 4G Communication Systems, Future Internet, and Trust and Reputation Systems in Mobile ad hoc and Sensor Networks. His current research interests include the Internet of Things, security in next generation networks, smart technologies, security engineering, mobile, and wireless communications.